%% file: mnras_template.tex
\documentclass[fleqn,usenatbib]{mnras}
\usepackage{newtxtext,newtxmath}
\usepackage[T1]{fontenc}

\DeclareRobustCommand{\VAN}[3]{#2}
\let\VANthebibliography\thebibliography
\def\thebibliography{\DeclareRobustCommand{\VAN}[3]{##3}\VANthebibliography}
\newcommand{\referee}[1]{{\color{black} #1}}

\usepackage{graphicx}	% Including figure files
\usepackage{amsmath}	% Advanced maths commands
\usepackage{longtable} % for 'longtable' environment
\usepackage{pdflscape} % for 'landscape' environment
\usepackage{ragged2e}
\usepackage{tikz}
\usetikzlibrary{shapes, arrows,positioning} 

\title[Finding lenses in a FLASH]{FLASH: Faint Lenses from Associated Selection with {\it Herschel}}

\author[Tom Bakx et al.]{Tom J. L. C. Bakx$^{1,2,3}$\thanks{E-mail: tom.bakx@chalmers.se (Chalmers University)},
Bethany S. Gray$^{4}$,
Joaquin González-Nuevo$^{5,6}$,
Laura Bonavera$^{5,6}$, \newauthor
Aristeidis Amvrosiadis$^7$, 
Stephen Eales$^8$,
Masato Hagimoto$^2$, and
Stephen Serjeant$^9$
.
\\
$^{1}$ Department of Space, Earth, \& Environment, Chalmers University of Technology, Chalmersplatsen 4 412 96 Gothenburg, Sweden \\
$^{2}$ Department of Physics, Graduate School of Science, Nagoya University, Aichi 464-8602, Japan\\
$^{3}$ National Astronomical Observatory of Japan, 2-21-1, Osawa, Mitaka, Tokyo 181-8588, Japan.\\
$^{4}$ Department of Architecture \& Civil Engineering, University of Bath, Claverton Down, Bath, BA2 7AY, UK \\
$^{5}$ Departamento de Fisica, Universidad de Oviedo, C. Federico Garcia Lorca 18, E-33007 Oviedo, Spain\\
$^{6}$ Instituto Universitario de Ciencias y Tecnologas Espaciales de Asturias (ICTEA), C. Independencia 13, E-33004 Oviedo, Spain\\
$^{7}$ Institute for Computational Cosmology, Durham University, South Road, Durham DH1 3LE, UK \\
$^{8}$ School of Physics and Astronomy, Cardiff University, The Parade, Cardiff, CF24 3AA, UK\\
$^{9}$ School of Physical Sciences, The Open University, Milton Keynes, MK7 6AA, UK\\
}
\date{Accepted 2023 November 27. Received 2023 November 22; in original form 2023 September 18}

\pubyear{2023}

\begin{document}
\label{firstpage}
\pagerange{\pageref{firstpage}--\pageref{lastpage}}
\maketitle

\begin{abstract}
We report the ALMA Band 7 observations of 86 \textit{Herschel} sources that likely contain gravitationally-lensed galaxies. These sources are selected with relatively faint 500~$\mu$m flux densities between 15 to 85~mJy in an effort to characterize the effect of lensing across the entire million-source {\it Herschel} catalogue. These lensed candidates were identified by their close proximity to bright galaxies in the near-infrared VISTA Kilo-Degree Infrared Galaxy Survey (VIKING) survey. Our high-resolution observations (0.15~arcsec) confirm 47 per cent of the initial candidates as gravitational lenses, while lensing cannot be excluded across the remaining sample. %relative to the 76~per cent of sources predicted to be lensed based on galaxy evolution models. 
We find average lensing masses ($\log M / M_{\odot} = 12.9 \pm 0.5 $) in line with previous experiments, although direct observations might struggle to identify the most massive foreground lenses across the remaining 53~per cent of the sample, particularly for lenses with larger Einstein radii. Our observations confirm previous indications that more lenses exist at low flux densities than expected from strong galaxy-galaxy lensing models alone, where the excess is likely due to additional contributions of cluster lenses and weak lensing. If we apply our method across the total 660~sqr. deg. H-ATLAS field, it would allow us to robustly identify 3000 gravitational lenses across the 660~square degree {\it Herschel} ATLAS fields.
\end{abstract}

\begin{keywords}
general
– galaxies: evolution
- submillimetre: galaxies
– galaxies: high-redshift
- gravitational lensing: strong
\end{keywords}

%%%%%%%%%%%%%%%%%%%%%%%%%%%%%%%%%%%%%%%%%%%%%%%%%%

%%%%%%%%%%%%%%%%% BODY OF PAPER %%%%%%%%%%%%%%%%%%

\section{Introduction}
% Matter tells space-time how to curve, and curved space-time tells matter (and light) how to move \citep{Wheeler1990}. As a result, c
Concentrated mass distributions, such as stars \citep{Dyson1920, Kelly2018,Welch2022}, galaxies \citep{Treu2010} and galaxy clusters \citep{Kneib2011, GN2012,GN2017,Bonavera2019,Crespo2022,Fernandez2022} can redirect light, extending the number of sightlines onto an object resulting in so-called gravitational lensing. Particularly in the case of strong gravitational lensing, defined as a magnification $\mu > 2$, these cases can offer a significant increase in spatial and observational sensitivity. This effect is determined by the foreground distribution of matter, and can thus provide a constraint on the mass distribution of our Universe \citep[][]{Kochanek1992,Kochanek1996,Grillo2008,Oguri2012,eales2015}.

Especially given the low angular resolution of sub-mm observations, the increase in angular resolution by gravitational lensing resulted in spectacular images of dust-obscured star-formation at cosmic noon \citep{SV2015, dye2015,Rybak2015,Tamura2015}.
Initial observations in the late 1990's had revealed a population of dust-obscured galaxies rivaling the total galaxy evolution seen in optical wavelengths \citep{Smail1997,Hughes1998,Ivison1998}. 
The brightest dusty star-forming galaxies (DSFGs) have observed star-formation rates in excess of 1000 M$_{\odot}$/yr, resulting in an unsustainable evolutionary phase through violent star-formation feedback \citep{Andrews2011,RowanRobinson2016}. 
The evolutionary pathway of these star-forming systems is still not adequately understood, as demonstrated by the prevalence estimates from DSFG models, which often predict three to four orders of magnitudes below what is observed \citep{Baugh2005}.
Because these galaxies are very rare (a few / deg$^2$), hydrodynamical models struggle to include enough volume to simulate these galaxies accurately in order to test the evolutionary pathways of these DSFGs \citep[e.g.,][]{narayanan2015}. As a consequence, the best path to understanding dusty star-forming galaxies is through direct observations of complete samples. 
Gravitational lensing offers an opportunity to study these DSFGs at high resolution. Meanwhile, observations to date have revealed a large source-to-source variation, with some sources showing stable rotation \citep{Dye2018,Rizzo2020}, while other sources appear to be in a state of rapid collapse (e.g., SDP.81; \citealt{dye2015,Rybak2015MNRAS.451L..40R,Tamura2015}). In order to capture this large variation of sources, large samples ($> 100$) of lensed DSFGs are needed to characterize the evolutionary pathway(s) of these extreme star-forming systems.

% Gravitational lensing is an a-chromatic phenomena that relies on the alignment between fore- and background galaxies. 
Although lensing is a rare phenomena, large area surveys in sub-mm and mm revealed a large population of ultra-bright sources, that upon further inspection were revealed to be gravitationally-lensed \citep{negrello2010,Negrello2014,negrello2017,Vieira2013}. %the large-area surveys at mm/submm have offered a unique possibility to select towards these unique cases. 
The steep bright-end of the luminosity function (i.e., brighter sources are increasingly rare; \citealt{Lapi2011}) means that the unlikely gravitational lensing magnification of fainter but more numerous sources are statistically-preferred to observing non-lensed intrinsically hyper-luminous sources.
As a result, in the sub-mm domain, the wide-field H-ATLAS \citep{eales2010} and HerMES \citep{oliver2012} surveys with the {\it Herschel Space Observatory} have revealed a population of dusty lensed sources by selecting sources at $S_{500} > 100$~mJy \citep{negrello2010}. 
Similarly, the large-area nature of CMB-studies with ground- and space-based telescopes means that these surveys are also well-suited towards lens selection, with the all-sky {\it Planck} survey showcasing exceptional lensing morphologies \citep{Kamieneski2023}, and perfectly-circular Einstein rings shown in the South Pole Telescope survey — revealed in high-resolution with ALMA and JWST \citep{Spilker2016, Rizzo2020}. 
Finally, while the mapping speed of ground-based observations at submm wavelengths is limited by the atmospheric transmission, the recent large-area SCUBA-2 Large eXtragalactic Survey (S2XLS) is bridging the border between lenses and intrinsically-bright sources \citep{Garratt2023}.
The large beam-width of these selection techniques, however, mean that only time-expensive follow-up observations of these sources can reveal the true nature of these galaxies \citep{bussmann13,Bussmann2015,Spilker2016} — and worse yet, the intrinsic properties of the sample as a whole \citep{Gruppioni2013}.

One way to circumvent these limitations is by a search for the foreground lensing systems at complementary wavelengths such as optical or near-infrared. These foreground galaxies might be detected in optical (SDSS; \citealt{GN2012,GN2017,GN2019,bourne2016}) or near-infrared (NIR; e.g., VIKING; \citealt{Fleuren2012,bakx2020,Ward2022}) surveys, while the dusty nature of these DSFGs mean the background galaxies are likely not detected in optical/NIR surveys. 
These tests vary in their sophistication, with several models simply identifying nearby foreground galaxies \citep{Roseboom2010,negrello2010}, to innovative mathematical techniques \citep{Fleuren2012,bourne2016} and statistical correlations accounting for redshifts and spatial distributions \citep{GN2019}, even including the additional spatial offsets due to gravitational lensing \citep{bakx2020}.

These methods can statistically characterize the prevalence of lensed sources across the full extent of the {\it Herschel} sample — near one million dusty sources \citep{valiante2016,furlanetto2018,maddox2018,Shirley2021,Ward2022} — however, they have not been tested experimentally. 
The best way of resolving gravitational lensing directly is through resolved sub-mm observations to reveal the lensing structures \citep{Spilker2016,Amvrosiadis2018,Dye2018,Kamieneski2023}. %which can be extended to cosmological studies of the matter distribution and to other surveys such as EUCLID and LSST.
While the easiest lenses to identify are at the brightest flux densities ($> 100$~mJy at 500~micron \citealt{negrello2010,Negrello2014,negrello2017}), both in terms of their apparent brightness and their likelihood to be lensed, the bulk of the {\it Herschel} population -- and thus also the lenses -- reside at the lower flux densities ($20$~mJy~$> S_{500} > 40$~mJy). A thorough test of the fidelity of a lens-selection method should thus focus on these low flux-density sources.

In this paper, we report on the observation of 86 galaxies selected using a method based on a VIKING (+ KiDS)-based analysis from \cite{bakx2020}.  In Section~\ref{sec:sec2}, we describe the selection method. Section~\ref{sec:sec3} details the observations, and Section~\ref{sec:sec4} describes the implications of this survey on lenses within {\it Herschel} samples. We conclude in Section~\ref{sec:conclusions}. Throughout this paper, we assume a flat $\Lambda$-CDM cosmology with the best-fit parameters derived from the \textit{Planck} results \citep{planck2018}, which are $\Omega_\mathrm{m} = 0.315$, $\Omega_\mathrm{\Lambda} = 0.685$ and $h = 0.674$.

\section{Lens identification and sample selection}
\label{sec:sec2}
In this section, we describe the method for finding lenses through combined near- and far-infrared surveys, as well as the selected galaxies for this pilot survey.

\subsection{Lens identification}
The lens identifying method is based on finding a VIKING galaxy which is statistically likely to be associated with a {\it Herschel} source. As a further requirement, the presumed foreground source should be at a lower redshift than the estimated redshift of the submm source -- estimated from the sub-mm colours of the source \citep{pearson13, bakx18} -- and is therefore highly likely to be a lens \citep{bakx2020}. 
The standard statistical way of finding associated sources is by finding galaxies close enough to the {\it Herschel} positions that they are unlikely to be there by chance \citep{bourne2016}. 
Our new method relies on the fact that most high-redshift ($z > 2$) submillimetre galaxies are not bright enough to be detected on wide-area optical and NIR surveys such as SDSS and VIKING \citep{Wright2019}, and so any galaxy that is close to the {\it Herschel} position on these images could likely be the lens \citep{GN2017}, although a subset of SMGs may not be NIR-faint \citep{GN2012}. 
The statistical tool used for identifying lenses is called the likelihood estimator \citep{Sutherland1992}. 
This likelihood estimator calculates an individual so-called likelihood of an association of a NIR source close to a {\it Herschel}-source given the magnitude distribution of the NIR magnitude and spatial offset. The likelihood reflects how unlikely each NIR-{\it Herschel} couple is through a measure of how many similar fields one would need to see before encountering a single one of these associations. The likelihood, $L$, of lensing candidates is often in excess of several hundreds or thousands, and is calculated as follows: \referee{
\begin{equation}
    L = \frac{q(m) f(r)}{n(m)}.
\end{equation}
In this equation, $q(m)$ represents the probability distribution of genuine counterparts at a magnitude $m$, $n(m)$ represents the background surface density distribution of unrelated objects (in units of arcsec$^{-2}$), and $f(r)$ represents the distribution of offsets between sub-mm and near-IR positions produced by both positional errors between both catalogues and gravitational lensing offsets (in units of arcsec$^{-2}$; see \citealt{bakx2020}).

In order to arrive at a single probability -- here called the reliability ($R_j$) -- the likelihoods of each nearby NIR source are added together, and include the possibility of the foreground source being too faint to be detected in the VIKING survey (i.e., $Q_0 = 0.82$; \citealt{Fleuren2012,bourne2016,bakx2020}):
\begin{equation}
    R_j = \frac{L_j}{\sum_i L_i + (1 - Q_0)}.
\end{equation}
In this equation, the reliability of each potential match, $j$, is calculated from the sum of the likelihoods of all nearby matches ($\sum_i L_i$) and the possibility that the foreground source is too faint to be detected.}
\cite{bakx2020} have found that the SDSS also misses about half the lenses, so an essential part of our method is that the search for the lens is carried out in the $K_S$-band in the VIKING survey, where the lens (often a massive ‘red-and-dead’ elliptical) is virtually always bright enough to be detected \citep{bakx2020}. 
On top of this, unlike other methods for finding associated sources, we calibrated our statistical estimator on a sample of gravitationally-lensed galaxies: \referee{ the {\it Herschel Bright Sources} (HerBS; \citealt{bakx18,Bakx2020Erratum})}. This resulted in the insight that the angular distribution of lensed galaxies is not simply described by a Gaussian distribution, but instead requires a distribution that accounts for an additional offset due to galaxy-galaxy and galaxy-cluster lenses. 
\referee{The likelihood estimator does introduce biases in the types of lenses we can find. The VIKING survey is deep enough to detect most lensing galaxies ($\rm M_* > 10^8~M_{\odot}$). Meanwhile, the radial probability distribution is non-zero out to $\approx 10$~arcseconds, allowing us to find most cases of galaxy-galaxy lensing ($< 10$~arcsec; \citealt{Amvrosiadis2018}). We do note, however, that the method becomes increasingly less sensitive towards larger offsets between the lensed galaxy and the deflector. }

\subsection{The FLASH sample: Faint lenses found through Associated Selection from Herschel}
\begin{figure}
    \centering
    \includegraphics[width=\linewidth]{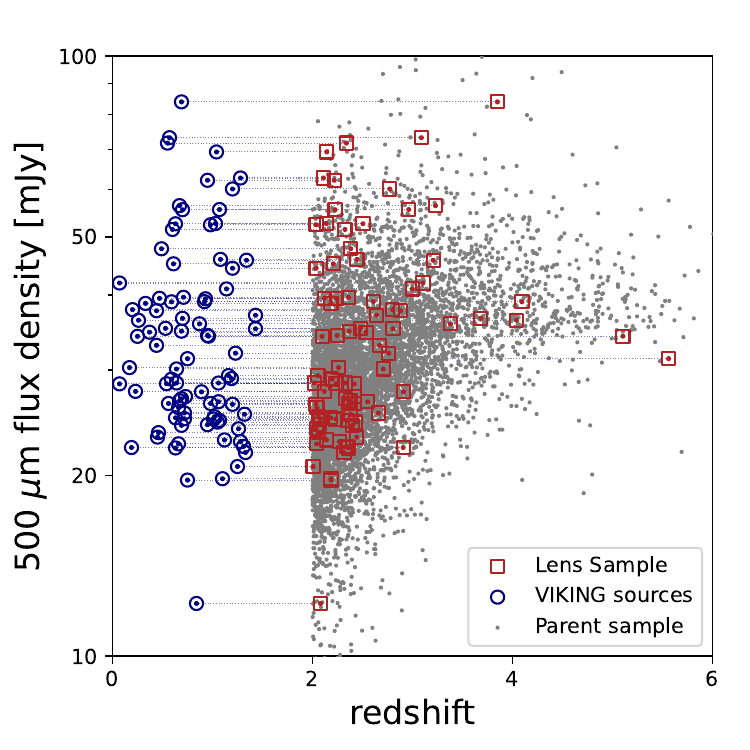}
    \caption{The photometric redshift distribution of the foreground ({\it blue circles}) VIKING and background ({\it red squares}) {\it Herschel} sources. The $z > 2$ {\it Herschel} sources from the equatorial GAMA 12 field are indicated in {\it grey points}. We link the associated points together with a {\it blue dashed line}. The redshift difference between the sources provides confidence in being different galaxies (or not the same galaxy) as the {\it Herschel} sources. }
    \label{fig:fluxdistribution}
\end{figure}

This survey is based on the H-ATLAS 12-hour equatorial field \citep{eales2010,valiante2016}, which has good coverage with the VIKING NIR and the optical KIDS survey. 
This field contains 35\,512 {\it Herschel} sources. 
The radio NRAO VLA Sky Survey (NVSS) survey was used to remove blazars from the sample. A total of 350 H-ATLAS sources fall within 10 arcseconds of an NVSS source, i.e., within the typical combined angular precision of {\it Herschel} ($\sim 2$~arcsec) and NVSS ($\sim 7$~arcsec). Here we note that this step could also remove bright DSFGs in our sample, which is not an important drawback, since we are mostly interested in the fainter DSFGs, and here choose sample purity over completeness. 
The photometric redshifts of the H-ATLAS sources is then estimated by fitting a modified black-body SED to the 250, 350 and 500~$\mu$m fluxes \citep{pearson13,bakx18}. 
Subsequently, we impose a redshift cut for all H-ATLAS sources, demanding $z_{\rm phot} > 2.0$. 
% This method also produces an error on the photometric redshift. 
These sources were then passed through the counterpart analysis of \cite{bakx2020}, which identifies counterparts on the VIKING $K_S$-band images that are likely to be statistically associated with the {\it Herschel} sources. This uses a standard likelihood estimator \citep{Sutherland1992,Fleuren2012,bourne2016,bakx2020}, which provides a probability for a VIKING counterpart to be genuine. A total of 7\,362 GAMA-12 {\it Herschel} sources have nearby VIKING counterparts, no nearby NVSS radio sources, and lie above $z_{\rm phot} > 2$. 
\cite{Wright2019} have used the nine photometric bands of VIKING and KIDS to produce photometric redshifts and stellar masses for the objects detected in the surveys. 
By comparing these redshift estimates and their errors, in combination with the photometric redshifts for the sub-mm sources and their errors (assuming $\Delta z = 0.13 (1+z)$; \citealt{pearson13}), we identify the systems for which there was only a 0.1~per cent chance ($\sim 3.1 \sigma$) that the {\it Herschel} source and the potential counterpart are actually at the same redshift, for a total of 6\,823 sources.
In order to test the evolution of the lensing probability with {\it Herschel} flux density, we identify the most likely sources to be gravitationally-lensed across a wide 500~\micron{} flux density range, selecting towards the highest reliabilities within each 10~mJy flux density region \referee{that can be observed in a single observation by ALMA. In order to efficiently observe these sources, we require each source to be within ten degrees from a single phase-centre to observe all targets within a minimum number of Scheduling Blocks.}
Above 40~mJy, fewer sources could be found in a single field that would be reliable candidates for gravitational lensing, and as a result most of the sources have $S_{500}$ between 20 and 40~mJy. \referee{As a consequence of the dearth of likely-lensed sources at the higher fluxes, several sources have stand-out properties such as large angular separations or lower reliabilities.}
We list the catalogue in Table~\ref{tab:flashcatalogue}, and show the redshift against the 500~\micron{} flux in Figure~\ref{fig:fluxdistribution}. The sources are sorted from lowest $S_{500}$ to highest.

\subsection{The statistics of the FLASH selection}
Although these sources are the most likely gravitationally-lensed candidates with $S_{500} = 10 - 90$~mJy, with very high individual probabilities, the large parent sample implies that there is a possibility for chance encounters. In an example as to why this is the case: even if there were no true lensing candidates, the size of our sample would by chance pick up sources as lensing candidates. 
\begin{figure}
    \centering
    \includegraphics[width=0.7\linewidth]{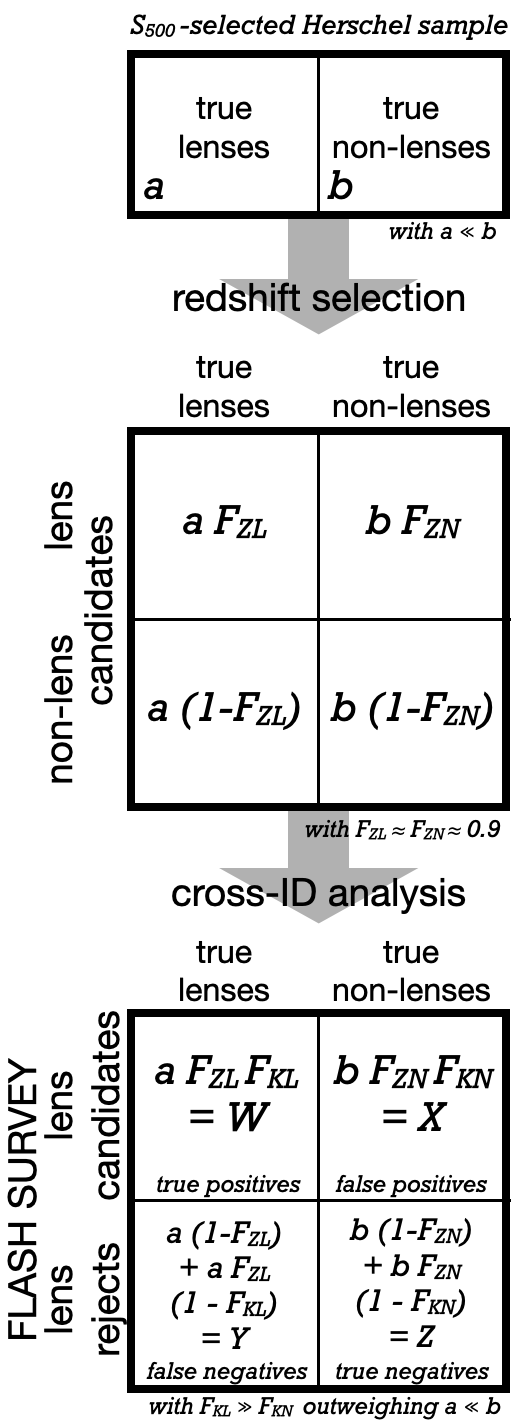}
    \caption{A schematic overview of the FLASH source selection provides insight in the total number of expected lenses in the FLASH sample and the total number of lenses in the {\it Herschel} sample. True lenses and non-lenses are initially classified by a redshift cut ($z > 2$), followed by the cross-identification analysis \citep{bakx2020} and a $3.1 \sigma$ redshift difference between VIKING and {\it Herschel} sources. The subsequent confusion matrix contains a measure for the true positives, the additional false positives, and the completeness of the FLASH method.}
    \label{fig:flowchart}
\end{figure}
Figure~\ref{fig:flowchart} shows a schematic overview of the FLASH source selection, starting from a perfect understanding of lensed ($a$) and non-lensed sources ($b$). The photometric selection ($z_{\rm phot} > 2$) reduces the fraction of lensed and non-lensed sources with $F_{ZL}$ and $F_{ZN}$, respectively. The subsequent cross-identification and removal of sources that are likely at the same redshift reduces the fraction by an additional $F_{KL}$ and $F_{KN}$, respectively. The sources can then be placed in four different categories according to a confusion matrix proportional to 
\begin{eqnarray*}
   {\rm true\ positives}\ \ \  =& W \ \ \ =& a F_{ZL} F_{KL}, \\
   {\rm false\ positives}\ \ \  =& X\ \ \  =& b F_{ZN} F_{KN}, \\
   {\rm false\ negatives}\ \ \  =& Y\ \ \  =& a (1 - F_{ZL} F_{KL}), {\rm and} \\
   {\rm true\ negatives}\ \ \  =& Z \ \ \ =& b (1 - F_{ZN} F_{KN}). 
\end{eqnarray*}
The objective of these ALMA observations is to identify the true positive sources, $W$, although it is not clear how many false positives are included in the FLASH selection. 
The ALMA observations will be able to identify the lens candidates from the sample, and provide a measure of $f_{\rm ALMA} = W / (W+X)$. 
Rewriting this equation, we find
\begin{equation}
    f_{\rm ALMA} = \frac{a}{a+Q^{-1}b}, \label{eq:lensprobability}
\end{equation}
with $Q$ a representative of the quality of our FLASH selection,
\begin{equation}
Q = \frac{F_{ZL}F_{KL}}{F_{ZN}F_{KN}}.  \label{eq:qvalue}
\end{equation}
This allows us to find a rough estimate for the number of lenses we can expect to detect with these ALMA observations.
The lensing fraction of sources decreases dramatically with decreasing flux density, which thus drives up the false-positive fraction at the lower fluxes. For instance, the lensing fraction predicted from cosmological models \citep{Cai2013} suggest that only one to two per cent of $z > 2$ sub-mm sources is lensed at $S_{500} = 30$~mJy, i.e., $a / (a+b) = 0.01 - 0.02$.
The selection effects of the cross-identification is roughly $F_{KL} = 0.82$ \citep{bakx2020}, under the assumption that there are no systematic differences between the brighter lens samples (i.e., $S_{500} > 80$~mJy) and the bulk of lenses in {\it Herschel}. This is in line with a brief comparison of the equatorial sources reported in both \cite{negrello2017} and \cite{bakx2020}: for eight sources in the lensed sample of \cite{negrello2017}, seven are strong lensing candidates in \cite{bakx2020}, or a $F'_{KL} = 0.88 \pm 0.12$.
The fraction of lower-redshift lenses excluded ($z < 2$) can be estimated from the fraction of such sources documented in \cite{negrello2017}. They report fifteen out of 80 lens candidates to have photometric redshifts below 2, or $F_{ZL} = 0.82$. By comparing the photometric redshifts of the {\it Herschel} catalogues \citep{valiante2016,furlanetto2018}, we find that around 46~per cent of sources in a flux-limited ($S_{500} > 15$~mJy) sample lie below $z < 2$; i.e., $F_{ZN} = 0.54$. 
%There would be scenarios where the VIKING and sub-mm source might be closer to one-another than $3.1 \sigma$ in error.
The main uncertainty of the method lies in our ability to remove false positives through the cross-identification analysis \citep{bakx2020}. As a lower limit, we can use the fact that {\it Herschel} and VIKING sources are excluded to be at the same redshift by $F_{KN} < 0.001$, although there are no direct measurements of $F_{KN}$ possible without observations. For an intrinsic lensing fraction of $a / (a+b) = 0.01$, we can expect a high lensing fraction of $f_{\rm ALMA} = 0.9$, and a $Q$-value of 1250. 
We compare the sources in our sample against the lensing fraction predicted from cosmological models \citep{Cai2013}, and find a total of 82 out of 86 sources are likely lensed based on the above predictions.

The method further provides insight in the total number of lenses in the {\it Herschel} samples. The completeness of the sample, $C$, is equal to the number of lenses our method is able to identify among all true lenses,
\begin{equation}
    C = \frac{W}{W+Y} = F_{KL}F_{ZL}. \label{eq:completeness}
\end{equation}
% In total, this additional cut reduced the number of sources in the FLASH selection from 7362 to 6823, although this would include false positives ($X$). 
As a result, an initial estimation of the completeness of the FLASH selection is around $C \approx 0.82 \times{} 0.8175 \approx 0.7 \pm 0.1$.

\include{tableSample}

\begin{twocolumn}

Figure~\ref{fig:distribution_of_properties} shows the distribution of properties of the FLASH sources. 
Unlike previous studies of gravitational lenses that focus on $S_{500} > 100$~mJy (\citealt{negrello2017}, or equivalent fluxes at longer wavelengths; \citealt{Vieira2013,Spilker2016,Harrington2021,Kamieneski2023}), this survey selects relatively low 500~\micron{} fluxes. 
The reliability, or the statistical association of the {\it Herschel} sources to VIKING galaxies, of these sources is high due to a combination of low angular separation and bright foreground galaxy selection. The angular separation is on the order of the typical astrometric uncertainties (e.g., \citealt{valiante2016}).
In fact, most association probabilities of FLASH sources are above the 99$^{\rm th}$ percentile. 
The foreground VIKING sources have information from both VISTA (Z, Y, J, H, and K$_S$) and from the KiDS survey (u, g, r, and i). Spectral fitting \citep{Wright2019} provides a stellar mass estimate of the foreground objects, which suggests massive ($M_{*} > 10^{10} M_{\odot}$) galaxy systems at lower redshifts.
\referee{As previously mentioned, the sample is drawn from a relatively small patch of sky to facilitate efficient ALMA observations. As a consequence}, each distribution appears to have one to three outliers source with a large separations ($\approx 6$"; FLASH-39, -83 and -84), a low reliability ($R \approx 0.8$; FLASH-79) or a low stellar mass ($M_{*} \approx 10^{7.8} M_{\odot}$; FLASH-67), although each such source is different, and do not suggest impurity of the sample. 

\begin{figure}
    \centering
    \includegraphics[width=\linewidth]{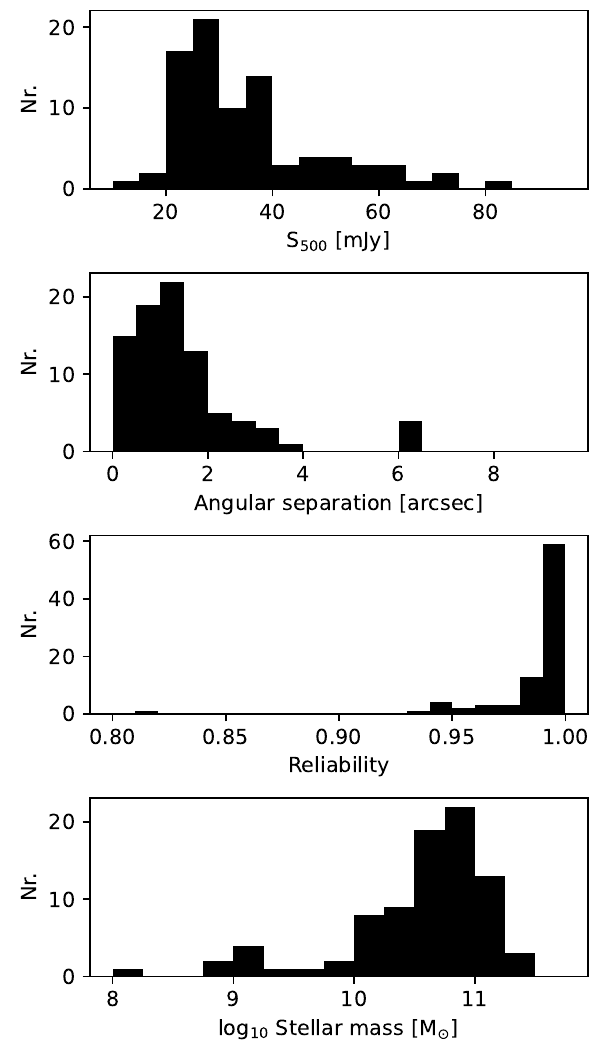}
    \caption{The FLASH sources are selected with {\it faint} 500~\micron{} flux densities, relative to the bulk of the lenses found by {\it Herschel} and other lensing surveys (i.e., the expected or observed $S_{500}$ of {\it Planck}, SPT, ACT are $> 100$~mJy). 
    They are found close to bright VIKING sources, with most {\it Herschel} and VIKING sources only 2 arcseconds removed from on the sky, within the typical astrometric uncertainties.
    As a result of the close location on the sky, the FLASH sources have high reliabilities (i.e., association probabilities) for each {\it Herschel} and VIKING-association, with most association probabilities above the 99$^{\rm th}$ percentile. 
    The foreground VIKING sources have information from both VISTA (Z, Y, J, H, and K$_S$) and from the KiDS survey (u, g, r, and i). Spectral fitting \citep{Wright2019} provides a stellar mass estimate of the foreground objects, which suggests massive ($M_{*} > 10^{10} M_{\odot}$) galaxy systems at lower redshifts.
    Interestingly, each distribution appears to have one to three {\it straggling} sources with a large separations ($\approx 6$"; FLASH-39, -83 and -84), a low reliability ($R \approx 0.8$; FLASH-79) or a low stellar mass ($M_{*} \approx 10^{7.8} M_{\odot}$; FLASH-67), although each such source is different, and it does not suggest impurity of the sample. 
    }
    \label{fig:distribution_of_properties}
\end{figure}

\section{ALMA observations, reduction and results}
\label{sec:sec3}
\subsection{ALMA observations and reduction}
We observe using Band~7 continuum observations to test whether these sources are actually lensed. The observation depth is based on a Cycle 2 ALMA program of 16 bright {\it Herschel} sources that showed that even short (2 minute) continuum observations were enough to reveal the lensed structure with enough signal-to-noise and resolution for a full lensing reconstruction \citep{Amvrosiadis2018,Dye2018}. In this study (2019.1.01784.S; P.I. Bakx), we have used the same resolution ($\sim 0.15$~arcsec) but scaled the integration times to allow for the fainter flux densities of the sources by 50~per cent deeper observations (see Table~\ref{tab:observationdetails}). The quasars J1058+0133 and J1256-0547 were used as bandpass calibrators, and quasars J1217-0029 and J1135-0428 were used as complex gain calibrators.

\begin{table*}
    \centering
    \caption{Parameters of the ALMA observations}
    \label{tab:observationdetails}
    \begin{tabular}{cccccc} \hline
UTC start time          & Baseline length & N$_{\rm ant}$ &  Frequency & T$_{\rm int}$ & PWV \\
$[$YYYY-MM-DD hh:mm:ss$]$  & [m] 			 & 			 	 &  [GHz]     & [min] &  [mm] \\ \hline
% \multicolumn{6}{c}{\textbf{Project 2019.1.01784.S}} \\
2021-05-10 03:08:31 & 14 -- 2492    & 44 &  343.484  &  49.0     &  0.96 \\
2021-05-16 03:21:37 & 14 -- 2517   &  47 &  343.484  &   48.8    &    0.96    \\
2021-05-17 00:05:58 & 14 -- 2517   & 47  &  343.484  &  48.9     &   0.65    \\
2021-05-17 01:55:18 & 14 -- 2517     & 48  &  343.484  &  49.0     &    0.64   \\
2021-05-17 02:43:09 & 14 -- 2517   &  48 &  343.484  & 21.8      & 0.45 \\
2021-05-18 00:13:21 & 14 -- 2517    & 49   & 343.484   & 42.3 & 0.62 \\
2021-05-18 04:03:52 & 14 -- 2517    & 49   & 343.484   & 42.3 & 0.39 \\ \hline
    \end{tabular}
\end{table*}

Data reduction was performed following the standard procedure and using the ALMA pipeline. Then, we use \textsc{CASA} for imaging the {\it uv}-visibilities using Briggs weighting with a robust parameter of 2.0 (to maximize the depth of the observations at the expense of slightly increasing the final synthesized beam size). The resulting beam size is 0\farcs18 by 0\farcs14 with a beam angle of $-71$ degrees at a continuum depth of 72~$\mu$Jy/beam. 

In order to test the effect of resolved observations and to facilitate aperture extraction, we also generate images with a tapering of 0.5~arcseconds. The resulting continuum maps have a beam size of 0\farcs60 by 0\farcs56 at the same beam angle of $-71$ degrees at a continuum depth of 137~$\mu$Jy/beam.

\subsection{ALMA photometry}
The source fluxes are extracted from the tapered image using the \textsc{CASA IMFIT} routine. For each source, the routine is repeated until no obvious sources exist in the residual image ($> 3 \sigma$). The resulting positions and fluxes are shown in Table~\ref{tab:catResults}. For significantly sources or sources where the lensing causes the emission to be spread across multiple components, we mention the individual extracted positions, as well as the combined flux of the source. We indicate these sources with italics. 

The resulting images are shown in Figure~\ref{fig:sources1}. These images, whose identification and FLASH-number are listed at the top, show the VIKING image in the background, with foreground contours from the high-resolution (white contours) and tapered (solid black contours). Inset images provide the high-resolution (0.15~arcsec resolution, red contours) for sources where zoom-ins are necessary. The images are centered on the VIKING position (which is also the ALMA phase center), and the orange cross indicates the {\it Herschel} position from \cite{valiante2016}. The red squares indicate the individual positions of the extracted fluxes.

\begin{figure*}
    \centering
\includegraphics[width=0.95\linewidth]{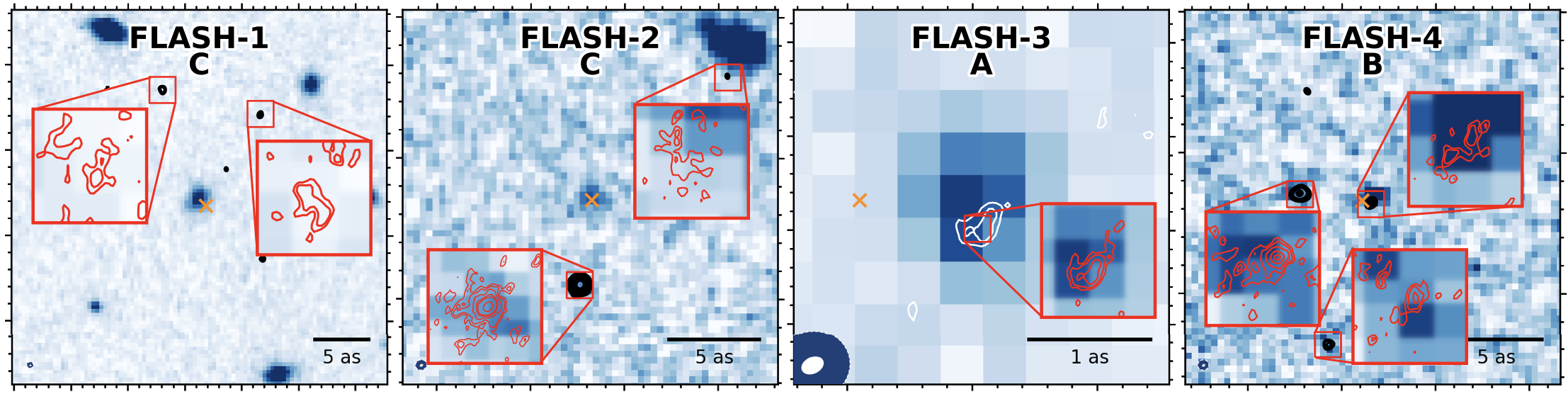}
\includegraphics[width=0.95\linewidth]{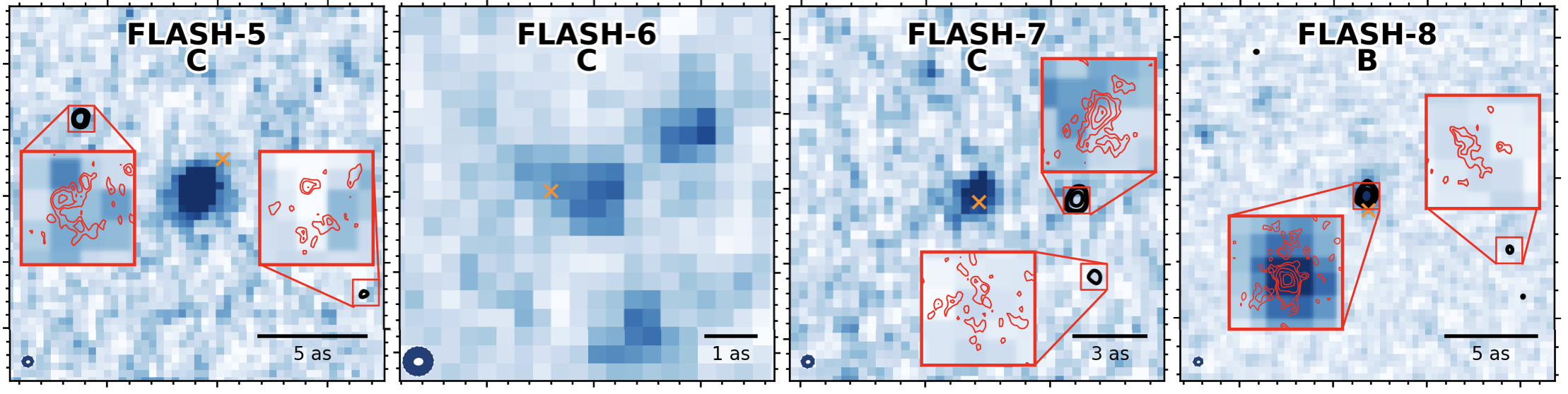}
\includegraphics[width=0.95\linewidth]{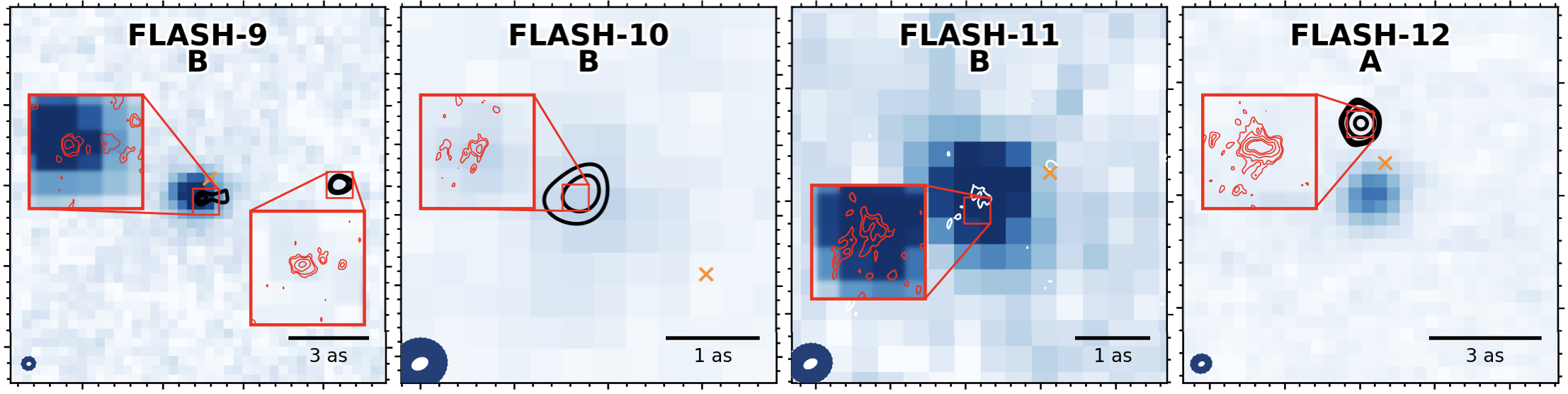}
\includegraphics[width=0.95\linewidth]{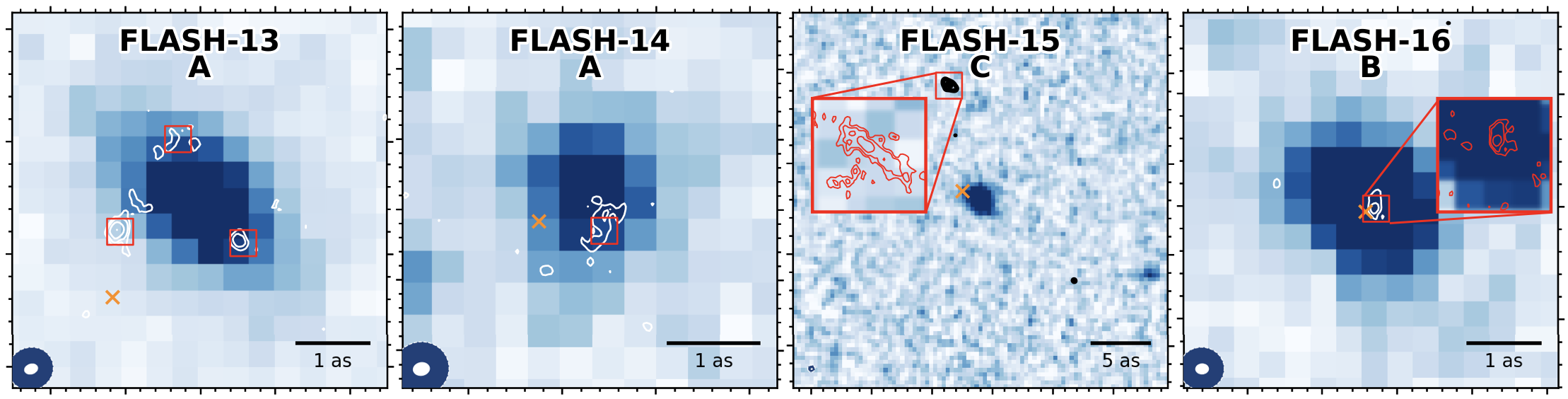}
    \caption{The VIKING images of the FLASH sources are shown in the {\it blue} background, with their FLASH-number and identification listed in the top of the figure. Contours in the central panels show either the robust parameter $= 2$ maps (\textit{white contours}; beam $\approx 0.15$~arcsec) or the tapered data (\textit{black contours}; beam $\approx 0.55$~arcsec). The contours are drawn at $3, 5, 8, 10$ and $20 \sigma$. The beams are shown in the lower-left of the panels, and the images are scaled to include all ALMA-identified galaxies. The angular scale is shown in the lower-right of each figure in units of arcseconds. The extraction positions of the sources in Table~\ref{tab:catResults} are indicated with {\it red boxes}, and where applicable, we provide insets of each source using red contours on the scale to capture the entire emission. In order to boost the fidelity of these insets, we lower the contour levels to $2, 3, 5, 10$~ and $20 \sigma$. The images are centered on the VIKING-position, and the {\it orange cross} indicates the {\it Herschel} position. The FLASH numbering is sorted by increasing 500~\micron{} flux, $S_{500}$. }
    \label{fig:sources1}
\end{figure*}\addtocounter{figure}{-1}
\begin{figure*}
    \centering
\includegraphics[width=0.95\linewidth]{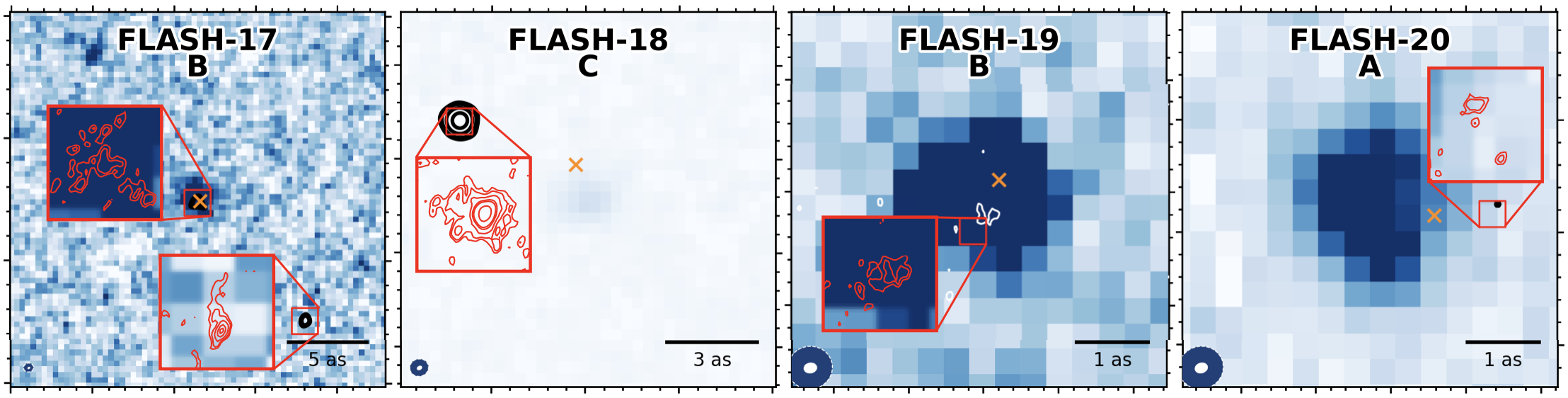}
\includegraphics[width=0.95\linewidth]{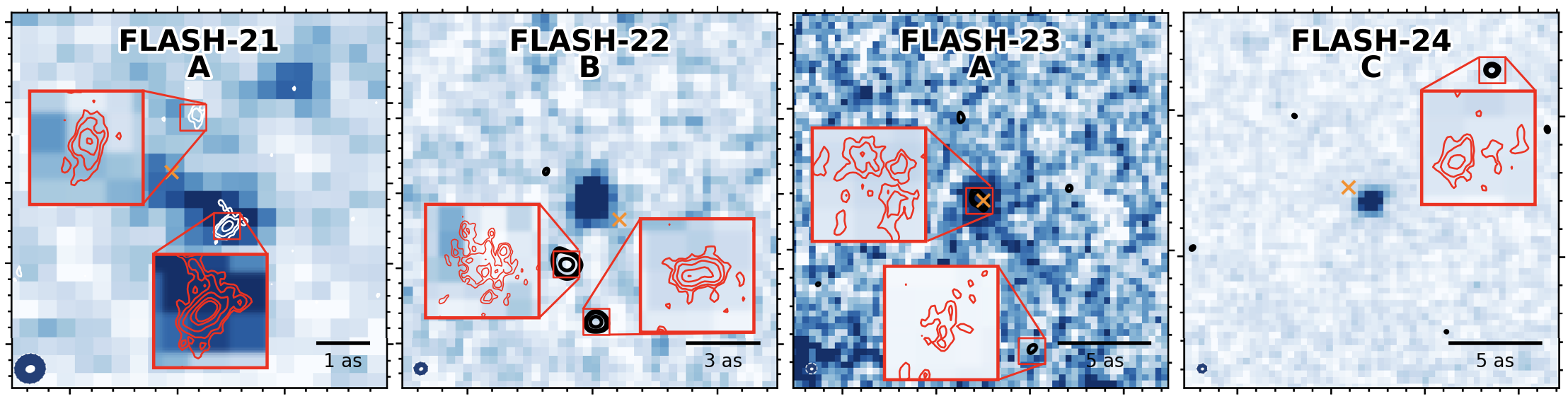}
    \includegraphics[width=0.95\linewidth]{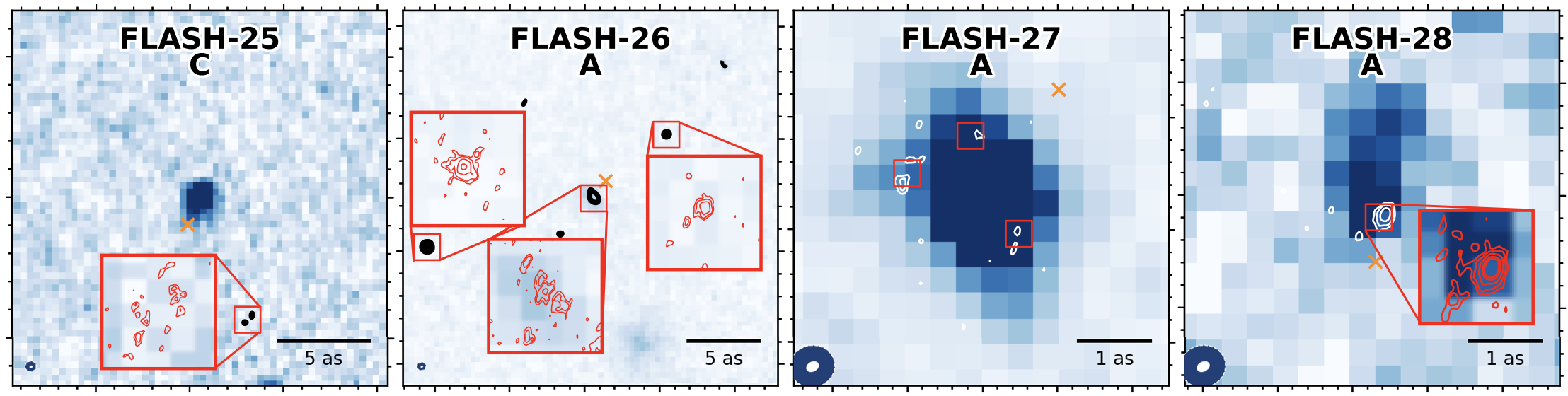}
\includegraphics[width=0.95\linewidth]{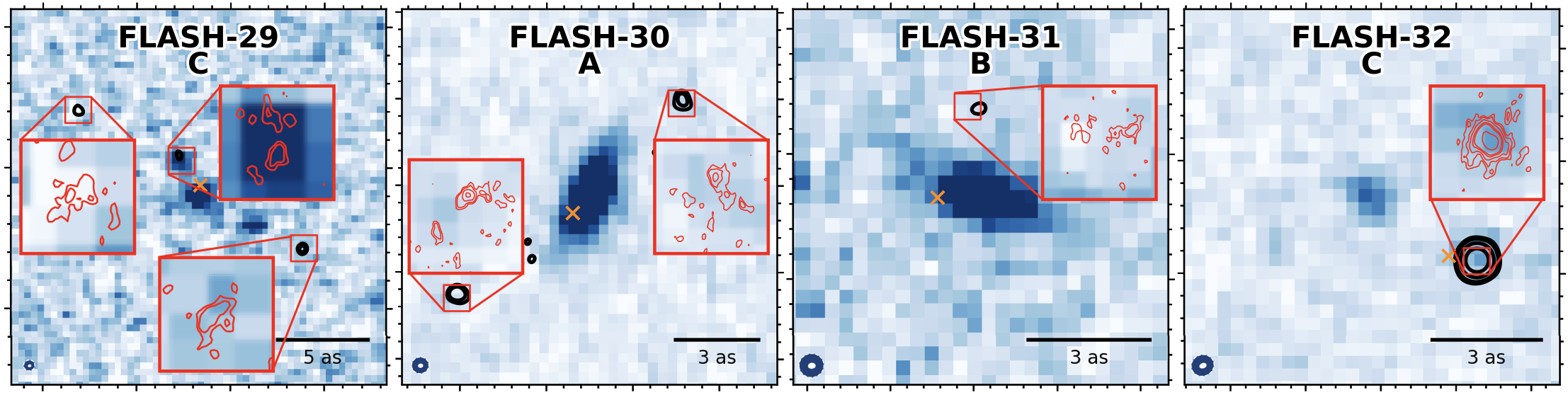}
\includegraphics[width=0.95\linewidth]{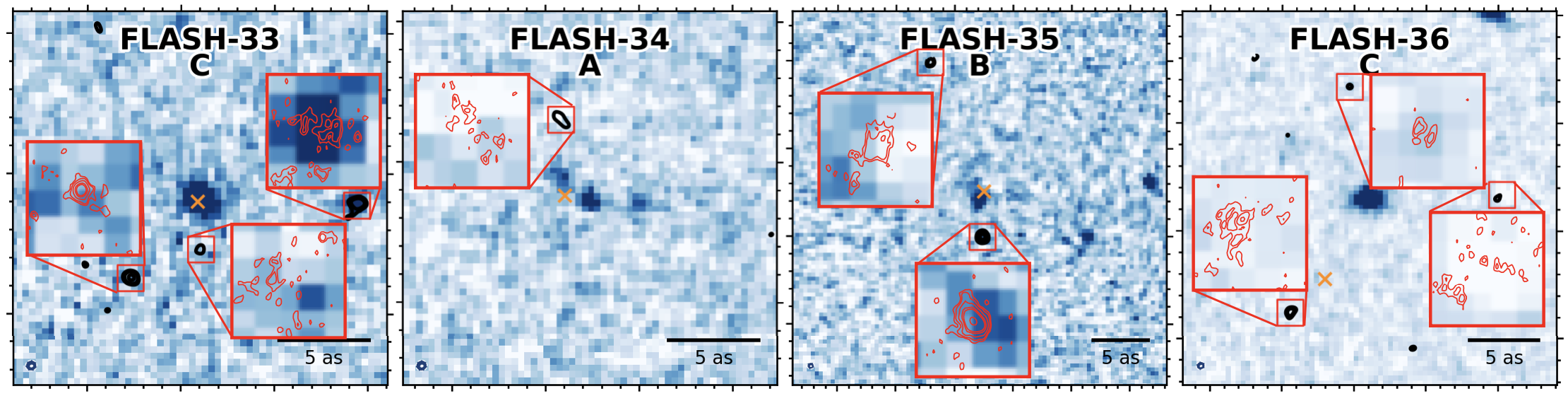}
    \caption{\it Continued.}
    \label{fig:sources2}
\end{figure*}\addtocounter{figure}{-1}
\begin{figure*}
    \centering
\includegraphics[width=0.95\linewidth]{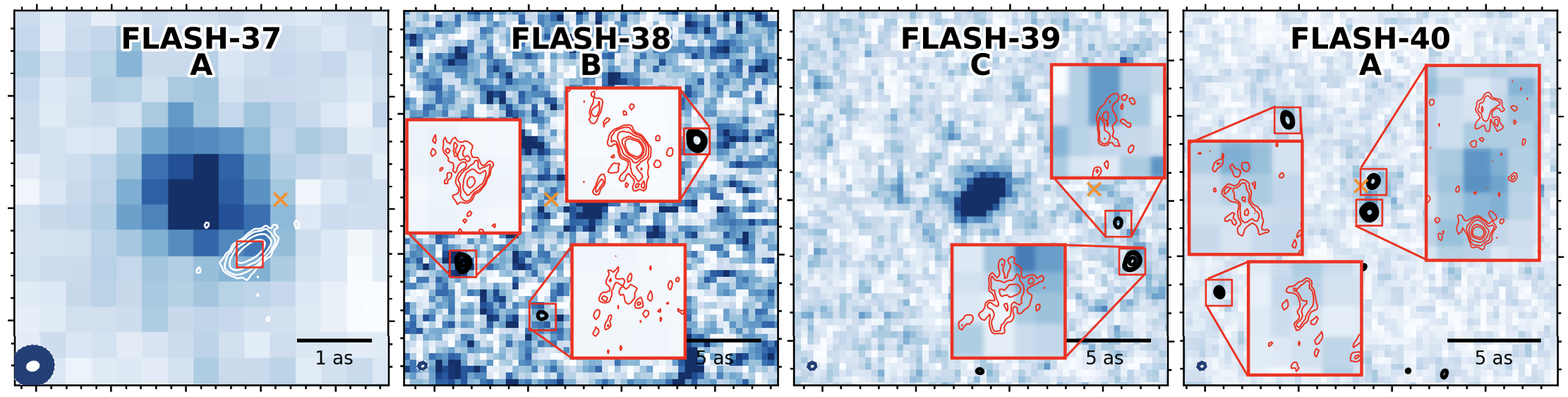}
\includegraphics[width=0.95\linewidth]{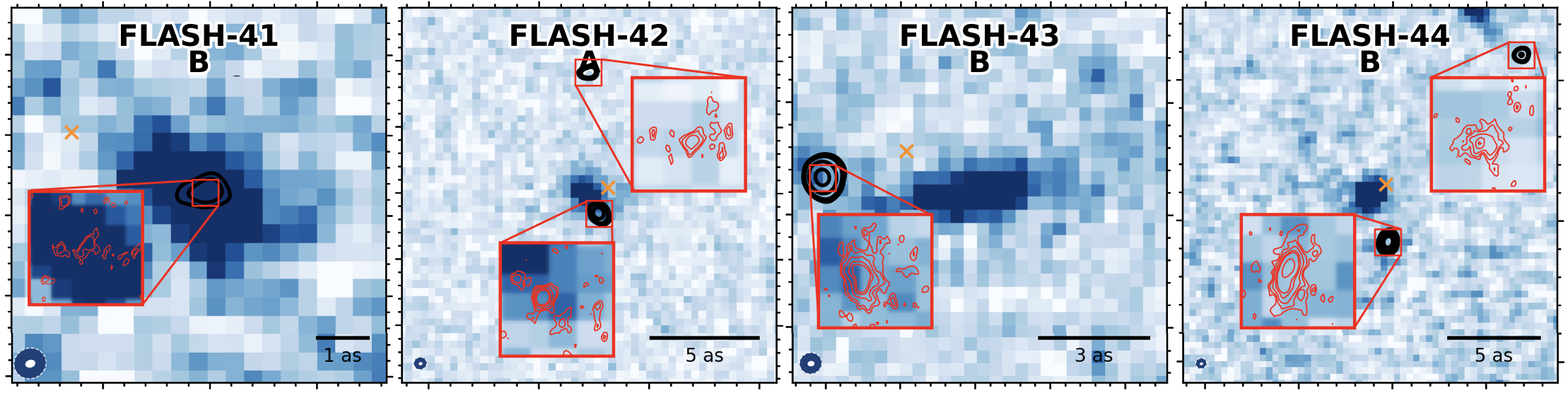}
    \includegraphics[width=0.95\linewidth]{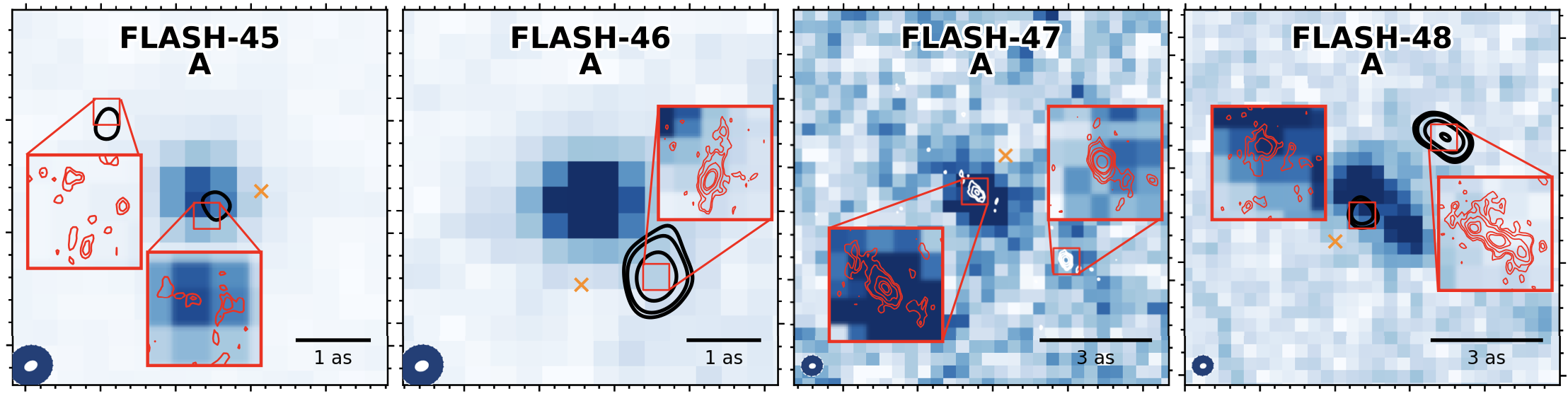}
\includegraphics[width=0.95\linewidth]{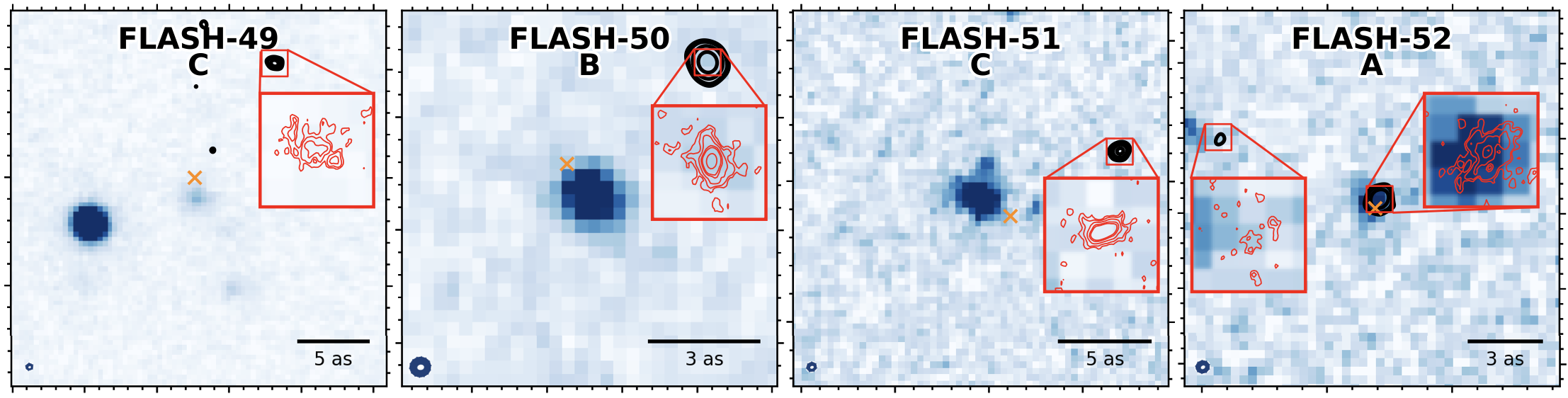}
\includegraphics[width=0.95\linewidth]{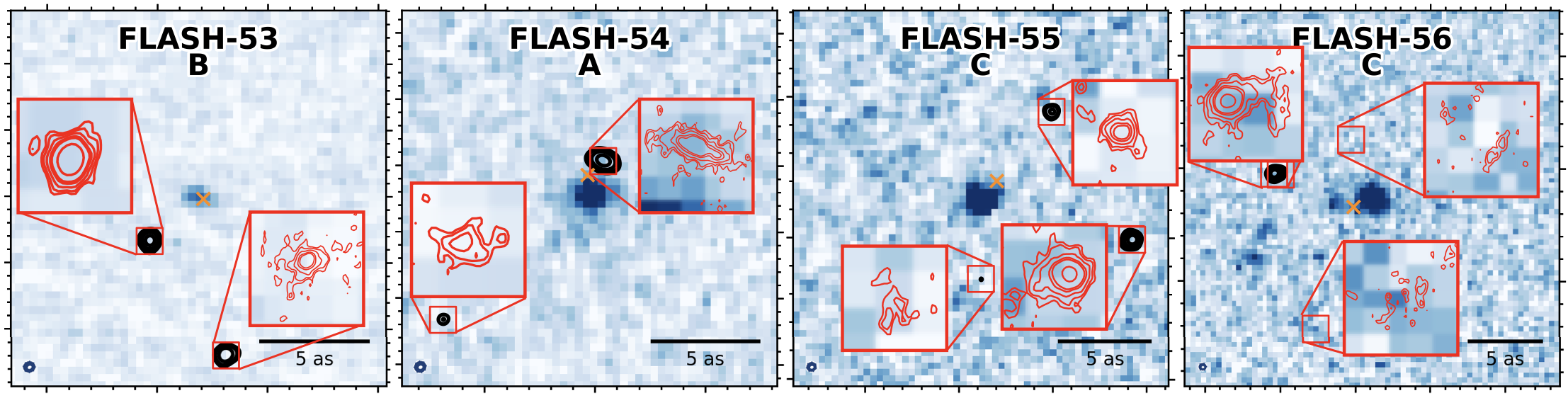}
    \caption{\it Continued.}
    \label{fig:sources3}
\end{figure*}\addtocounter{figure}{-1}
\begin{figure*}
    \centering
\includegraphics[width=0.95\linewidth]{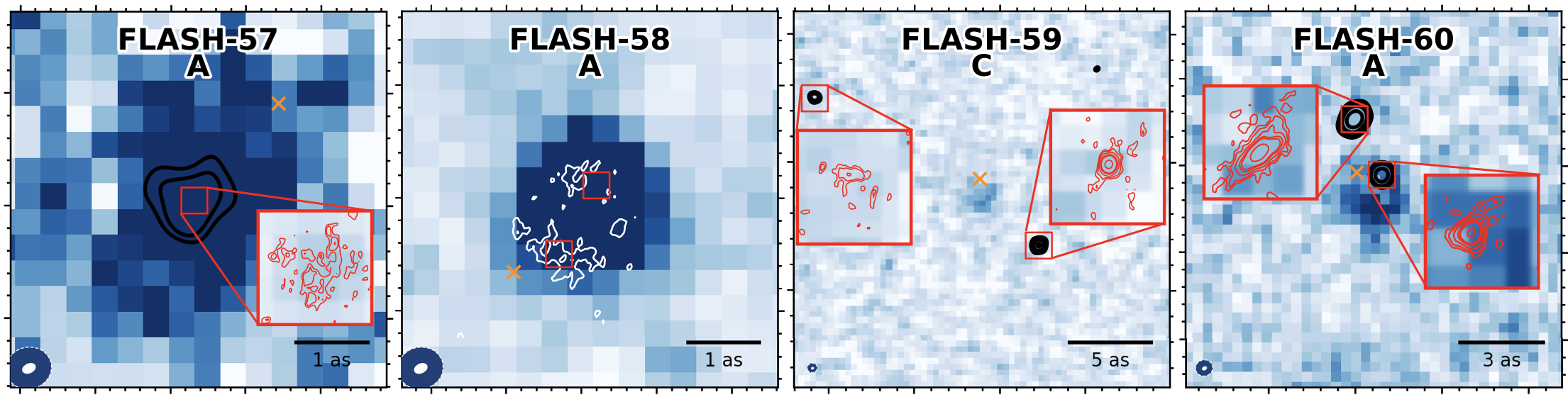}
\includegraphics[width=0.95\linewidth]{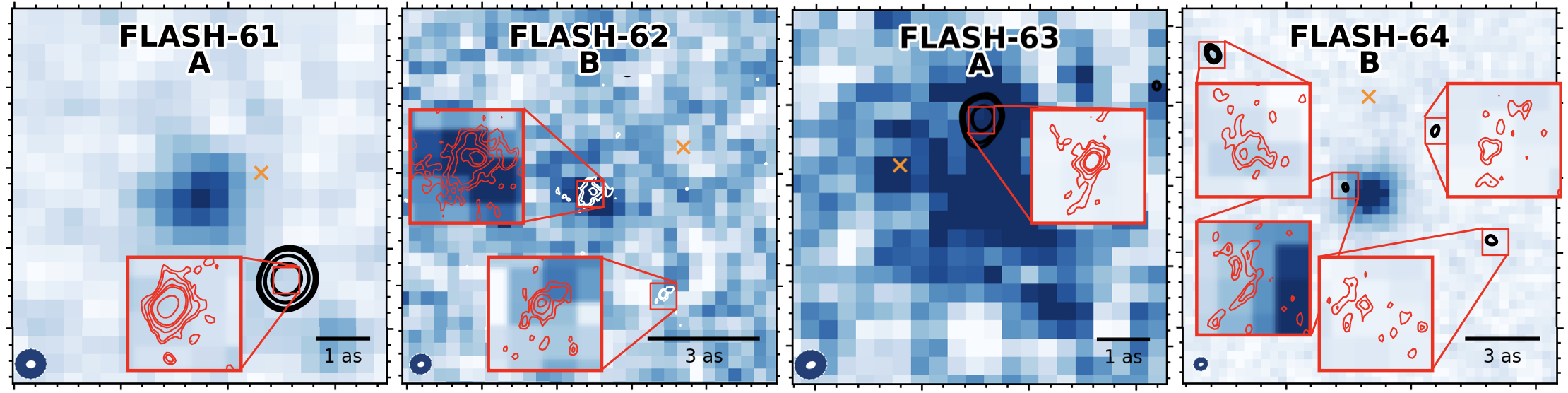}
    \includegraphics[width=0.95\linewidth]{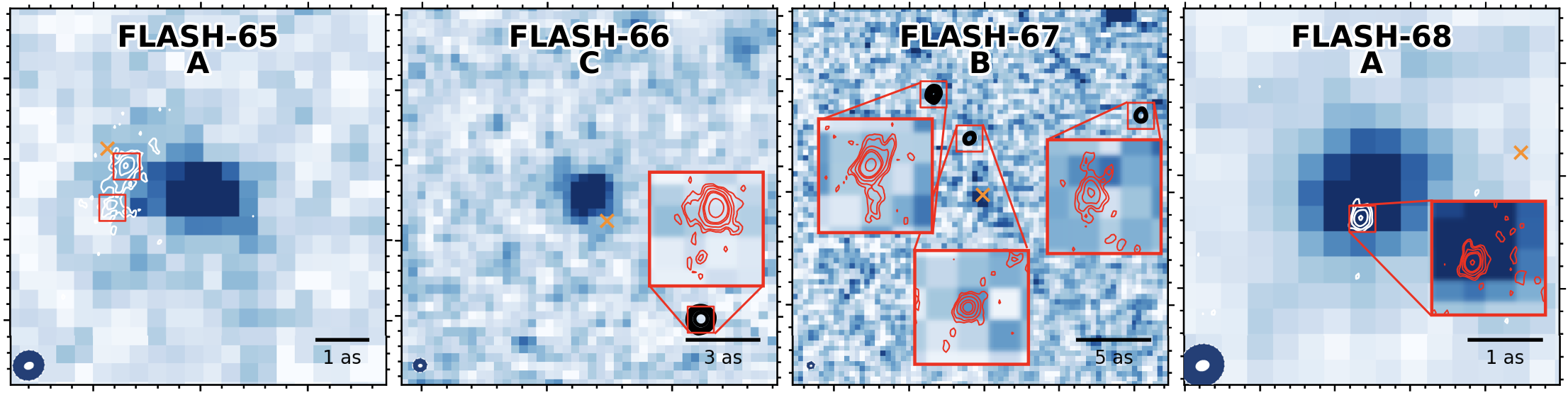}
\includegraphics[width=0.95\linewidth]{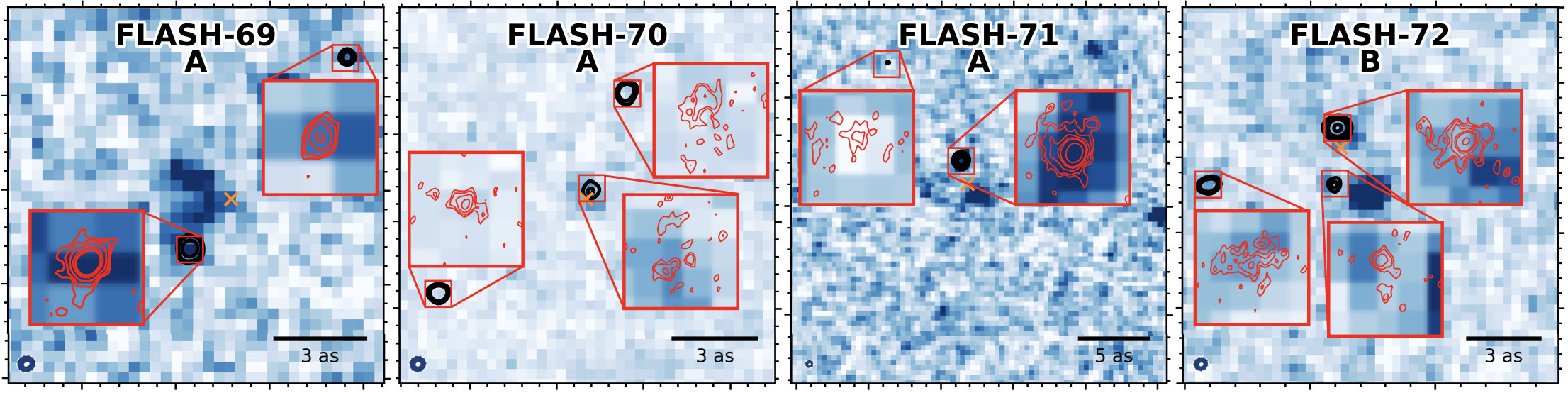}
\includegraphics[width=0.95\linewidth]{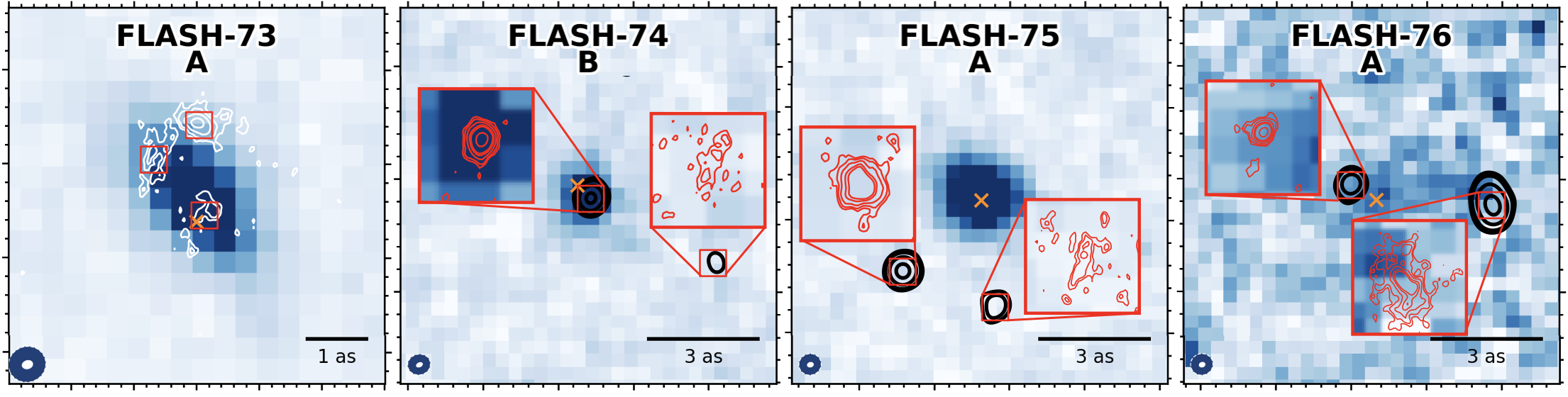}
    \caption{\it Continued.}
    \label{fig:sources4}
\end{figure*}\addtocounter{figure}{-1}
\begin{figure*}
    \centering
\includegraphics[width=0.95\linewidth]{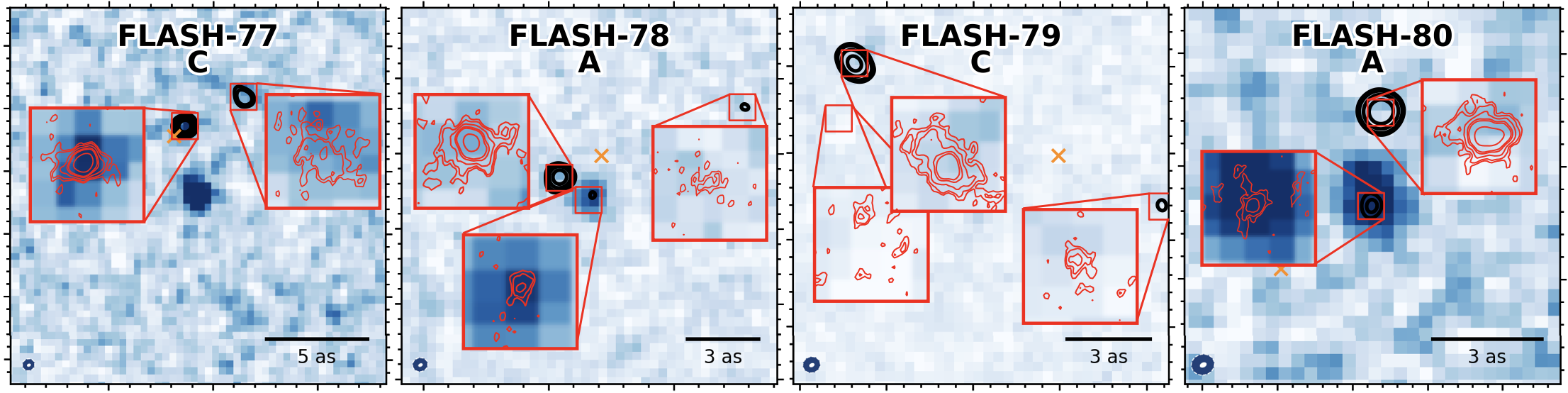}
\includegraphics[width=0.95\linewidth]{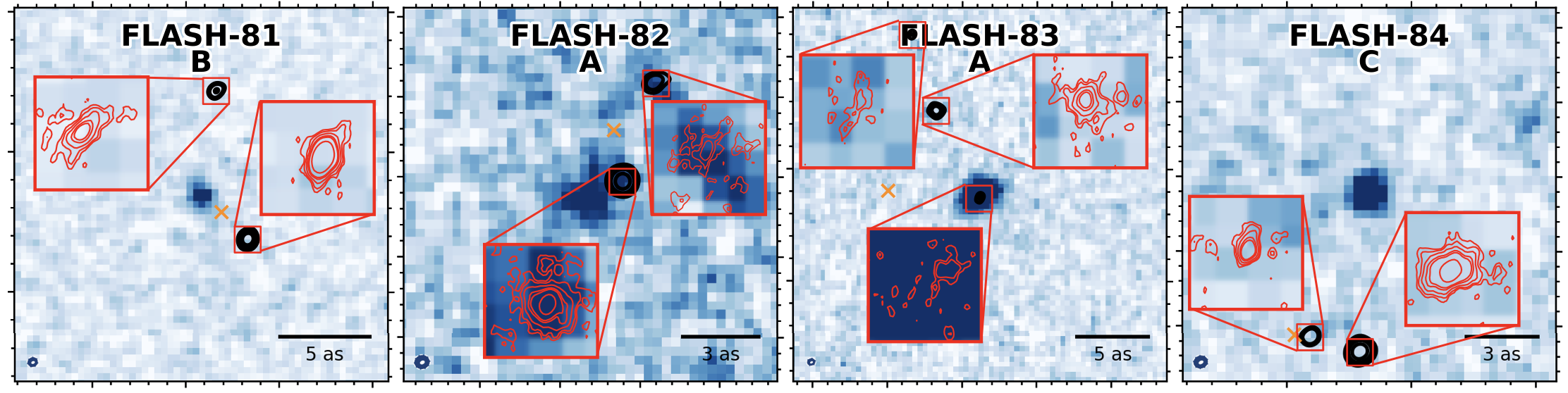}
\includegraphics[width=0.95\linewidth]{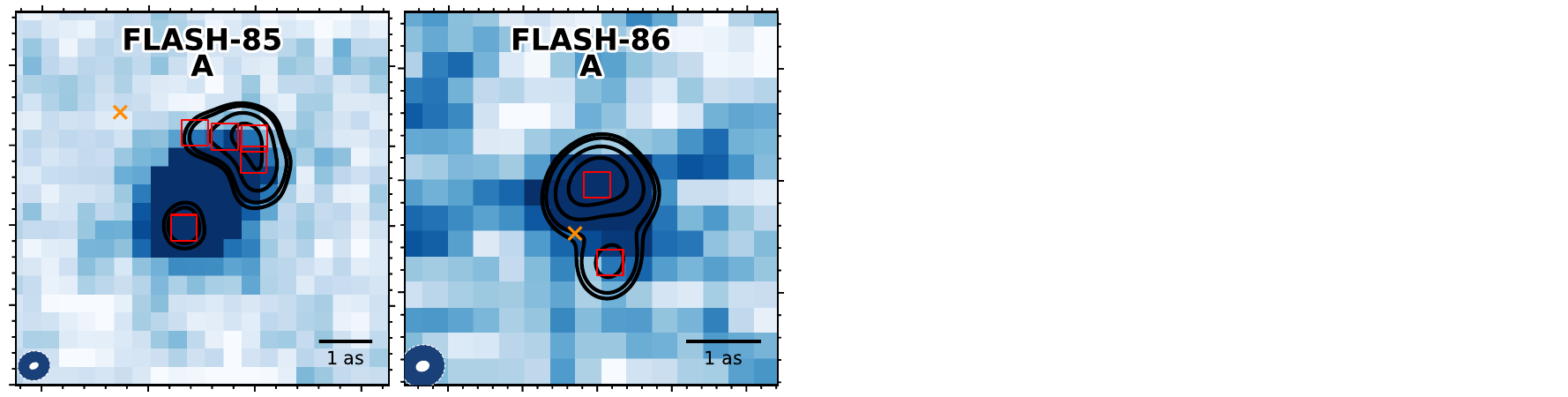}
    \caption{\it Continued.}
    \label{fig:sources5}
\end{figure*}

\end{twocolumn}
\include{tableResults}
\begin{twocolumn}

\subsection{ALMA observation completion}
\begin{figure}
    \centering
    \includegraphics[width=\linewidth]{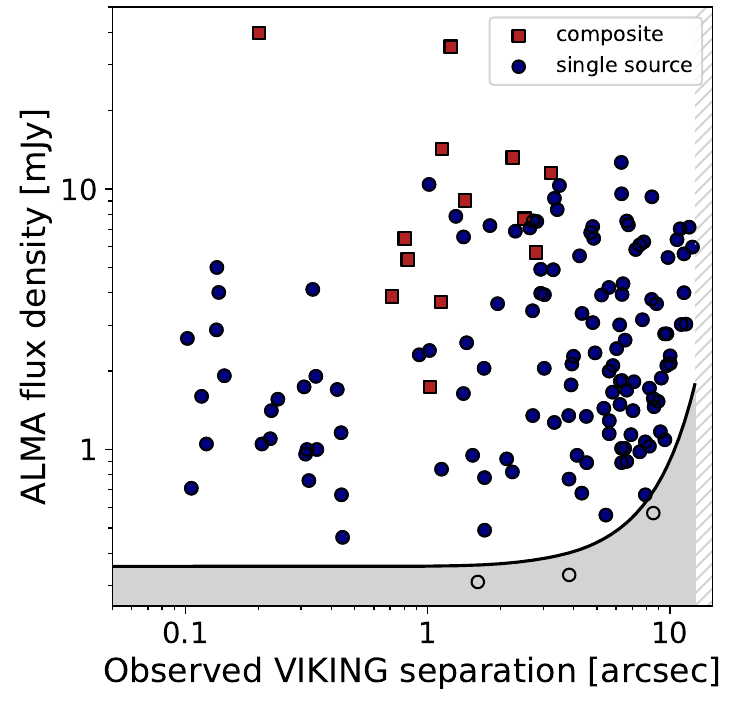}
    \caption{The observed ALMA flux densities ({\it solid blue circles} for single sources; {\it red squares} for composite sources, where composite sources are shown in italics in Table~\ref{tab:catResults}) are shown against the angular offset from the phase centre, i.e., the VIKING source position. The {\it solid black line and fill} indicate the ALMA primary beam sensitivity down to $3 \sigma$. 
    The typical flux densities of the sources are higher than the observation depth. Although most sources are detected away from the edge of the ALMA beam, we cannot guarantee that sources could lie outside of the Field-of-View. Similarly, several sources scatter below the detection threshold, and particularly since lensed sources could be extended, we cannot guarantee the sample observations are complete.}% It is thus likely that we only missed a small portion of the sources with ALMA that we expected to see, and those would likely not be lensed.}
    \label{fig:DetectionThreshold_VIKING}
\end{figure}
Several sources do not have bright emission in their reduced images. Notably FLASH-6 ($S_{500} = 22.2$~mJy) does not show any emission in the ALMA imaging. Here we explore the reasons for these non-detections. The ALMA observations were centered on the VIKING position, in order to accurately probe the lensing structure expected around the foreground lens. The field-of-view (FoV) of ALMA is limited to $\sim 15$~arcseconds, however, and there exists the possibility of sources falling outside of the primary beam -- although this is increasingly unlikely to be due to galaxy-galaxy lensing \citep{Amvrosiadis2018}. 

Figure~\ref{fig:DetectionThreshold_VIKING} shows the observed offset between the ALMA-identified sources and the VIKING central source. The black line shows the $3 \sigma$ detection limit based on the 0.5~arcsecond tapered image used for source extraction. The majority of sources have detected emission within the FoV of ALMA, and importantly, the typical source flux lies a factor of two or more above the detection limit. 
However, the individual sources approach the end of the FoV of the Band~7 observations by ALMA, even though the selection towards large values of the reliability means that these sources have a small distance between the VIKING and {\it Herschel}-estimated position for the source. Even a shift of five arcseconds -- common across the sample -- could push sources into a less sensitive part of the primary beam, and result in non-detections.
The fact that lenses can be extended across the source further increases the detection threshold, further complicates this matter.
On the whole, we have a large detection fraction for most sources,  although current observations cannot exclude faint, extended, or cluster lenses to be completely accounted for in the ALMA observations.

\section{Lensing in \textit{Herschel} samples}
\label{sec:sec4}
In this section we discuss the lensing nature of FLASH sources based on the ALMA images. Here, we differentiate obvious strong lenses, investigate more difficult sources which could be lensed or not lensed, and explore the effects of selection biases in the sample. Finally, we zoom out to the complete perspective of lenses to be found in the {\it Herschel} samples with the FLASH method.

\subsection{Lensing nature of FLASH sources}
\label{sec:identificationOfLenses}
The ALMA observations reveal a large spread in the observed morphologies (Figure~\ref{fig:sources1}). Some \textit{Herschel} sources are easy to visually identify as gravitational lenses, showing morphological features such as arcs, multiple images and even near-complete Einstein rings. Other sources have multiple nearby counterparts, making interpretation of their lensing nature more difficult. These systems could be chance alignments, a situation where the {\it Herschel} source and VIKING galaxy are the same source, or cases of cluster lensing, where foreground clusters provide a speckled ALMA field with multiple sources, as well as the possibility of (proto-)cluster environments where overdensities in the cosmic web are seen through an excess of ALMA sources. We summarize our knowledge on the lensing nature of each source by a grade ranging from A (secure lens identification) via B (some evidence for lensing) to C (no indications for lensing, or the lack thereof). 

\referee{In brief, we identify A-grade lenses as sources where robust ALMA emission shows lensing features such as arcs or rings, or where the ALMA emission is between 0.2 to 2~arcseconds removed from the central VIKING galaxy. B-grade sources consist of sources with emission either removed further than 2~arcseconds -- but morphologically appears to be consistent with lensing -- or is within 0.2~arcseconds of the VIKING galaxy -- where we cannot exclude the ALMA observation of the VIKING galaxy. Sources without any of these features are categorized as C-grade.} Below, the lens identification criteria are discussed in detail, and we summarize the results in Table~\ref{tab:flashresults}.

\begin{table}
    \centering
    \caption{The classification of FLASH sources}
    \label{tab:flashresults}
    \begin{tabular}{cccccc} \hline \hline
S$_{500}$ [mJy] & Nr. & A-grade & B-grade & C-grade  \\ \hline
% 10 - 85 & 86 &  66 $\pm$  9     & 40 & 23 & 23 \\
% 10 - 85 & 86 &  76 $\pm$  5 \% & 46.6  \% & 26.7 \% & 26.7  \% \\ \hline
10 - 25 & 21 &  29 	\%	& 33 \% 	& 38 \% \\
25 - 35 & 30 &  43 	\%	& 30 \% 	& 27 \% \\
35 - 45 & 17 &  53 	\%	& 24 \% 	& 24 \% \\
45 - 55 & 8 &  75 	\%	& 0  \%	& 25 \% \\
55 - 65 & 6 &  50 	\%	& 33 \% 	& 17 \% \\
65 - 75 & 3 &  67 	\%	& 33 \% 	& 0 \% \\
75 - 85 & 1 &  100 	\%	& 0  \%	& 0 \% \\ \hline
    \end{tabular} 
    \raggedright \justify \vspace{-0.2cm}
\textbf{Notes:} 
Col. 1: The 500~$\mu$m flux bin. %The top two rows indicate the full sample (i.e., $10 < S_{500} < 85$~mJy), with the top row showing the absolute numbers, and the bottom row indicating the separate percentages.
Col. 2: The number of sources contributing to each flux bin.
Col. 3: The expected number of lenses based on the false-positive considerations in the Sample selection (Section~\ref{sec:sec2}). 
Col. 4 -- 6: The distribution of sources in each bin.
\end{table}

\subsubsection{Identifying lenses in FLASH}
We investigate the lensing features of sources visually, \referee{identifying A-grade lenses by their} extended or arced ALMA emission close to the central VIKING sources (i.e., $< 2$~arcsec). 
These sources were selected with a small spatial separation between the {\it Herschel} and VIKING positions. The combined source-to-source angular separation of {\it Herschel} and VIKING sources, particularly at the lower-significance levels, is on the order of one or two arcseconds. 
For sources without obvious lensing features such as arcs or rings (e.g., FLASH-3), we interpret ALMA emission offset from the central VIKING source by more than 0.2 but less than 2~arcseconds as indications of strong gravitational lensing. At these separations, the emission is likely not originating from the near-infrared emitting VIKING galaxy given the accurate photometry of ALMA and VIKING ($< 0.1$~arcsec; \citealt{Wright2019}), but instead is lensed by the foreground source. On the other hand, if there exists bright VIKING emission at the position of the ALMA emission, we exclude the source as a lensing candidate, \referee{and award the source a B-grade}.
These provide us with a first-pass estimate of the number of A-grade lens candidates in the FLASH sample, for a total of 37 sources.

Since the sensitivity of our observations is not guaranteed to detect extended lensing features for all sources, we measure the curve-of-growth of ALMA emission through multiple annuli at different widths (0.15, 0.3 and 0.5). 
We calculate the signal-to-noise ratio for all pixels $i$ within the annulus using the following equation,
\begin{equation}
    {\rm SNR} = \sum_i \frac{S_i}{\sigma \sqrt{N_{\rm pix} N_{\rm beam}}}. 
\end{equation}
In this explicit equation, the per-beam flux density, $S_i$ [Jy / beam], is converted to the per-pixel flux density by dividing by the number of pixels per beam, $N_{\rm beam}$. Subsequently, the per-pixel flux is summed over all pixels in the annulus, $N_{\rm pix}$. and is normalized to the per-field standard-deviation, $\sigma$. The pixels are cross-correlated on the scale of a beam, so this standard-deviation needs to be corrected by the square-root of the number of pixels per beamsize, as well as correct for the reduced uncertainty for the larger aperture, i.e., $\sigma / \sqrt{N_{\rm pix} N_{\rm beam}}$, resulting in a noise profile with a unit variance centered around zero. 

Figure~\ref{fig:COGs} shows the annuli-based curve-of-growth analyses for the three sources where additional $> 5 \sigma$ ring-like features were found below the ordinary detection threshold: \referee{FLASH-30, FLASH-34 and FLASH-75.} These graphs show the 5.5~arcsecond surroundings of the VIKING sources, and fit annuli with three different widths (0.15, 0.3, and 0.5~arcseconds) in an effort to reveal lensing features. The bottom panels show the signal-to-noise ratio as a function of angular distance. Direct observations of lensing features is explicitly less sensitive to larger lensing arcs, and all these features indicate Einstein radii below 1 arcsecond. Several more sources have $\sim 4 \sigma$, although deeper observations are necessary to confirm these sources to exclude false-positives.

\begin{figure*}
    \centering
    \includegraphics[width=0.25\linewidth]{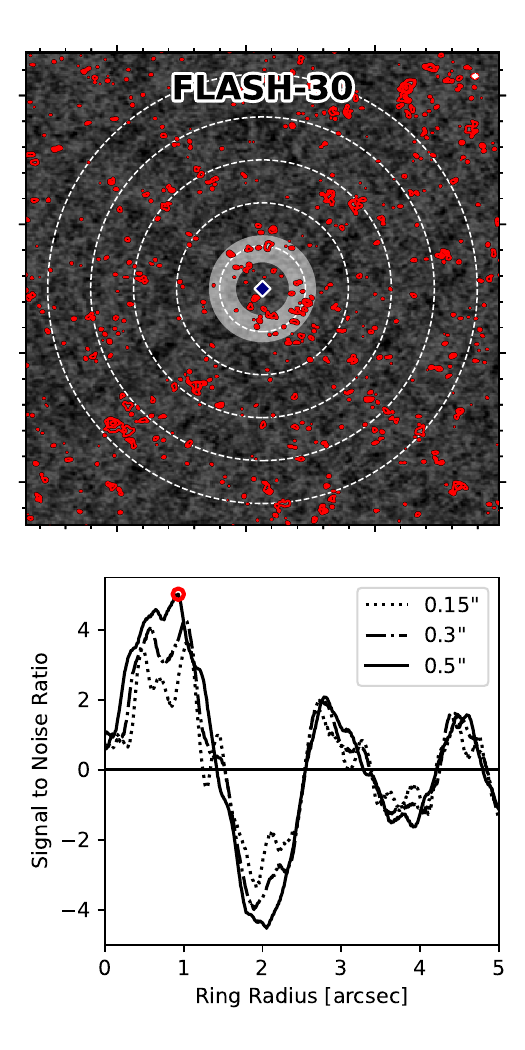}
    \includegraphics[width=0.25\linewidth]{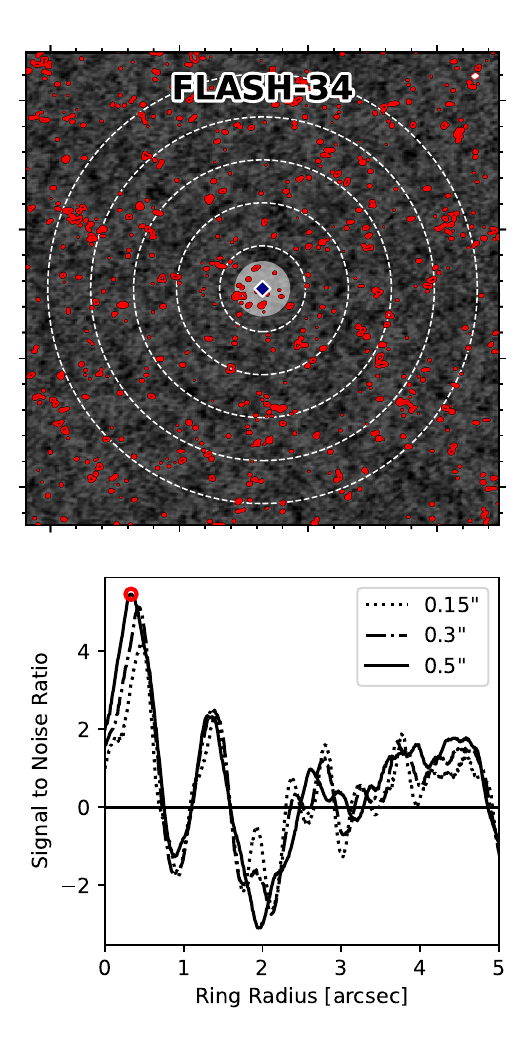}
    \includegraphics[width=0.25\linewidth]{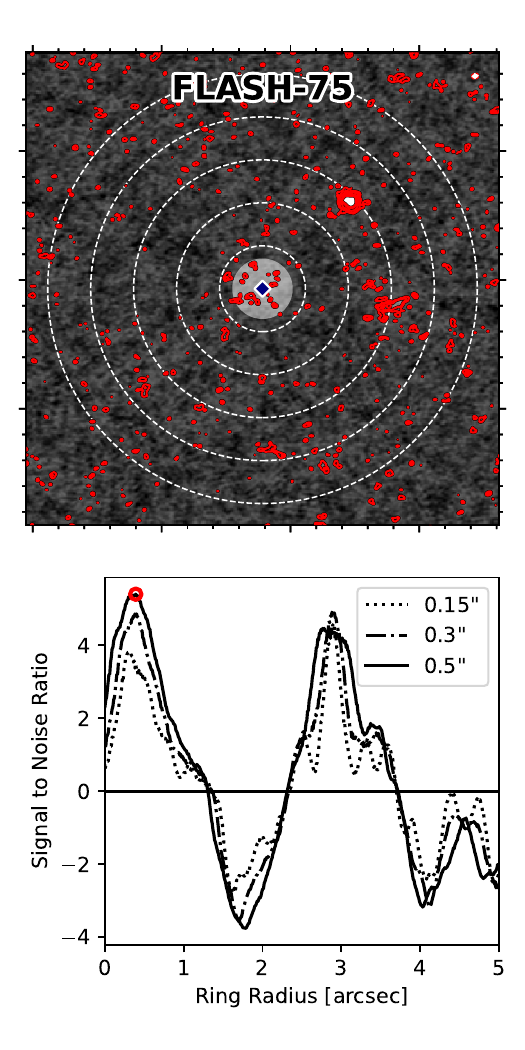}
    \caption{Top panels show a 11 by 11 arcsecond view of the high-resolution (0.15~arcsec) ALMA continuum, with {\it red contours} at 2, 3, 5, 8, 10 and $20 \sigma$, with the dashed contours indicating negative continuum at the same levels. The {\it blue diamond} shows the VIKING position, with centred circles at 1, 2, 3, 4, and 5~arcseconds in radius. The filled contour indicates the most significant emitting annulus.
    The bottom panels show the annulus-integrated signal-to-noise of the sources as a function of the ring radius for three different annulus widths. These three sources are the only sources where lensing was not already identified through direct observations, and the signal-to-noise in one of the three annuli exceeds $5 \sigma$. }
    \label{fig:COGs}
\end{figure*}

\subsubsection{Confusion between fore- and background sources}
Our source selection included a low probability ($< 0.1$~\%) of overlap between the redshift probabilities of VIKING galaxies and {\it Herschel} sources. However, without spectroscopic observations, both of the VIKING galaxies and {\it Herschel} sources, we cannot exclude the possibility of observing the same object in VIKING as in \textit{Herschel}. 
Since these sources are selected from roughly 50~square degrees, our method could instead be an effective way for finding near-infrared bright DSFGs (to be VIKING-detected) with cold dust (resulting in a vast over-estimation of the sub-mm photometric redshift). 

\begin{figure}
    \centering
    \includegraphics[width=\linewidth]{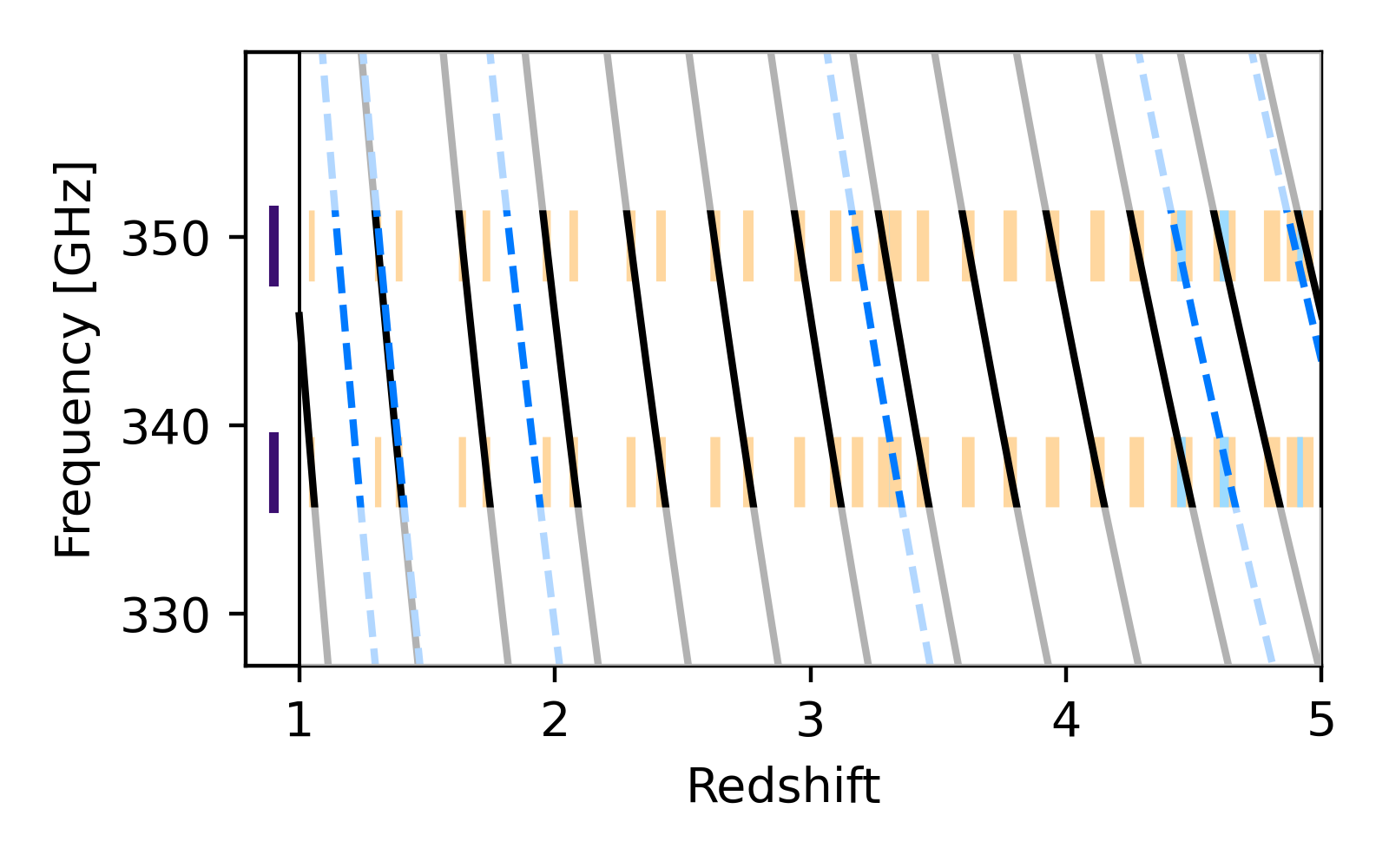}
    \caption{The ALMA observations have the ability to detect spectroscopic features. As galaxies red-shift, different lines come into view. The {\it black lines} indicate the CO lines, starting at CO(3-2) on the left-hand side. The {\it blue lines} indicate alternative lines we can expect to detect, such as [\textsc{C\,i}], H$_2$O and [\textsc{N\,ii}]. The graph was made using the redshift-search-graph tool by \citet{Bakx2022}, which highlights redshift regions where single CO lines are targeted in {\it orange}, and where multiple lines are detected in {\it blue}. Since no blue regions are seen, no redshift regions would expect more than a single CO line detection, although combinations of [\textsc{C\,i}], H$_2$O and CO lines are still possible.}
    \label{fig:RSG}
\end{figure}
While the main goal of the ALMA observations was to unravel the morphologies of these galaxies, the observations also offer spectral information on these sources. Given the spectral coverage, we could detect carbon monoxide (CO), atomic Carbon ([\textsc{C\,i}]) or atomic lines of galaxies in the sample, see Figure~\ref{fig:RSG}. For most of these solutions, the uncertainty in the photometric redshift is too large to use just a single line for a robust redshift identification, particularly since several of these sources were extracted at low flux densities.

Using the tapered data cubes, we initially inspect the galaxies at the extraction position. These tapered data cubes are created by effectively down-weighing the long baselines of ALMA. This allows us to extract positions and fluxes with higher fidelity, since tapering results in higher-significance detections at a moderate cost in resolution for resolved sources. After a visual inspection of the spectra, we perform per-source based aperture photometry to extract the emission line across the source. 

In total, five sources show line emission at their ALMA position. We note that these are tentative spectroscopic redshift solutions, and require confirmation. \\
{\bf FLASH-28} ({\it A-grade}) has a line feature at 348~GHz, with an extension at 348.3 GHz. The source is identified as a gravitational lens, with an expected VIKING-lens redshift of $z = 0.9$ to 1.07. The background photometric redshift is $z_{sub} = 2.36$, and a potential redshift solution could be $z = 2.312$ from CO(10-9). \\
{\bf FLASH-33C} ({\it C-grade}) has a line feature at 350.2~GHz. This source could be associated with the foreground VIKING source, and while the background redshift ($z_{sub} = 2.91$), the VIKING system has a photometric redshift between $z = 0.85$ to 0.93. The source is offset by 2.7~arcseconds, and is likely not associated to the foreground system. The curve-of-growth analysis finds a ringed system surrounding FLASH-33, on top of the multiple components identified by direct imaging. It is thus likely that it is a background source at $z_{\rm spec} = 2.62$ for CO(10--9) for example. \\ %A likely solution could be CO(6-5) line at redshift $z = 0.975$. \\
{\bf FLASH-49} ({\it A-grade}) has a line feature at 351~GHz. This ALMA-detected source, with $z_{sub} = 2.25$, is likely unrelated to the foreground VIKING source between $z = 0.91$ to 0.98 given its spatial offset. A potential solution would be the CO(10-9) line at redshift $z = 2.284$. \\
{\bf FLASH-76W} ({\it A-grade}) shows a line feature at 338.0 GHz with an additional feature at 339~GHz. The {\it Herschel} source is expected to lie at $z_{sub} = 2.14$, with the VIKING source between $z = 0.95$ to 1.13. The ALMA morphology suggests it is a lensed system, although no line emission was seen in the weaker counter-image. A potential redshift solution could be that these are the CO(7-6) and [\textsc{C\,i}](2-1) emission lines at $z = 1.387$. The fidelity of the second line is currently still too low to exclude any other solutions. \\
{\bf FLASH-86N} ({\it A-grade}) shows an absorption feature at 336.2~GHz. The source at $z = 3.85$ is lensed by a foreground VIKING source between $z = 0.63$ to 0.73. The only bright absorption feature is the CH$^+$(2-1) absorption line, confirming its redshift to be $z = 3.965$.

Although four of these five sources had already been confirmed to be lensing systems through the visual identification methods, none of the five sources with line observations provide indications of confusion between the foreground and background source. At least, this provides some confidence in the redshift cut between the fore- and background source. 

In Figure~\ref{fig:histogramRedshifts}, we show the distribution of the photometric redshift difference between sources with different lensing classifications. Simplistically, the assumption could be that the confusion between fore- and background sources would show up as a lower average difference between the two sources. Instead, there does not appear to be a significant difference between the redshift difference of A-grade sources and the B- and C-grade classifications. 
There could thus be a small contribution of confusion between fore- and background systems, although so far there are no concrete indications from spectroscopy or from the redshift difference distributions. The redshift selection criterion in the selection technique appears robust.

\begin{figure}
    \centering
    \includegraphics[width=\linewidth]{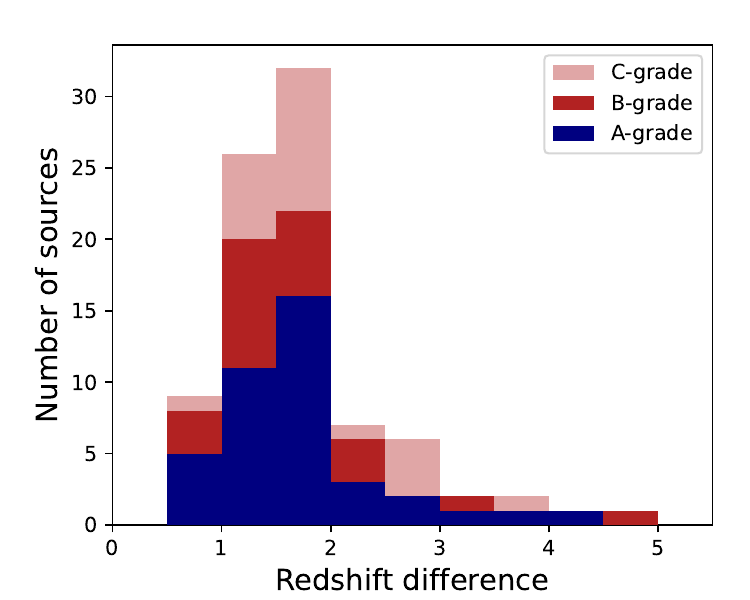}
    \caption{The distributions of the redshift difference between sources with different lensing classifications. The redshift difference distributions do not appear different between the A-, B- and C-grade sources, suggesting there is no large contribution of confusion between fore- and background sources. }
    \label{fig:histogramRedshifts}
\end{figure}

\subsubsection{Cluster lenses and protoclusters in the FLASH sample}
The sources in the FLASH sample are selected from \textit{Herschel} catalogues, which are extracted from the $\approx 18$~arcsecond-wide SPIRE 250~\micron{} point-spread functions. This has increased the possibility of \textit{source confusion}, where multiple sources are confused as a single emitting source. While the FLASH sources are selected as singular \textit{Herschel} sources, they could instead be resolved into multiple sources by ALMA. Particularly at fluxes of $S_{500} = 20 - 40$, the confusion fraction can be around 40~per cent \citep{Scudder2016MNRAS.460.1119S} or higher (see \citealt{Bendo2023} for a more complete discussion). Based on number counts from a hydrodynamical model by \cite{Lagos2020}, about half of the fields is predicted to contain an additional emitter at $3 \sigma$, although the {\it Herschel} source pre-selection increases this probability. 47 fields contain more than one emitter (excluding multiple images from lenses), in line with the prediction from random pointings. 
This does not mean that there is no indication of excess sources. Sixteen fields contain more than two sources, of which FLASH-40 and -64 contain four ALMA sources, far above the expected number of fields with multiplicity. The FLASH-40 system has a A-grade lens, while FLASH-64 has a nearby source that could indicate a lensed galaxy, suggesting that source multiplicity is not the only driver of such sources, but that gravitationally-lensed sources could also trace environments with multiple sources \citep{Overzier2016}.

On the other hand, the foreground imaging from VIKING reveals around twelve fields with multiple NIR bright sources. Our selection method aims towards galaxy-galaxy lensing, but might also pick up galaxy-cluster lensing. Using a visual identification, we identify two A-grade sources (FLASH-21 and -82), five B-grade sources (FLASH-4, -35, -43, -44, and -62) and five C-grade sources (FLASH-1, -6, -29, -56, and -77) with additional bright VIKING galaxies. The grade identification relies on low angular separations, in line with high-magnification galaxy-galaxy lenses, however our method might not be as good at identifying cluster lenses with larger separations, or galaxy-galaxy lenses with large separations and lower magnifications, perhaps in the range of weak lensing.

\subsection{Effectiveness of the FLASH method}
We robustly identify 40 lensed sources (A-grade), at 47~per cent of the total sample. In 23 cases, there are some tentative indications of lensing, which cannot be confirmed with current observations \referee{(B-grade)}. For the remaining 23 cases, the ALMA observations provide no indications of gravitational lensing \referee{(C-grade)}.
Several stand-out sources show-case lensing in near-complete Einstein rings, such as FLASH-13, -27, -58, -73, -85 and -86 \referee{(see Appendix Figure \ref{fig:compilation_stronglenses})}. Particularly the lensing nature of FLASH-13, selected at a mere $S_{500} = 23.6$~mJy, confirms that our method is capable of selecting lenses at four or five times lower apparent flux densities than previous methods \citep{negrello2010,Negrello2014,negrello2017,Vieira2013}. \referee{Meanwhile, several A-grade sources do not have apparent counter-images in the ALMA observations, which could indicate weaker lensing below the strong-lensing regime of $\mu > 2$.}

\begin{figure}
    \centering
    \includegraphics[width=\linewidth]{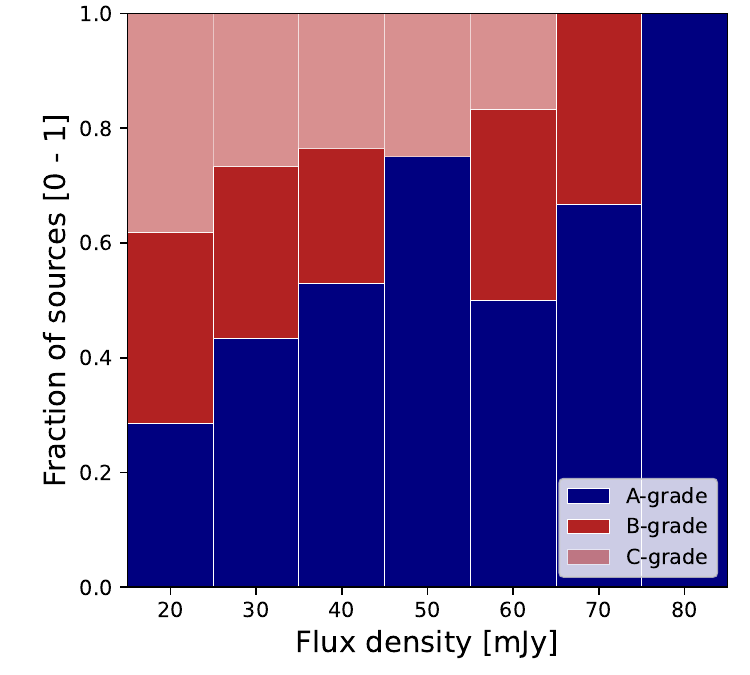}
    \caption{The fraction of sources in each classification is shown for each flux density interval at $S_{500}$. The different categories of sources are coloured in {\it blue} and {\it red}, with A-grade lens candidates in {\it blue}, B-grade candidates in {\it dark red}, and C-grade candidates in {\it light red}. %The {\it white error bars} indicate the expected uncertainty on the method, based on potential false positives in the method from equation~\ref{eq:lensprobability}. 
    The number of confirmed lens candidates (A-grade) is highest for the brightest sources, although robust lensed sources are seen at all flux densities. 
    }
    \label{fig:successratio}
\end{figure}
Figure~\ref{fig:successratio} shows the distribution of source types as a function of their flux density. % We compare this against the expected number of lenses, based on the inclusion of potential false positives based on equation~\ref{eq:lensprobability}, which includes the lensing predictions from the galaxy evolution model of \cite{Cai2013}. 
Equation~\ref{eq:lensprobability} predicted a high lensing fraction among the sample ($f_{\rm ALMA} = 0.9$), however we are only able to confirm lensing for 47~per cent of sources (A-grade) through the ALMA observations. 
Meanwhile, the method appears to be most successful at the highest flux-densities, as expected from previous lensing searches that focus on sources with a higher probability to be gravitationally lensed, such as \cite{negrello2010}.

\referee{
\begin{figure}
    \centering
    \includegraphics[width=\linewidth]{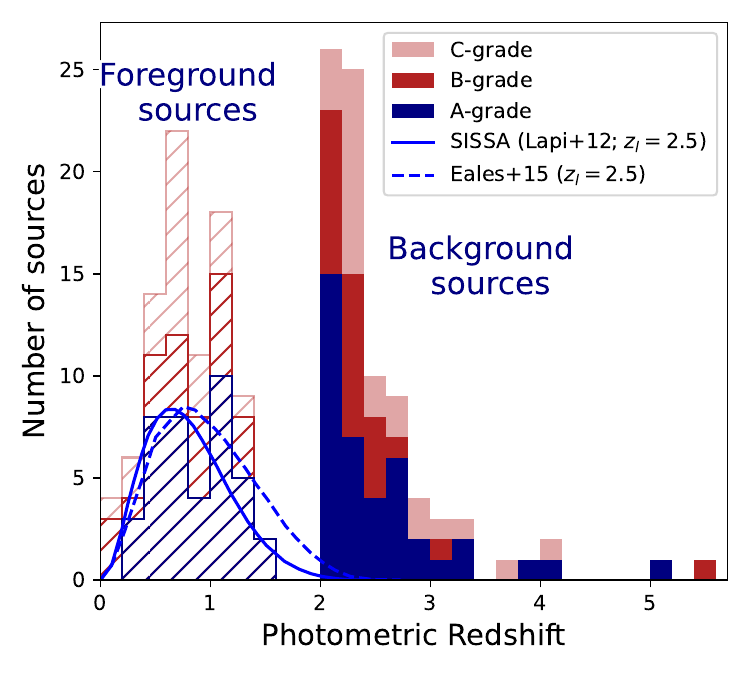}
    \caption{\referee{The fore- (\textit{hatched histograms}) and background (\textit{filled histograms}) redshift distribution of sources in A- (\textit{blue}), B- (\textit{dark red}) and C-grade (\textit{light-red}). The foreground redshift distribution appears similar to the ones predicted from cosmological models (\citet{Lapi2012,eales2015}, assuming $z_s = 2.5$). This suggests there is no obvious redshift bias in the foreground lensing distribution.}}
    \label{fig:redshift_of_low_and_high_redshift_sources}
\end{figure}
The selection of foreground galaxies through VIKING could introduce a bias in the redshift selection. In Figure~\ref{fig:redshift_of_low_and_high_redshift_sources}, the redshifts of the fore- and background sources of each grade are shown against predictions from \cite{Lapi2012,Cai2013} and \cite{eales2015} for the foreground distributions. The models are dependent on the background source distribution. Here, we assume a lensed source redshift of $z_{l} = 2.5$, in line with the average redshift of our sample ($z_{\rm l, FLASH} = 2.5 \pm 0.6$). The redshift distribution of the foreground sources in FLASH is similar to the distributions predicted by the models, and there thus does not appear to be a preferential selection to either low- or high-redshift lenses in the FLASH method. 
}

\begin{figure}
    \centering
    \includegraphics[width=\linewidth]{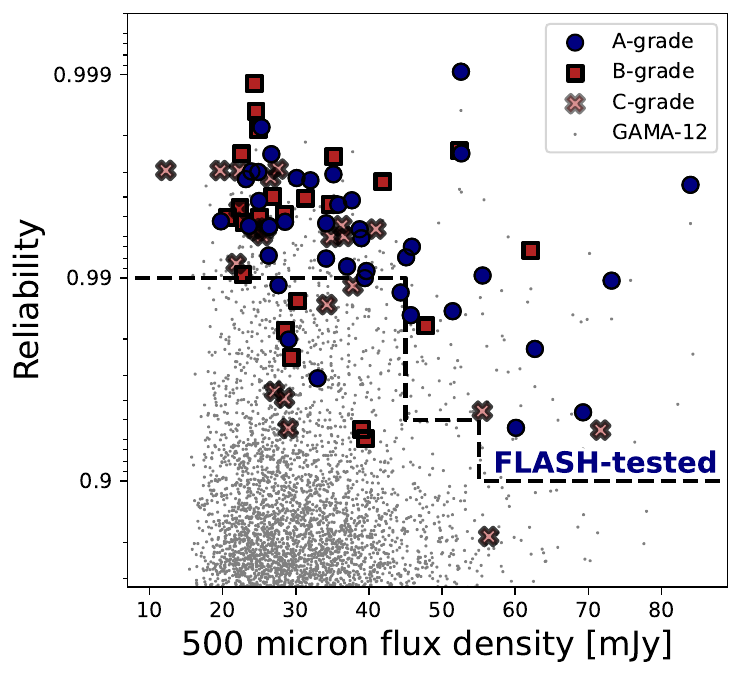}
    \caption{The distribution of sources as a function of their flux density and reliability for different source grades. There does not appear to be a clear correlation between the reliability and the nature of the source at these high reliabilities, likely because these sources pre-select towards {\it Herschel} positions scattered close to VIKING sources, which could pose a fundamental limit to the highest-fidelity lens selection based on positions. Meanwhile, the method can accurately select lenses, even at lower reliabilities, enabling large lensing samples in the future. We show the underlying population of GAMA-12 sources with a similar selection function as the FLASH survey, as well as a line indicating the region where the FLASH observations are representative of the underlying population.
    }
    \label{fig:LHRvsFlux}
\end{figure}
In Figure~\ref{fig:LHRvsFlux}, the reliability of each source is compared against their 500~$\mu$m flux density, highlighting the different nature of the sources accordingly. There does not appear to be a clear split in reliability between the C- and A-grade sources throughout the sample, with several C-grade sources at reliabilities $R \sim 0.995$. The effect of the false-positives might thus be less than expected from previous work \citep{bakx2020}, which should increase rapidly for decreasing reliabilities. FLASH targeted the most likely lens candidates, and is representative of the sources with the highest reliabilities among the GAMA-12 sources, with reliabilities between 0.9 and 0.99.

The angular offset of the ALMA sources from the {\it Herschel} position is between 0.5 to 2~arcseconds even for A-grade lens candidates. This indicates that one of the core ingredients in the lensing identification method, namely the angular offset, could be more uncertain than predicted. There thus exists an additional uncertainty in the likelihood ratio, resulting in scatter in the reliabilities of sources at the high end of the reliabilities. %The likelihood of an association between VIKING and {\it Herschel} sources is determined by the VIKING magnitude and its separation to a {\it Herschel} galaxy. The angular separation between the {\it Herschel} position and the ALMA position is frequently offset. 
Instead, there could be a fundamental limit to the reliability of fainter sources, and consequently, there could be a certain level of false-positives that statistical estimators for gravitational lenses are likely to include also in future works. The method from \cite{bakx2020}, as well as other methods such as SHALOS \citep{GN2019}, offer the ability to include the effect of additional angular offset, however it is likely that the highest reliability sources ($R > 0.99$) will always be those scattered close to the nearby VIKING source. Conversely, the ability of this method to select lenses even at lower reliabilities suggests that it is useful to target lower-reliability sources, enabling large lens samples in the near future.

\subsection{The lensed galaxies of the FLASH sample}
The properties of a galaxy-galaxy lensing event are described perfectly in the geometric terms of general relativity, as a function of the distances between the foreground and background galaxy, their individual distances towards our telescope, and the mass distribution of the source. For a Single Isothermal Sphere mass profile, the resulting angular separation between the centre-of-mass of the foreground source and the dust emission, $\theta_E$, can be simplified to the equation 

\begin{equation}
    \theta_{\rm E} = 4\pi \left(\frac{\sigma_v}{c}\right)^2 \frac{D_{\rm LS}}{D_{\rm S}}. \label{eq:thetaE}
\end{equation}
In this equation, the velocity dispersion $\sigma_{\rm v}$ is taken to be $\sqrt{GM / 2r_h}$, with the halo definition ratio ($r_h$) take as the $r_{200} = 3 M / (4 \pi \ 200 * \rho_{\rm crit}(z_l) )$, where $\rho_{\rm crit}(z_l)$ is the critical density of the Universe at redshift $z_l$ \citep{Bartelmann2001}. \referee{$M$ is the total lensing mass, $D_{LS}$ is the angular distance between the sub-mm source and the lens, and $D_S$ represents the angular distance of the sub-mm source.}

% \begin{equation}
%     \theta_E = \left( \frac{4 G M}{c^2} \frac{D_{LS}}{D_L D_S} \right)^{1/2}. \label{eq:thetaE}
% \end{equation}
% In this equation,  \citep{Narayan1996}. 

The VIKING catalogue provides a stellar mass based on a fit to the optical and NIR fluxes \citep{Wright2019}. In the case of galaxy-galaxy lensing, these sources are assumed to be massive cuspy systems, typically red-and-dead elliptical galaxies with little dust obscuration. 
Although the gas fraction of these galaxies is likely low -- and the baryonic mass is thus locked up in the stars -- the lensing mass of galaxies is dominated by the dark matter content of these sources. Cosmological models predict roughly a stellar-to-halo mass ratio of $\sim10 - 1000$, depending on the halo mass \citep{Girelli2020}.

\begin{figure}
    \centering
    \includegraphics[width=\linewidth]{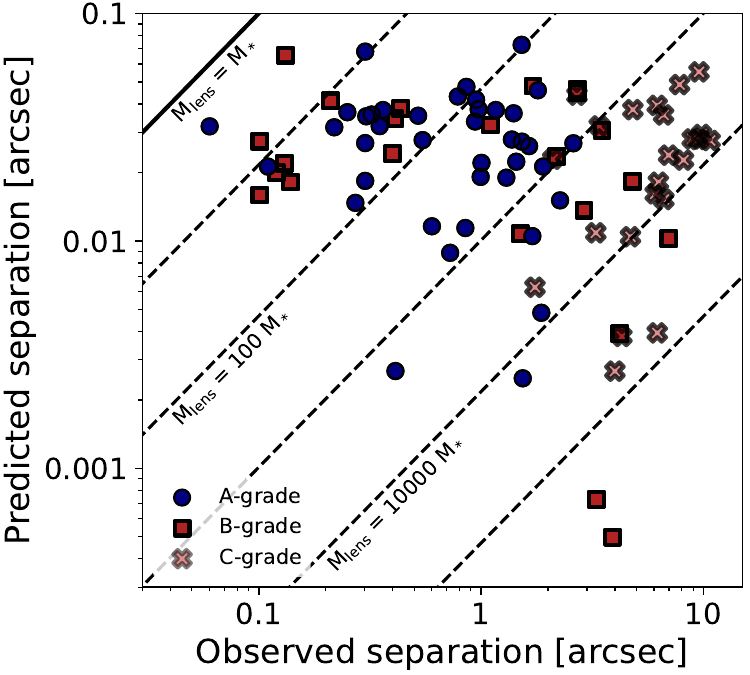}
    \caption{The expected separation based on the stellar mass estimates of the VIKING sources compared for the different source types. The lensing nature of the foreground sources depends on the total mass, including the enclosed dark-matter halo. The A-grade sources are located at lower lensing-to-stellar mass ratios. B-grade sources are split in two groups, with one group at low separation, while the other group suggests higher lensing-to-stellar mass ratios. C-grade sources are found at the highest separations, although they could still be lensed through galaxy-cluster lensing at larger separations. }
    \label{fig:lensingangles}
\end{figure}

The ALMA observations provide a high-resolution view at the lensing geometry, and thus an estimate for $\theta_E$. In Figure~\ref{fig:lensingangles} we compare the observed separation against the one predicted from the equation~\ref{eq:thetaE}. The observed separation are taken from the weighted average of the positions of B- and C-grade lens candidates. For A-grade lens candidates, we select the weighted average of the angular separation of the sources that are the lensed counterpart (see Table~\ref{tab:catResults}). 
There is a clear distribution of sources based on their observed separation, although we note that this is part of the lensing grade identification. The A-grade sources are distributed below 2~arcseconds. The B-grade sources are roughly distributed in two clumps, one group lies below 0.3 arcseconds separation, where it is not possible to clearly differentiate between the foreground and background source, and the other group lies at separations above one arcsecond. The C-grade sources are seen above 2 arcseconds.
A-grade sources are scattered around and below the predicted separation (i.e., $M_{\rm lens} = 100 M_*$; see for example \citealt{Crespo2022,Fernandez2022}), with masses between 10 and 1000 times the solar mass of the foreground system. Sources with B- and C-grades at higher separations could be more massive ($M_{\rm lens} > 1000 M_*$) lensing events that are difficult to confirm with our current data.
\referee{It is important to note that, although the Einstein radius used in equation \ref{eq:thetaE} is correct for an SIS, the observed angular separation from the VIKING source to the ALMA-observed emission is an upper limit for $\theta_{\rm E}$. This measure also includes an additional factor that accounts for the source plane impact parameter, which can reduce the necessary mass to produce an observed offset. As a result, the most conservative approach would be to take the extreme offsets as upper limits.
} 

We further compare the relative mass estimates of the sources in Figure~\ref{fig:lensingmass}, where we show the mass of the foreground system, based on the angular offset between the ALMA and VIKING source, and the stellar mass of the foreground source. 
The average mass for the A-grade lens candidates is $\log_{10} M / M_{\odot} = 12.9 \pm 0.5$, and lies approximately one order of magnitude below the C-grade sources.
\referee{The uncertainty in the lensing mass is well below one order of magnitude, and results from a combination in uncertainty in redshift and stellar mass. The effect of redshift has been studied in \cite{serjeant2012}, who reports a relatively minor variation ($< 50$~\%) for a large deflector redshift variation between $z_d = 0.3$ and $z_d = 1.5$ for a lensed source at redshift $z_l = 2.5$. The uncertainty in the stellar masses is also relatively low, however we should consider that these observations target a very specific galaxy population, which could introduce a systemic uncertainty in stellar masses. That said, the stellar masses are unlikely to exceed much beyond $10^{11} M_{\odot}$, and due to the square-root coefficient in equation~\ref{eq:thetaE}, the resulting uncertainty will also be below 50~\% ($< 0.2$~dex; \citealt{Wright2019}). }
We estimate the halo-to-stellar mass ratio of our A-grade lens candidates, which is around 10$^{2.2 \pm 0.1} M_{\odot}$, with a source-to-source variation on the order of 0.9~dex. These values are in line with previous works from \citep{Amvrosiadis2018,Crespo2022,Fernandez2022}. Halo-to-stellar mass ratios in excess of 100 are high for dark matter haloes, although the profile taken in equation~\ref{eq:thetaE} assumes all the mass to be located solely at $\theta_E$. %\referee{ This assumption ignores the potential for extended mass profiles, which would mis-interpret the necessary mass to produce this observed deflection.} 
Similarly, galaxies could be lensed by a group ($N < 5$) of sources, which would not be included in the stellar mass estimate from \cite{Wright2019}. The mass ratio is around 0.5~dex. higher than those predicted in models from \cite{Girelli2020} for halo masses around $10^{12.9}$~M$_{\odot}$ around $z = 0 - 1$. 

In part, this could be because of weak lensing affecting the sample \referee{(see further discussion of this in Section~\ref{sec:totalnumberoflenses})}. Our lens identification method skews towards high-magnification, galaxy-galaxy lenses with cuspy profiles, however the VIKING images appear to indicate several fields with multiple galaxies, producing galaxy-cluster lensing missed in this analysis. Empirically, \cite{Dunne2020Lensing} found that weak lensing boosted the selection of even nearby ($z = 0.35$) galaxies. As a result of using direct ALMA observations to identify lensing, we are likely missing weak lensing events. As a consequence, higher-mass haloes are likely contributing to the B- and C-grade sources, as shown in Figure~\ref{fig:lensingmass}, \referee{ although we note that it is necessary to account for the additional effect of the impact parameter of the sources, which is not perfectly represented by using the observed angular separation as $\theta_E$ in equation~\ref{eq:thetaE}, as discussed above.}

\begin{figure}
    \centering
    \includegraphics[width=\linewidth]{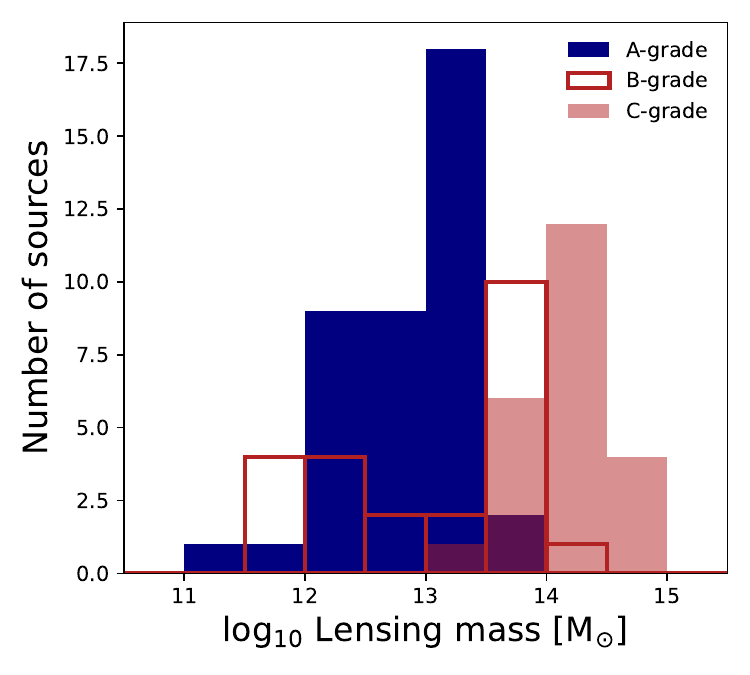}
    \caption{A histogram of the logarithm of the lensing mass expected from the observed angular separation between the foreground and background source, and the stellar mass derived from the VIKING photometry \citep{Wright2019}. It shows that the average mass of an A-grade lens candidates is $\log_{10} M / M_{\odot} = 12.9 \pm 0.5$. The B-grade candidates are more broadly distributed, and C-grade candidates lie at larger lensing masses if their ALMA emission is lensed by foreground VIKING galaxies, necessitating galaxy-cluster lensing events. }
    \label{fig:lensingmass}
\end{figure}

\subsection{Total number of lenses in the \textit{Herschel} catalogue}
\label{sec:totalnumberoflenses}
This is not the first search for gravitational lenses among the {\it Herschel} sample, however it is one of the first study to comprehensively test the method across the {\it Herschel} fluxes using high-resolution sub-mm observations. 
In Figure~\ref{fig:sourcecounts}, we compare the source counts of sub-mm galaxies at 500~$\mu$m for both purely-lensed candidate samples and non-differentiated samples. We compare these against the source counts of the lensed sources found by the FLASH method across the 53.56~sqr. deg. GAMA-12 field. We use the success ratio of the FLASH observations across the regime where the observations are representative for the underlying GAMA-12 field (see Figure~\ref{fig:LHRvsFlux}), and calculate the number of lensed sources that the current FLASH configuration will be able to identify. At the faintest flux end, we include an adjustment to show the effect of the 250~\micron{} selection in light red: in the original {\it Herschel} catalogues, sources are extracted by their $S_{\rm 250}$ flux density, removing some of the highest-redshift galaxies. We adjust this by comparing the fraction of sources with low redshift ($z < 2$) to high redshift ($z > 2$) for the lowest two flux bins when compared to the highest flux bins.

The lensed candidates from the SHALOS method from \cite{GN2019} and the VIKING-based selection from \cite{Ward2022} are compared against the non-differentiated selections from \textit{Planck} \citep{Trombetti2021}, the brightest {\it Herschel} galaxies from \cite{negrello2017}, and the recent discovery of a lens among the STUDIES sample \citep{Pearson2023}. Surveys from SCUBA-2 at 450~$\mu$m from \cite{Casey2013,Chen2013} and \cite{Zavala2017} explore the lower flux-density regime. Note that all lensed candidate samples are based on unresolved predictions.
We compare the sources against the different known 500~$\mu$m-bright emitters, namely late-type local galaxies at the brightest end, radio sources, lensed galaxies from the galaxy evolution model of \cite{Cai2013} -- assuming a maximum lensing magnification of $\mu_{\rm max} = 15$ -- and finally unlensed DSFGs.

The predicted number of lenses found with the method described in \cite{bakx2020} is in agreement with the predicted values from \cite{Ward2022} and with those found in the SDSS-based estimates by \cite{GN2019}, even though the SDSS  have been shown to be incomplete for the highest-redshift lenses \citep{bakx2020}. 
Similar to previous statistical studies, we now robustly confirm the elevated number of lensed sources starting at $\sim 60$~mJy when compared to strong-lensing models. These models only account for galaxy-galaxy lensing, which misses galaxy-cluster lenses that have already been shown to dominate at lower flux densities through unresolved statistical studies \citep{GN2012, GN2014}. 
Although these galaxy-cluster lenses are an important contributor to the total number of lensed sources, identifying these systems is difficult given the requirement for deeper observations at larger fields-of-view.
As noted in \cite{Ward2022}, the likelihood estimator is not well-suited for the large separations of galaxy-cluster lenses ($\sim 20$~arcseconds), and even misses the majority of galaxy-galaxy sources with separations above 8~arcsec. 
Our research is unable to quantify the galaxy-cluster lensing beyond a tentative visual tally of twelve fields with excess VIKING sources, particularly in B- and C-grade fields. Meanwhile, these sources would still not be able to account for the 0.7~dex excess relative to the predicted models. The most likely explanation is the contribution of weak lensing, since the galaxy evolution models have a minimum classification of strong lensing at $\mu > 2$, and some of our sources might be only weakly magnified. 
\referee{
This is corroborated by a visual inspection of our sources finds some sources with lensing identification (i.e., grade A) without multiple imaging or a counter-image resolved in the ALMA imaging. Although this does not exclude strong lensing. The contribution of weak lensing from our sub-sample of eighteen sources without multiple imaging (FLASH-3, -12, -14, -20, -28, -30, -34, -37, -46, -47, -52, -54, -60, -61, -65, -68, -69, -71) would nearly double ($18 / 40 = 45$~\%) the number of strongly-lensed sources when compared to the $\mu > 2$ strong-lensing criterion adopted by \cite{Cai2013}. This is on the same order as the excess of lenses seen between the lenses found by FLASH against those predicted in the galaxy evolution model of \cite{Cai2013}.
}

For submillimeter galaxies with a redshift around 2.5, the likelihood of the flux at 250 micron being close to the detection limits increases as the flux density at 500 micron becomes fainter. This is a consequence of the detection strategy employed to construct the official H-ATLAS catalogues, as previously mentioned. Consequently, some of these sources are detected primarily due to lensing amplification, even with relatively small amplification factors ranging from 5~per cent to 20~per cent. This phenomenon, known as magnification bias, has received considerable attention in recent years and has been the subject of detailed measurement and analysis (\citealt{Bonavera2022} provides a concise overview of the topic).
 
Notably, the halo masses of the lenses, derived from the analysis of the cross-correlation function, exhibit a strong agreement with those estimated directly from individual lensing events in the current study, yielding an approximate range of 10$^{12-12.5}$ \citep{GN2017,GN2021,Bonavera2019,Bonavera2021AnA,Cueli2021,Cueli2022}. Furthermore, these studies have concluded that the majority of magnification bias arises not from isolated galaxies, but from small groups of galaxies featuring one or two dominant members and a few satellites (see also \citealt{Fernandez2022,Crespo2022}). Therefore, it is likely that some of the lensing events observed in the current sample are also caused by small groups of galaxies. While there are indications in the images, verifying these observational findings would require data beyond the scope of this current study.
 
Finally, similar to the discoveries made by \cite{Dunne2020Lensing}, a few of the lensing events examined in this work can be considered as direct observations of magnification bias, which is typically studied only at a statistical level.
% The visual inspection -- and selection of these sources specifically -- is best-suited to find galaxy-galaxy lenses given the narrow ALMA beam. However, one could expect to see observational evidence of this large contribution across these sources as well, as these fidelity-corrected counts exceed the estimates of \cite{Cai2013}. 
In the study by \cite{Dunne2020Lensing}, weak lensing modestly biased their fluxes even in the low redshift Universe, although further investigations of the cluster-lensing population is necessary to see whether the excess is indeed due to galaxy-cluster lenses or whether an excess of lensed sources exist -- an important point indeed, since this would require additional masses to exist beyond the ones predicted in our current cosmological paradigm \citep[e.g.,][]{eales2015}. %{\bf Note: I do not think we yet know whether weak lensing can explain the 0.4 dex excess of sources, and I think we should advocate for more observations, instead of claiming we see weak lensing.}

Accounting for the efficiency and the strict selection criteria of our sample, a total of 3000 lenses are expected to be observable across the 660~square degree H-ATLAS survey. This assumes the existence of complete VIKING-level observations, which are non-existent in the Northern field to date. This is large relative to the SHALOS method described in \cite{GN2019}, that provides a sample -- adjusted for the total 660 square degrees of H-ATLAS -- of $\sim 870$ robust lensed candidates. The southern-field study of \cite{Ward2022} has already demonstrated efficient selection of lensed sources down to $\sim 30$~mJy at 500~$\mu$m, and they report the ability to find 13\,730 lensed sources across the entire H-ATLAS fields, which suggests it is worthwhile to test the FLASH method more comprehensively down to lower reliabilities as shown in Figure~\ref{fig:LHRvsFlux}.

At the lowest 500~$\mu$m flux densities, the H-ATLAS survey thins out significantly due to the prior selection at 250~$\mu$m in the catalogues. Lensed sources, located at redshifts above 2.5, would have lower fluxes at 250~$\mu$m. Our estimate based on the photometric redshift estimates of the entire {\it Herschel} catalogues, the expected number of sources across the full {\it Herschel} catalogues could be expanded to 7000 sources (a $2.35 \times{}$ increase if we remove our current $\sigma_{250} \approx 7$~mJy criterion). These sources would be worthwhile to include in future lensing models, since (i) these sources have a larger cosmic volume for foreground lenses to magnify the background population, and more importantly, (ii), higher-redshift galaxies are rarer in 500~$\mu$m-selected samples. Their steeper luminosity function (\citealt{Gruppioni2013}) ensures that more of the apparently-bright high-redshift population is instead fainter gravitationally-lensed sources. Future works could further improve lens selection by investigating samples that explicitly overcome this 250~$\mu$m selection, either through re-extraction \citep{ivison16,Oteo2017} or through 500~$\mu$m $-$ 250~$\mu$m difference maps \citep{Asboth2016,Duivenvoorden2018}. That said, the current methods are already powerful enough to enable large sub-mm selected lens samples (e.g., \citealt{eales2015}), and an ALMA survey of such large lens samples could become an important cosmological tool in the near future.

\begin{figure}
    \centering
    \includegraphics[width=\linewidth]{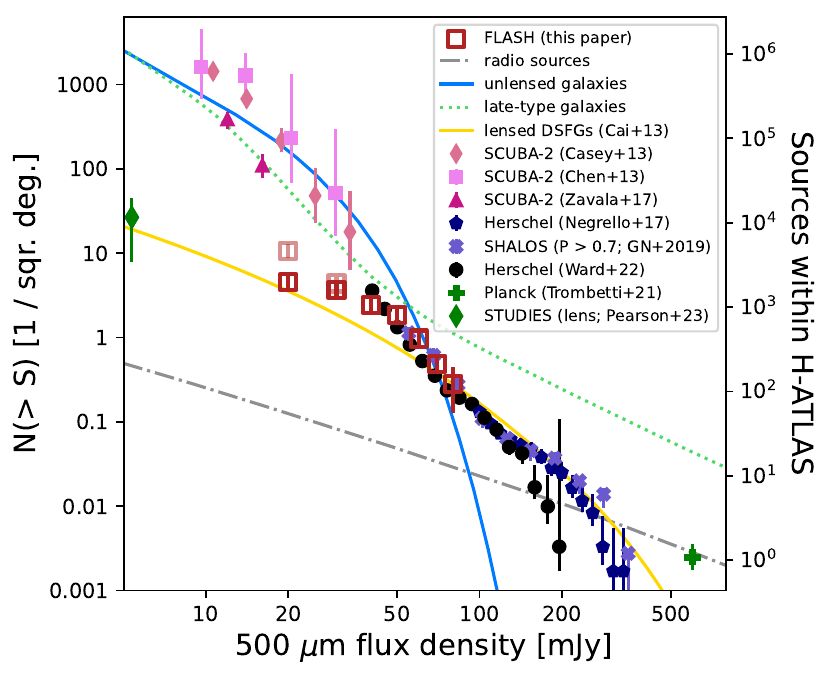}
    \caption{Cumulative number counts at 500~$\mu$m for the lensed sources in the FLASH sample and other samples -- both pre-selected to contain lenses and without lens-preselections. 
    {\it Red squares} indicate the number counts for all {\it Herschel} sources based on the FLASH-selection and {\it light-red squares} show the source counts including an additional correction for the redshift-incompleteness towards lower flux densities.
The counts of late-type, normal, and starburst galaxies and of unlensed dusty star-forming galaxies (DSFGs), interpreted as proto-spheroidal galaxies in the process of forming the bulk of their stars, are from the \citet{Cai2013} model, as well as the {yellow line} indicating the source counts for lensed sources with a magnification cut-off at $\mu_{\rm max} = 15$. As seen in previous studies, these counts exceed the predicted number counts for lensed sources in the $10 - 60$~mJy regime, as these models only account for \referee{ strong galaxy-galaxy lensing, and do not account for galaxy-cluster lensing or weak lensing events potentially identified by the FLASH method}. 
    }
    \label{fig:sourcecounts}
\end{figure}

% \subsection{Unlensed sources}

% \subsection{Cosmic star-formation rates from Herschel}

% \subsection{The future of large lens samples}

\section{Conclusions}
\label{sec:conclusions}
The {\it Herschel} Space Observatory detected near-one million sources across 1000 square degrees from low to high redshift. In this study, we observationally test the validity of the selection method through resolved observations with ALMA, by targeting a sample of 86 likely lensed sources identified by close, bright VIKING counterparts.
We find:
\begin{itemize}
\renewcommand\labelitemi{\tiny \textbf{$\blacksquare{}$}}
\item The ALMA observations are able to confirm 40 (47~per cent) of these sources to be strong lenses (A-grade lens candidates). For an additional 23 (27~per cent) sources, there are tentative indications of lensing, however our ALMA observations are not able to conclusively indicate lensing (B-grade lens candidates). For the final 23 (27~per cent), it remains unclear whether these sources are lensed (C-grade lens candidates).
\item The number of robust lensed sources is below what is expected from false-positive estimations, although we note that our current false-positive estimations might not be a reliable estimator, and future tests can focus on a more comprehensive study of $R > 0.9$ sources to verify the FLASH method and increase the number of lensed \textit{Herschel} sources we can identify.
\item Although we do not find direct indication for sources where the VIKING galaxy and {\it Herschel} source are the same object, near-infrared spectroscopic confirmation of the foreground objects and sub-mm spectroscopic confirmation of the background sources is important to exclude such sources, particularly for the B-grade lens candidates below 1~arcsecond separation.
\item Most of the lensing features would require a total lensing mass between 10 and 1000 times that of the stellar emission reported for these VIKING sources, with typical lensing masses of $\log_{10} M / M_{\odot} = 12.9 \pm 0.5$, in line with previous observations \citep{Amvrosiadis2018} and above what is predicted from models \citep{Girelli2020}. The ALMA identification method likely misses several sources at larger angular separations, potentially due to weak gravitational lensing.
\item Our method will be able to find $\sim 3000$~lensed sources over the entire H-ATLAS field, in excess of what is expected from the galaxy-galaxy strong lensing predicted by the galaxy evolution models of \cite{Cai2013}. 
\end{itemize}

\section*{Acknowledgements}
This work was supported by NAOJ ALMA Scientific Research Grant Nos. 2018-09B and JSPS KAKENHI No.~17H06130, 22H04939, 22J21948, and 22KJ1598.
JGN and LB acknowledge the PID2021-125630NB-I00 project funded by MCIN/AEI/10.13039/501100011033/FEDER. LB also acknowledges the CNS2022-135748 project funded by MCIN/AEI/10.13039/501100011033 and by the EU “NextGenerationEU/PRTR”.
SS was partly supported by the ESCAPE project; ESCAPE -- The European Science Cluster of Astronomy and Particle Physics ESFRI Research Infrastructures has received funding from the European Union’s Horizon 2020 research and innovation programme under Grant Agreement No.~824064. SS also thanks the Science and Technology Facilities Council for financial support under grant ST/P000584/1. 
This paper makes use of the following ALMA data: ADS/JAO.ALMA\#2019.1.01784.S.
\referee{We would like to kindly thank the anonymous referee for their insightful comments and suggested additions.}

%%%%%%%%%%%%%%%%%%%%%%%%%%%%%%%%%%%%%%%%%%%%%%%%%%
\section*{Data Availability}
The reduced, calibrated and science-ready
ALMA data are available from the ALMA Science Archive
at {\tt https://almascience.eso.org/asax/}.



\bibliographystyle{mnras}
\bibliography{example} % if your bibtex file is called example.bib

%\begin{thebibliography}{99}
%\end{thebibliography}

%%%%%%%%%%%%%%%%%%%%%%%%%%%%%%%%%%%%%%%%%%%%%%%%%%

%%%%%%%%%%%%%%%%% APPENDICES %%%%%%%%%%%%%%%%%%%%%

\appendix

\referee{
\section{Compilation of strong lenses}
We present a graphical depiction of strong lenses detected across the spread in 500~\micron{} flux density in Figure~\ref{fig:compilation_stronglenses}, showing the capability of FLASH to detect strong lenses down to $S_{500} = 20$~mJy.
\begin{figure*}
    \includegraphics[width=0.7 \textwidth]{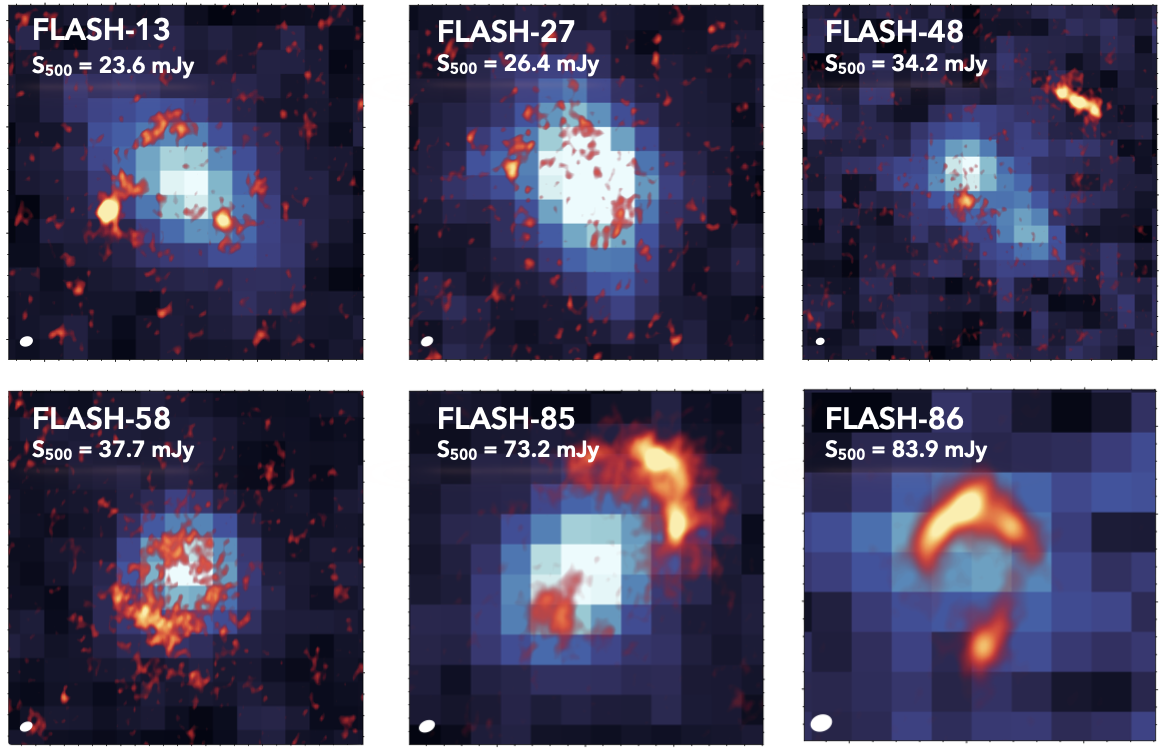}
    \caption{These six ALMA images ({\it foreground red}) indicate the ability of the FLASH method to select select strong lenses by matching \textit{Herschel} and VIKING images (\textit{Background images}) from low- to high 500~\micron{} flux densities. The images are 5~arcseconds across, except for FLASH-48 (7~arcseconds), FLASH-85 (4~arcseconds) and -86 (3~arcseconds). The \textit{red foreground} is drawn at $1\sigma$ and beyond for FLASH-13, -27 and -58, and it is drawn starting at $2 \sigma$ for FLASH-48, -85 and -86 to best show the lensing behaviour and extent. }
    \label{fig:compilation_stronglenses}
\end{figure*}
}

%%%%%%%%%%%%%%%%%%%%%%%%%%%%%%%%%%%%%%%%%%%%%%%%%%

\end{twocolumn}
% Don't change these lines
\bsp	% typesetting comment
\label{lastpage}
\end{document}

%% file: tableSample.tex
\begin{onecolumn}
\begin{landscape}
\begin{longtable}{lccccccccccc}
\caption{FLASH catalogue} \label{tab:flashcatalogue} \\
\hline \hline
Name & {\it Herschel} Position & $S_{250\micron}$ & $S_{350\micron}$ & $S_{500\micron}$ & $z_{\rm phot,submm}$ & VIKING Position & $\theta$ & z$_{\rm phot,VIK}$ & $K_S$ & LHR & $M_*$ \\ 
& [hms dms] & mJy & mJy & mJy & & [hms dms] & [''] & & mag$_{\rm AB}$ & [0 - 1] & M$_{\odot}$\\
\hline
\endfirsthead

\multicolumn{12}{l}%
{{\tablename\ \thetable{} -- {\it continued from previous page}}} \\
\hline \hline
Name & {\it Herschel} Position & $S_{250\micron}$ & $S_{350\micron}$ & $S_{500\micron}$ & $z_{\rm phot,submm}$ & VIKING Position & $\theta$ & z$_{\rm phot,VIK}$ & $K_S$ & Rel. & $\log_{10} M_*$ \\ 
& [hms dms] & mJy & mJy & mJy & & [hms dms] & [''] & & mag$_{\rm AB}$ & [0 - 1] & $\log_{10} M_{\odot}$\\ \hline
\endhead
\multicolumn{12}{r}{{\it Continued on next page...}} \\ \hline \hline
\endfoot
\hline \hline
\multicolumn{12}{p{21cm}}{\textbf{Notes:} Col. 1: Source name, sorted from lowest 500~$\mu$m flux to highest.
Col. 2: RA and DEC positions in hms and dms units, respectively.
Col. 3 - 5: The {\it Herschel} photometry  
Col. 6: Photometric redshift of the {\it Herschel} source based on the fitting of the modified black-body template from \citet{pearson13}. Typical errors on this fit are on the order of $0.13 (1 + z)$.
Col. 7: The position of the VIKING-identified source from the catalogues described in \citet{Wright2019}
Col. 8: The angular offset in units of arcseconds between the {\it Herschel} and VIKING source.
Col. 9: The photometric redshift of the VIKING source based on both VISTA and KiDS photometry best-fit \citep{Wright2019}.
Col. 10: The \textsc{AB}-magnitude in the $K_S$-band photometry.
Col. 11: The reliability (i.e., the probability) of the {\it Herschel} and VIKING sources to be associated. Note that this does not account for the inclusion of false positives in a large-area survey.
Col. 12: The stellar mass estimate of the VIKING source identified from the KIDS and VISTA photometry, as described in \citet{Wright2019}. This table is available in machine-readable form in the supplementary material.}
\endlastfoot
FLASH-1 & 12:14:36.2 -01:24:06.9 & 23.0 $\pm$ 5.9 & 32.4 $\pm$ 7.4 & 12.2 $\pm$ 7.7 & 2.08 & 12:14:36.2 -01:24:06.3 & 0.93  & $0.84_{-0.04}^{+0.03}$ & $18.87 \pm 0.02$ & 0.997 & 11.21 \\ 
FLASH-2 & 11:36:31.9 +00:40:21.7 & 30.8 $\pm$ 7.5 & 42.5 $\pm$ 8.1 & 19.7 $\pm$ 8.7 & 2.18 & 11:36:31.9 +00:40:21.8 & 0.16  & $0.75_{-0.05}^{+0.11}$ & $20.63 \pm 0.10$ & 0.997 & 10.07 \\ 
FLASH-3 & 11:46:51.9 -00:00:44.1 & 24.6 $\pm$ 6.1 & 18.4 $\pm$ 7.4 & 19.8 $\pm$ 7.8 & 2.18 & 11:46:51.9 -00:00:45.1 & 0.97  & $1.1_{-0.05}^{+0.12}$ & $20.26 \pm 0.05$ & 0.995 & 11.05 \\ 
FLASH-4 & 11:54:08.8 -01:44:16.1 & 33.2 $\pm$ 7.4 & 34.5 $\pm$ 8.2 & 20.7 $\pm$ 8.7 & 2.01 & 11:54:08.8 -01:44:16.6 & 0.57  & $1.25_{-0.13}^{+0.20}$ & $20.75 \pm 0.12$ & 0.995 & 10.95 \\ 
FLASH-5 & 12:21:23.7 +00:28:34.7 & 28.7 $\pm$ 7.3 & 33.9 $\pm$ 8.2 & 21.9 $\pm$ 8.8 & 2.32 & 12:21:23.6 +00:28:35.9 & 1.96  & $1.33_{-0.03}^{+0.03}$ & $18.94 \pm 0.03$ & 0.991 & 11.18 \\ 
FLASH-6 & 12:16:56.5 -02:37:41.2 & 29.0 $\pm$ 7.3 & 39.1 $\pm$ 8.2 & 22.2 $\pm$ 8.5 & 2.36 & 12:16:56.5 -02:37:41.9 & 0.67  & $0.63_{-0.05}^{+0.04}$ & $20.60 \pm 0.11$ & 0.995 & 9.677 \\ 
FLASH-7 & 11:59:32.4 +00:02:17.9 & 21.3 $\pm$ 5.7 & 32.7 $\pm$ 7.4 & 22.3 $\pm$ 7.7 & 2.91 & 11:59:32.4 +00:02:17.9 & 0.36  & $0.19_{-0.04}^{+0.03}$ & $20.24 \pm 0.05$ & 0.997 & 8.983 \\ 
FLASH-8 & 12:19:11.4 -00:30:36.9 & 29.4 $\pm$ 7.3 & 38.4 $\pm$ 8.0 & 22.4 $\pm$ 8.5 & 2.34 & 12:19:11.4 -00:30:36.8 & 0.9   & $1.31_{-0.04}^{+0.04}$ & $19.86 \pm 0.06$ & 0.995 & 10.72 \\ 
FLASH-9 & 12:15:14.0 -01:59:52.8 & 36.2 $\pm$ 7.0 & 39.9 $\pm$ 7.8 & 22.6 $\pm$ 8.2 & 2.04 & 12:15:14.0 -01:59:52.3 & 0.76  & $0.66_{-0.04}^{+0.02}$ & $18.76 \pm 0.01$ & 0.998 & 10.83 \\ 
FLASH-10 & 11:45:49.3 +00:20:38.0 & 27.6 $\pm$ 6.5 & 20.3 $\pm$ 7.5 & 22.8 $\pm$ 7.7 & 2.27 & 11:45:49.3 +00:20:39.2 & 1.51 & $1.28_{-0.08}^{+0.05}$ & $20.60 \pm 0.07$ & 0.99 & 10.67 \\ 
FLASH-11 & 11:34:08.3 +00:27:47.9 & 30.2 $\pm$ 7.5 & 25.3 $\pm$ 8.2 & 22.9 $\pm$ 9.3 & 2.14 & 11:34:08.3 +00:27:48.8 & 0.99 & $1.12_{-0.02}^{+0.41}$ & $19.85 \pm 0.04$ & 0.995 & 11.25 \\ 
FLASH-12 & 11:39:05.7 -01:10:30.7 & 27.8 $\pm$ 7.5 & 33.2 $\pm$ 8.2 & 23.2 $\pm$ 8.6 & 2.44 & 11:39:05.6 -01:10:30.3 & 0.93 & $0.45_{-0.04}^{+0.03}$ & $19.08 \pm 0.03$ & 0.997 & 10.41 \\ 
FLASH-13 & 12:01:18.9 -02:23:28.8 & 35.5 $\pm$ 7.2 & 35.6 $\pm$ 8.1 & 23.6 $\pm$ 8.7 & 2.06 & 12:01:18.9 -02:23:30.0 & 1.74 & $0.46_{-0.03}^{+0.02}$ & $18.19 \pm 0.01$ & 0.994 & 10.74 \\ 
FLASH-14 & 12:17:00.0 -00:35:14.8 & 34.2 $\pm$ 7.4 & 29.8 $\pm$ 8.3 & 23.9 $\pm$ 8.9 & 2.04 & 12:17:00.0 -00:35:15.3 & 0.59 & $1.26_{-0.10}^{+0.09}$ & $19.83 \pm 0.04$ & 0.997 & 11.23 \\ 
FLASH-15 & 11:59:30.8 -01:04:52.5 & 29.1 $\pm$ 6.4 & 29.0 $\pm$ 7.3 & 24.3 $\pm$ 7.6 & 2.4 & 11:59:30.8 -01:04:54.0 & 1.65  & $0.69_{-0.03}^{+0.03}$ & $18.78 \pm 0.02$ & 0.994 & 10.70 \\ 
FLASH-16 & 12:19:34.2 +00:22:15.4 & 31.5 $\pm$ 7.4 & 21.4 $\pm$ 8.1 & 24.3 $\pm$ 8.6 & 2.07 & 12:19:34.2 +00:22:15.4 & 0.17 & $0.95_{-0.04}^{+0.03}$ & $18.67 \pm 0.02$ & 0.999 & 11.38 \\ 
FLASH-17 & 11:36:41.7 -01:17:26.6 & 29.5 $\pm$ 7.3 & 34.3 $\pm$ 8.2 & 24.6 $\pm$ 8.4 & 2.43 & 11:36:41.8 -01:17:26.4 & 0.2  & $1.05_{-0.05}^{+0.03}$ & $19.51 \pm 0.02$ & 0.998 & 11.04 \\ 
FLASH-18 & 11:44:18.6 +00:04:54.3 & 30.0 $\pm$ 6.5 & 28.0 $\pm$ 7.5 & 24.7 $\pm$ 7.7 & 2.35 & 11:44:18.5 +00:04:53.9 & 1.18 & $1.07_{-0.06}^{+0.04}$ & $19.60 \pm 0.04$ & 0.994 & 11.01 \\ 
FLASH-19 & 11:56:00.4 -01:49:55.2 & 32.7 $\pm$ 7.5 & 31.9 $\pm$ 8.2 & 24.8 $\pm$ 8.6 & 2.23 & 11:56:00.4 -01:49:54.9 & 0.35 & $1.01_{-0.05}^{+0.03}$ & $19.28 \pm 0.03$ & 0.998 & 11.02 \\ 
FLASH-20 & 12:02:08.7 -02:34:13.6 & 37.1 $\pm$ 7.2 & 37.9 $\pm$ 8.1 & 24.9 $\pm$ 8.6 & 2.08 & 12:02:08.7 -02:34:12.7 & 0.89 & $0.72_{-0.04}^{+0.02}$ & $18.87 \pm 0.02$ & 0.997 & 10.67 \\ 
FLASH-21 & 12:14:25.6 -00:43:15.9 & 35.4 $\pm$ 6.5 & 39.0 $\pm$ 7.5 & 25.0 $\pm$ 7.7 & 2.18 & 12:14:25.6 -00:43:16.4 & 0.73 & $0.63_{-0.31}^{+0.02}$ & $20.45 \pm 0.09$ & 0.996 & 9.923 \\ 
FLASH-22 & 11:50:34.9 -00:54:26.1 & 37.7 $\pm$ 7.4 & 37.8 $\pm$ 8.1 & 25.1 $\pm$ 8.5 & 2.06 & 11:50:35.0 -00:54:24.9 & 1.42 & $1.02_{-0.04}^{+0.03}$ & $18.98 \pm 0.02$ & 0.995 & 11.20 \\ 
FLASH-23 & 11:57:23.2 +01:13:12.5 & 30.1 $\pm$ 7.3 & 26.8 $\pm$ 8.1 & 25.3 $\pm$ 8.6 & 2.38 & 11:57:23.2 +01:13:12.6 & 0.15 & $1.32_{-0.07}^{+0.06}$ & $20.20 \pm 0.10$ & 0.998 & 10.75 \\ 
FLASH-24 & 11:56:49.2 +00:35:49.1 & 26.7 $\pm$ 6.4 & 23.4 $\pm$ 7.4 & 25.4 $\pm$ 7.6 & 2.66 & 11:56:49.2 +00:35:47.9 & 1.39 & $0.72_{-0.04}^{+0.02}$ & $19.55 \pm 0.03$ & 0.994 & 10.65 \\ 
FLASH-25 & 11:50:55.9 -01:33:54.9 & 33.8 $\pm$ 7.3 & 22.3 $\pm$ 8.0 & 26.0 $\pm$ 8.6 & 2.04 & 11:50:56.0 -01:33:55.5 & 1.56 & $0.66_{-0.04}^{+0.02}$ & $19.11 \pm 0.06$ & 0.994 & 10.59 \\ 
FLASH-26 & 11:46:26.8 -01:45:41.4 & 38.4 $\pm$ 6.5 & 34.9 $\pm$ 7.4 & 26.3 $\pm$ 7.8 & 2.04 & 11:46:26.7 -01:45:40.4 & 1.51 & $1.2_{-0.10}^{+0.09}$ & $20.01 \pm 0.06$ & 0.992 & 11.07 \\ 
FLASH-27 & 11:53:05.8 -00:37:46.0 & 31.8 $\pm$ 7.4 & 35.4 $\pm$ 8.3 & 26.4 $\pm$ 8.8 & 2.42 & 11:53:05.7 -00:37:45.0 & 1.78 & $0.56_{-0.04}^{+0.03}$ & $17.98 \pm 0.01$ & 0.994 & 10.97 \\ 
FLASH-28 & 11:48:36.3 +01:01:10.7 & 30.8 $\pm$ 7.3 & 23.5 $\pm$ 8.1 & 26.4 $\pm$ 8.7 & 2.36 & 11:48:36.4 +01:01:10.7 & 0.85 & $0.98_{-0.08}^{+0.09}$ & $20.79 \pm 0.12$ & 0.994 & 10.36 \\ 
FLASH-29 & 12:04:29.4 -00:42:43.4 & 29.7 $\pm$ 7.1 & 33.7 $\pm$ 7.8 & 26.6 $\pm$ 8.2 & 2.55 & 12:04:29.4 -00:42:43.3 & 0.63 & $1.06_{-0.06}^{+0.06}$ & $19.88 \pm 0.05$ & 0.997 & 10.64 \\ 
FLASH-30 & 12:14:41.5 +00:24:08.1 & 32.5 $\pm$ 7.3 & 34.1 $\pm$ 8.2 & 26.7 $\pm$ 8.8 & 2.38 & 12:14:41.6 +00:24:07.5 & 0.81 & $0.67_{-0.03}^{+0.03}$ & $18.47 \pm 0.01$ & 0.998 & 10.92 \\ 
FLASH-31 & 12:01:37.1 -00:04:19.1 & 34.8 $\pm$ 7.3 & 41.1 $\pm$ 8.2 & 26.9 $\pm$ 8.7 & 2.33 & 12:01:37.1 -00:04:20.2 & 1.03 & $0.69_{-0.04}^{+0.02}$ & $19.29 \pm 0.04$ & 0.996 & 10.56 \\ 
FLASH-32 & 11:59:37.8 -00:06:25.8 & 32.6 $\pm$ 6.5 & 35.5 $\pm$ 7.4 & 27.1 $\pm$ 7.7 & 2.42 & 11:59:37.9 -00:06:23.8 & 2.58 & $0.73_{-0.06}^{+0.04}$ & $20.45 \pm 0.07$ & 0.964 & 10.06 \\ 
FLASH-33 & 12:01:28.9 -01:10:15.9 & 26.3 $\pm$ 6.3 & 24.8 $\pm$ 7.4 & 27.6 $\pm$ 7.7 & 2.91 & 12:01:28.9 -01:10:16.0 & 0.21 & $0.89_{-0.04}^{+0.04}$ & $19.28 \pm 0.03$ & 0.997 & 10.96 \\ 
FLASH-34 & 12:25:09.9 -00:18:06.0 & 38.5 $\pm$ 7.3 & 35.8 $\pm$ 8.1 & 27.6 $\pm$ 8.6 & 2.12 & 12:25:09.9 -00:18:07.4 & 1.34 & $0.23_{-0.02}^{+0.47}$ & $20.59 \pm 0.09$ & 0.989 & 9.012 \\ 
FLASH-35 & 12:06:08.8 -00:34:57.5 & 34.1 $\pm$ 7.0 & 36.1 $\pm$ 7.9 & 28.4 $\pm$ 8.3 & 2.42 & 12:06:08.7 -00:34:57.2 & 0.69 & $0.07_{-0.03}^{+0.03}$ & $20.31 \pm 0.10$ & 0.995 & 8.116 \\ 
FLASH-36 & 11:54:37.3 +00:59:37.0 & 37.5 $\pm$ 7.3 & 37.0 $\pm$ 8.3 & 28.5 $\pm$ 8.9 & 2.23 & 11:54:37.7 +00:59:33.9 & 6.41 & $0.54_{-0.04}^{+0.04}$ & $18.79 \pm 0.01$ & 0.961 & 10.63 \\ 
FLASH-37 & 12:17:23.4 -02:06:00.3 & 43.3 $\pm$ 7.2 & 41.4 $\pm$ 8.2 & 28.5 $\pm$ 8.7 & 2.02 & 12:17:23.4 -02:05:59.2 & 1.06 & $1.06_{-0.05}^{+0.06}$ & $19.71 \pm 0.03$ & 0.995 & 10.66 \\ 
FLASH-38 & 11:49:25.7 -02:07:21.2 & 36.5 $\pm$ 7.2 & 40.9 $\pm$ 8.3 & 28.5 $\pm$ 8.7 & 2.33 & 11:49:25.7 -02:07:23.4 & 2.14 & $0.64_{-0.05}^{+0.03}$ & $20.18 \pm 0.09$ & 0.982 & 10.00 \\ 
FLASH-39 & 12:19:50.5 +00:33:35.5 & 39.4 $\pm$ 7.4 & 39.9 $\pm$ 8.2 & 29.0 $\pm$ 8.9 & 2.2 & 12:19:50.5 +00:33:41.6 & 6.08  & $0.59_{-0.04}^{+0.02}$ & $19.01 \pm 0.03$ & 0.945 & 10.51 \\ 
FLASH-40 & 11:59:41.3 +00:02:41.7 & 39.3 $\pm$ 6.5 & 38.5 $\pm$ 7.4 & 29.0 $\pm$ 7.7 & 2.19 & 11:59:41.2 +00:02:41.2 & 0.8  & $1.19_{-0.14}^{+0.05}$ & $20.59 \pm 0.07$ & 0.98 & 10.34 \\ 
FLASH-41 & 12:13:48.9 -01:03:11.4 & 43.0 $\pm$ 6.5 & 40.5 $\pm$ 7.4 & 29.4 $\pm$ 7.7 & 2.05 & 12:13:48.8 -01:03:13.8 & 2.65 & $1.16_{-0.06}^{+0.04}$ & $19.70 \pm 0.06$ & 0.975 & 10.80 \\ 
FLASH-42 & 11:59:10.0 -01:20:58.1 & 31.1 $\pm$ 6.6 & 27.0 $\pm$ 7.5 & 30.1 $\pm$ 7.8 & 2.71 & 11:59:10.0 -01:20:57.3 & 0.92 & $0.64_{-0.03}^{+0.03}$ & $19.04 \pm 0.04$ & 0.997 & 10.47 \\ 
FLASH-43 & 12:23:05.3 -01:13:10.2 & 40.2 $\pm$ 7.4 & 43.6 $\pm$ 8.1 & 30.3 $\pm$ 8.6 & 2.26 & 12:23:05.2 -01:13:12.1 & 2.27 & $0.17_{-0.03}^{+0.03}$ & $19.35 \pm 0.04$ & 0.987 & 9.236 \\ 
FLASH-44 & 12:00:47.6 -00:40:19.6 & 11.1 $\pm$ 4.7 & 23.1 $\pm$ 7.4 & 31.3 $\pm$ 7.7 & 5.56 & 12:00:47.6 -00:40:18.8 & 1.03 & $0.75_{-0.04}^{+0.04}$ & $19.18 \pm 0.04$ & 0.996 & 10.85 \\ 
FLASH-45 & 11:36:53.6 +01:16:32.9 & 32.2 $\pm$ 7.2 & 24.8 $\pm$ 8.2 & 32.0 $\pm$ 8.8 & 2.76 & 11:36:53.6 +01:16:33.7 & 0.82 & $1.23_{-0.04}^{+0.02}$ & $19.19 \pm 0.01$ & 0.997 & 10.99 \\ 
FLASH-46 & 12:14:12.1 -00:52:13.6 & 34.6 $\pm$ 6.5 & 33.7 $\pm$ 7.5 & 33.0 $\pm$ 7.7 & 2.67 & 12:14:12.2 -00:52:13.7 & 1.17 & $0.44_{-0.04}^{+0.02}$ & $18.56 \pm 0.01$ & 0.969 & 10.57 \\ 
FLASH-47 & 12:20:19.6 -00:39:12.3 & 50.8 $\pm$ 7.3 & 53.3 $\pm$ 8.2 & 34.1 $\pm$ 8.5 & 2.1 & 12:20:19.5 -00:39:11.7 & 1.29  & $0.25_{-0.06}^{+0.05}$ & $20.68 \pm 0.11$ & 0.992 & 8.972 \\ 
FLASH-48 & 12:01:15.4 -01:27:22.0 & 14.6 $\pm$ 5.0 & 20.5 $\pm$ 7.4 & 34.2 $\pm$ 7.8 & 5.1 & 12:01:15.5 -01:27:23.0 & 1.53  & $0.96_{-0.04}^{+0.03}$ & $19.38 \pm 0.03$ & 0.995 & 11.19 \\ 
FLASH-49 & 11:58:05.5 -01:55:47.6 & 43.2 $\pm$ 6.8 & 36.2 $\pm$ 7.6 & 34.3 $\pm$ 7.9 & 2.25 & 11:58:05.4 -01:55:47.9 & 1.45 & $0.95_{-0.04}^{+0.03}$ & $19.23 \pm 0.02$ & 0.986 & 10.78 \\ 
FLASH-50 & 12:05:31.3 -00:37:00.5 & 39.1 $\pm$ 7.0 & 43.8 $\pm$ 7.8 & 34.7 $\pm$ 8.3 & 2.54 & 12:05:31.3 -00:37:01.1 & 1.1  & $0.37_{-0.03}^{+0.17}$ & $19.34 \pm 0.03$ & 0.996 & 10.29 \\ 
FLASH-51 & 12:01:41.9 -02:20:06.2 & 41.2 $\pm$ 7.4 & 32.5 $\pm$ 8.2 & 34.8 $\pm$ 8.5 & 2.37 & 12:01:41.9 -02:20:04.6 & 1.85 & $0.69_{-0.04}^{+0.03}$ & $18.41 \pm 0.02$ & 0.994 & 11.03 \\ 
FLASH-52 & 11:50:09.4 -00:36:51.9 & 34.7 $\pm$ 7.4 & 40.2 $\pm$ 8.3 & 35.1 $\pm$ 8.8 & 2.8 & 11:50:09.4 -00:36:51.8 & 0.42  & $1.43_{-0.08}^{+0.09}$ & $20.24 \pm 0.10$ & 0.997 & 11.02 \\ 
FLASH-53 & 11:49:21.6 -01:03:02.1 & 40.6 $\pm$ 7.1 & 41.8 $\pm$ 8.0 & 35.2 $\pm$ 8.4 & 2.48 & 11:49:21.6 -01:03:01.9 & 0.23 & $0.53_{-0.06}^{+0.10}$ & $20.20 \pm 0.08$ & 0.997 & 10.14 \\ 
FLASH-54 & 12:08:06.2 +00:45:10.3 & 27.5 $\pm$ 6.5 & 38.9 $\pm$ 7.4 & 35.8 $\pm$ 7.7 & 3.38 & 12:08:06.1 +00:45:10.3 & 1.05 & $0.87_{-0.04}^{+0.04}$ & $19.54 \pm 0.05$ & 0.996 & 10.95 \\ 
FLASH-55 & 11:55:00.7 -00:07:22.0 & 22.4 $\pm$ 6.8 & 31.8 $\pm$ 8.3 & 36.3 $\pm$ 8.7 & 4.04 & 11:55:00.6 -00:07:21.2 & 1.26 & $0.26_{-0.03}^{+0.03}$ & $19.51 \pm 0.05$ & 0.994 & 9.235 \\ 
FLASH-56 & 12:16:54.3 -01:27:29.9 & 24.8 $\pm$ 7.1 & 40.9 $\pm$ 8.2 & 36.6 $\pm$ 8.8 & 3.68 & 12:16:54.4 -01:27:31.1 & 1.35 & $0.7_{-0.03}^{+0.03}$ & $19.93 \pm 0.04$ & 0.994 & 10.19 \\ 
FLASH-57 & 11:46:10.1 -00:50:28.4 & 39.4 $\pm$ 6.5 & 41.7 $\pm$ 7.4 & 37.0 $\pm$ 7.8 & 2.64 & 11:46:10.0 -00:50:27.4 & 1.67 & $1.43_{-0.06}^{+0.13}$ & $19.93 \pm 0.06$ & 0.991 & 10.99 \\ 
FLASH-58 & 11:43:59.8 -00:16:00.1 & 36.0 $\pm$ 6.5 & 41.0 $\pm$ 7.5 & 37.7 $\pm$ 7.8 & 2.88 & 11:43:59.9 -00:16:01.1 & 1.4  & $0.44_{-0.03}^{+0.04}$ & $18.46 \pm 0.03$ & 0.996 & 10.72 \\ 
FLASH-59 & 12:22:11.4 -01:41:53.8 & 37.6 $\pm$ 7.3 & 39.4 $\pm$ 8.1 & 37.8 $\pm$ 8.5 & 2.8 & 12:22:11.4 -01:41:53.9 & 1.19  & $0.2_{-0.04}^{+0.03}$ & $20.70 \pm 0.12$ & 0.989 & 9.216 \\ 
FLASH-60 & 11:50:55.0 -00:44:06.1 & 52.7 $\pm$ 7.5 & 51.6 $\pm$ 8.1 & 38.7 $\pm$ 8.6 & 2.18 & 11:50:54.9 -00:44:06.6 & 1.09 & $0.33_{-0.05}^{+0.05}$ & $20.31 \pm 0.09$ & 0.994 & 9.426 \\ 
FLASH-61 & 12:14:27.1 -02:24:46.6 & 24.7 $\pm$ 7.0 & 25.3 $\pm$ 8.1 & 39.0 $\pm$ 8.5 & 4.09 & 12:14:27.1 -02:24:45.5 & 1.23 & $0.59_{-0.05}^{+0.03}$ & $19.98 \pm 0.06$ & 0.994 & 10.15 \\ 
FLASH-62 & 12:14:02.6 -01:43:07.4 & 42.7 $\pm$ 6.5 & 57.1 $\pm$ 7.4 & 39.0 $\pm$ 7.6 & 2.61 & 12:14:02.6 -01:43:04.9 & 2.83 & $0.92_{-0.07}^{+0.37}$ & $20.87 \pm 0.19$ & 0.944 & 10.35 \\ 
FLASH-63 & 11:44:39.9 +00:54:32.4 & 48.2 $\pm$ 6.5 & 35.4 $\pm$ 7.4 & 39.4 $\pm$ 7.7 & 2.24 & 11:44:39.9 +00:54:30.9 & 1.61 & $0.93_{-0.10}^{+0.06}$ & $20.31 \pm 0.07$ & 0.99 & 10.63 \\ 
FLASH-64 & 11:58:50.0 -00:57:08.3 & 57.4 $\pm$ 6.5 & 59.6 $\pm$ 7.4 & 39.5 $\pm$ 7.6 & 2.12 & 11:58:49.8 -00:57:08.3 & 3.97 & $0.47_{-0.03}^{+0.02}$ & $18.28 \pm 0.01$ & 0.938 & 10.71 \\ 
FLASH-65 & 12:10:58.0 -00:44:38.3 & 50.8 $\pm$ 6.5 & 63.0 $\pm$ 7.5 & 39.7 $\pm$ 7.8 & 2.36 & 12:10:58.0 -00:44:40.0 & 1.91 & $0.71_{-0.06}^{+0.03}$ & $19.42 \pm 0.04$ & 0.991 & 10.66 \\ 
FLASH-66 & 12:13:58.5 +01:10:47.6 & 37.1 $\pm$ 7.2 & 46.0 $\pm$ 8.1 & 40.9 $\pm$ 8.9 & 3.0 & 12:13:58.5 +01:10:48.3 & 1.2   & $1.14_{-0.06}^{+0.04}$ & $19.71 \pm 0.04$ & 0.994 & 10.89 \\ 
FLASH-67 & 12:24:46.0 -01:52:39.9 & 36.8 $\pm$ 7.3 & 40.4 $\pm$ 8.3 & 41.9 $\pm$ 8.7 & 3.11 & 12:24:46.0 -01:52:39.7 & 0.2  & $0.07_{-0.04}^{+0.02}$ & $20.86 \pm 0.13$ & 0.997 & 7.860 \\ 
FLASH-68 & 12:18:20.7 +01:13:46.5 & 64.6 $\pm$ 7.4 & 58.5 $\pm$ 8.1 & 44.3 $\pm$ 8.8 & 2.04 & 12:18:20.7 +01:13:48.5 & 2.07 & $1.2_{-0.09}^{+0.05}$ & $19.76 \pm 0.03$ & 0.988 & 11.15 \\ 
FLASH-69 & 11:44:40.4 +00:26:49.2 & 59.1 $\pm$ 6.4 & 54.4 $\pm$ 7.4 & 45.1 $\pm$ 7.7 & 2.21 & 11:44:40.4 +00:26:50.4 & 1.14 & $0.61_{-0.06}^{+0.03}$ & $20.47 \pm 0.12$ & 0.992 & 10.02 \\ 
FLASH-70 & 12:13:57.7 +00:09:11.0 & 38.9 $\pm$ 7.5 & 41.9 $\pm$ 8.2 & 45.7 $\pm$ 8.7 & 3.21 & 12:13:57.7 +00:09:11.0 & 0.15 & $1.34_{-0.11}^{+0.19}$ & $20.94 \pm 0.11$ & 0.985 & 10.48 \\ 
FLASH-71 & 12:24:07.6 -00:33:31.8 & 52.9 $\pm$ 7.4 & 42.8 $\pm$ 8.2 & 45.9 $\pm$ 8.4 & 2.44 & 12:24:07.6 -00:33:32.6 & 1.09 & $1.08_{-0.09}^{+0.14}$ & $20.46 \pm 0.08$ & 0.993 & 10.77 \\ 
FLASH-72 & 12:06:32.2 -00:33:11.2 & 58.0 $\pm$ 7.0 & 58.5 $\pm$ 7.8 & 47.8 $\pm$ 8.2 & 2.38 & 12:06:32.1 -00:33:12.4 & 2.23 & $0.49_{-0.06}^{+0.03}$ & $19.84 \pm 0.08$ & 0.983 & 9.985 \\ 
FLASH-73 & 12:05:53.3 -01:51:18.0 & 67.7 $\pm$ 7.3 & 83.9 $\pm$ 8.3 & 51.4 $\pm$ 8.6 & 2.32 & 12:05:53.4 -01:51:18.0 & 0.42 & $0.6_{-0.04}^{+0.03}$ & $18.24 \pm 0.01$ & 0.985 & 10.92 \\ 
FLASH-74 & 11:39:32.4 -01:54:58.5 & 74.4 $\pm$ 7.4 & 63.3 $\pm$ 8.1 & 52.4 $\pm$ 8.8 & 2.04 & 11:39:32.4 -01:54:58.8 & 0.4  & $0.98_{-0.04}^{+0.05}$ & $19.88 \pm 0.04$ & 0.998 & 10.84 \\ 
FLASH-75 & 12:07:00.2 -01:13:03.3 & 59.1 $\pm$ 7.5 & 54.0 $\pm$ 8.2 & 52.6 $\pm$ 8.9 & 2.51 & 12:07:00.2 -01:13:03.2 & 0.13 & $0.63_{-0.03}^{+0.02}$ & $18.45 \pm 0.01$ & 0.999 & 10.72 \\ 
FLASH-76 & 12:02:23.8 +01:03:17.3 & 73.0 $\pm$ 7.4 & 69.1 $\pm$ 8.1 & 52.6 $\pm$ 8.6 & 2.14 & 12:02:23.8 +01:03:17.5 & 0.18 & $1.03_{-0.08}^{+0.10}$ & $20.44 \pm 0.10$ & 0.998 & 10.28 \\ 
FLASH-77 & 12:05:43.1 -00:38:33.1 & 75.9 $\pm$ 7.1 & 82.8 $\pm$ 8.0 & 55.5 $\pm$ 8.6 & 2.23 & 12:05:42.9 -00:38:34.3 & 3.07 & $0.7_{-0.06}^{+0.06}$ & $19.55 \pm 0.05$ & 0.955 & 10.78 \\ 
FLASH-78 & 11:39:01.2 -02:14:12.1 & 51.2 $\pm$ 7.4 & 61.4 $\pm$ 8.1 & 55.5 $\pm$ 8.7 & 2.96 & 11:39:01.1 -02:14:11.6 & 1.66 & $1.07_{-0.07}^{+0.12}$ & $20.20 \pm 0.07$ & 0.99 & 10.89 \\ 
FLASH-79 & 11:45:15.4 -00:08:44.9 & 46.6 $\pm$ 6.5 & 57.7 $\pm$ 7.5 & 56.4 $\pm$ 7.7 & 3.23 & 11:45:15.3 -00:08:42.3 & 3.02 & $0.67_{-0.09}^{+0.24}$ & $20.76 \pm 0.13$ & 0.812 & 10.24 \\ 
FLASH-80 & 11:38:16.0 -01:31:18.1 & 60.1 $\pm$ 7.4 & 74.8 $\pm$ 8.0 & 60.1 $\pm$ 8.5 & 2.77 & 11:38:16.1 -01:31:20.5 & 3.11 & $1.2_{-0.10}^{+0.07}$ & $20.45 \pm 0.08$ & 0.945 & 10.76 \\ 
FLASH-81 & 11:57:14.8 -00:28:41.9 & 84.2 $\pm$ 7.3 & 88.1 $\pm$ 8.1 & 62.1 $\pm$ 8.7 & 2.22 & 11:57:14.8 -00:28:40.8 & 1.45 & $0.95_{-0.05}^{+0.04}$ & $19.91 \pm 0.05$ & 0.993 & 10.85 \\ 
FLASH-82 & 11:52:51.5 -01:52:20.0 & 91.4 $\pm$ 7.3 & 93.7 $\pm$ 8.1 & 62.7 $\pm$ 8.5 & 2.11 & 11:52:51.4 -01:52:19.1 & 2.56 & $1.28_{-0.06}^{+0.05}$ & $20.41 \pm 0.07$ & 0.978 & 10.76 \\ 
FLASH-83 & 11:44:04.4 +00:38:49.9 & 96.8 $\pm$ 6.6 & 94.7 $\pm$ 7.5 & 69.3 $\pm$ 7.7 & 2.14 & 11:44:04.4 +00:38:43.8 & 6.14 & $1.04_{-0.03}^{+0.06}$ & $18.42 \pm 0.01$ & 0.954 & 11.43 \\ 
FLASH-84 & 11:36:16.6 +00:48:54.7 & 90.6 $\pm$ 7.3 & 99.2 $\pm$ 8.1 & 71.7 $\pm$ 8.7 & 2.34 & 11:36:17.0 +00:48:51.7 & 6.39 & $0.55_{-0.03}^{+0.04}$ & $19.50 \pm 0.05$ & 0.944 & 10.34 \\ 
FLASH-85 & 11:37:39.2 -02:44:43.5 & 63.2 $\pm$ 7.4 & 85.8 $\pm$ 8.2 & 73.2 $\pm$ 8.7 & 3.09 & 11:37:39.1 -02:44:45.0 & 2.23 & $0.57_{-0.05}^{+0.03}$ & $18.65 \pm 0.01$ & 0.99 & 10.78 \\ 
FLASH-86 & 11:54:33.7 +00:50:42.3 & 53.9 $\pm$ 7.4 & 85.8 $\pm$ 8.1 & 83.9 $\pm$ 8.6 & 3.85 & 11:54:33.7 +00:50:42.1 & 0.51 & $0.69_{-0.06}^{+0.04}$ & $20.26 \pm 0.13$ & 0.997 & 10.57 \\ 
\end{longtable}
\end{landscape}
\end{onecolumn}

%% file: tableResults.tex
\begin{onecolumn}
% \begin{landscape}
\begin{longtable}{lccrcc}
\caption{ALMA observations catalogue} \label{tab:catResults} \\
\hline \hline
Name & Lens? & ALMA Position & \multicolumn{1}{c}{$S_{\rm 1.1mm}$} & $\theta_{\rm ALMA,VIK}$ & $\theta_{{\rm ALMA,}Herschel}$ \\ 
& & [hms dms] & \multicolumn{1}{c}{mJy}   & [''] & ['']\\
\hline
\endfirsthead

\multicolumn{3}{l}%
{{\tablename\ \thetable{} -- {\it continued from previous page}}} \\
\hline \hline
Name & Lens? & ALMA Position & \multicolumn{1}{c}{$S_{\rm 1.1mm}$} & $\theta_{\rm ALMA,VIK}$ & $\theta_{{\rm ALMA,}Herschel}$ \\ 
& & [hms dms] & \multicolumn{1}{c}{mJy}   & [''] & [''] \\
\hline
\endhead
\multicolumn{6}{r}{{\it Continued on next page}} \\ \hline \hline
\endfoot
\hline \hline
\multicolumn{6}{p{\textwidth}}{\textbf{Notes:} 
Col. 1: Source name. We identify the fluxes of the individually-extracted fluxes (red boxes in Figure~\ref{fig:sources1}), and for extended or lensed sources, we provide the total flux and weighted signal-to-noise ratios of the lensed components in the sources indicated with italics. The additional letters in the naming convention refer to N, E (left in figures), S, W (right in figures) for North, East, South and West. C refers to Centre, for sources closer to the centre than others.
Col. 2: The identification of the source, as discussed in Section~\ref{sec:identificationOfLenses}.
Col. 3: The RA and DEC position of the ALMA positions in units of hms and dms respectively. We do not provide the position for the combined lensed sources, as these do not represent any physical position.
Col. 4: The 1.1~mm flux density followed by the signal-to-noise ratio of the detection in brackets. 
Col. 5: The angular offset between the ALMA and VIKING position in units of arcseconds.
Col. 5: The angular offset between the ALMA and {\it Herschel} position in units of arcseconds. This table is available in machine-readable form in the supplementary material.}
\endlastfoot
FLASH-1NE & C-grade & 12:14:36.9 -01:24:03.0 & 2.14 $\pm$  0.66 (3.3$\sigma$) & 10.0 & 10.9 \\ 
FLASH-1NW & ... & 12:14:36.7 -01:24:11.7 & 1.17 $\pm$  0.4 (2.9$\sigma$) & 9.1 & 9.4 \\ 
FLASH-2S & C-grade & 11:36:31.5 +00:40:22.4 & 6.81 $\pm$  0.3 (22.4$\sigma$) & 4.7 & 4.6 \\ 
FLASH-2NW & ... & 11:36:32.3 +00:40:14.5 & 2.1 $\pm$  0.7 (3.0$\sigma$) & 9.7 & 9.8 \\ 
FLASH-3 & A-grade & 11:46:51.9 -00:00:45.0 & 0.96 $\pm$  0.19 (5.0$\sigma$) & 0.3 & 1.0 \\ 
FLASH-4E & B-grade & 11:54:08.9 -01:44:12.8 & 2.13 $\pm$  0.32 (6.7$\sigma$) & 3.9 & 3.3 \\ 
FLASH-4C & ... & 11:54:08.8 -01:44:16.6 & 0.67 $\pm$  0.19 (3.5$\sigma$) & 0.4 & 0.5 \\ 
FLASH-4S & ... & 11:54:08.3 -01:44:14.3 & 1.72 $\pm$  0.58 (3.0$\sigma$) & 8.2 & 7.9 \\ 
FLASH-5E & C-grade & 12:21:23.8 +00:28:41.2 & 1.83 $\pm$  0.4 (4.6$\sigma$) & 6.2 & 6.7 \\ 
FLASH-5W & ... & 12:21:23.3 +00:28:28.2 & 1.53 $\pm$  0.57 (2.7$\sigma$) & 8.9 & 8.9 \\ 
FLASH-6 & C-grade & 12:16:56.6 -02:37:42.2 & 0.49 $\pm$  0.28 (1.8$\sigma$) & 1.7 & 2.0 \\ 
FLASH-7W & C-grade & 11:59:32.4 +00:02:14.0 & 2.28 $\pm$  0.25 (9.1$\sigma$) & 4.0 & 3.9 \\ 
FLASH-7SW & ... & 11:59:32.2 +00:02:13.3 & 1.66 $\pm$  0.53 (3.1$\sigma$) & 5.8 & 5.5 \\ 
FLASH-8C & B-grade & 12:19:11.4 -00:30:36.8 & 4.01 $\pm$  0.32 (12.7$\sigma$) & 0.1 & 0.8 \\ 
FLASH-8W & ... & 12:19:11.2 -00:30:44.4 & 1.03 $\pm$  0.48 (2.2$\sigma$) & 8.2 & 7.8 \\ 
FLASH-9W & B-grade & 12:15:14.0 -01:59:57.6 & 1.44 $\pm$  0.28 (5.1$\sigma$) & 5.3 & 4.8 \\ 
FLASH-9C & ... & 12:15:14.0 -01:59:52.6 & 1.16 $\pm$  0.37 (3.1$\sigma$) & 0.4 & 0.9 \\ 
FLASH-10 & B-grade & 11:45:49.3 +00:20:39.3 & 1.05 $\pm$  0.28 (3.7$\sigma$) & 0.1 & 1.6 \\ 
FLASH-11 & B-grade & 11:34:08.3 +00:27:48.8 & 1.1 $\pm$  0.34 (3.3$\sigma$) & 0.2 & 1.1 \\ 
% FLASH-11 & ... & 11:34:08.5 +00:27:48.1 & -0.54 $\pm$  -0.4 (1.4$\sigma$) & 3.2 & 2.8 \\ 
FLASH-12 & A-grade & 11:39:05.7 -01:10:30.0 & 3.63 $\pm$  0.23 (15.5$\sigma$) & 1.9 & 1.2 \\ 
\textit{FLASH-13} & A-grade &  & 3.68 $\pm$  0.31 (11.78$\sigma$) &  &  \\ 
FLASH-13E & ... & 12:01:18.9 -02:23:28.9 & 2.4 $\pm$  0.21 (11.5$\sigma$) & 1.1 & 0.9 \\ 
FLASH-13N & ... & 12:01:18.9 -02:23:30.6 & 0.98 $\pm$  0.21 (4.6$\sigma$) & 0.8 & 1.9 \\ 
FLASH-13SW & ... & 12:01:19.0 -02:23:29.7 & 0.3 $\pm$  0.09 (3.2$\sigma$) & 0.9 & 2.3 \\ 
FLASH-14 & A-grade & 12:17:00.0 -00:35:15.4 & 1.7 $\pm$  0.3 (5.6$\sigma$) & 0.4 & 0.7 \\ 
FLASH-15 & C-grade & 11:59:31.4 -01:04:51.4 & 5.46 $\pm$  0.73 (7.4$\sigma$) & 9.8 & 8.8 \\ 
FLASH-16 & B-grade & 12:19:34.2 +00:22:15.3 & 0.71 $\pm$  0.19 (3.8$\sigma$) & 0.1 & 0.1 \\ 
FLASH-17SW & B-grade & 11:36:41.3 -01:17:33.0 & 2.29 $\pm$  0.52 (4.4$\sigma$) & 10.0 & 9.8 \\ 
FLASH-17C & ... & 11:36:41.7 -01:17:26.4 & 1.41 $\pm$  0.32 (4.3$\sigma$) & 0.2 & 0.2 \\ 
FLASH-18 & C-grade & 11:44:18.7 +00:04:58.0 & 7.18 $\pm$  0.32 (22.2$\sigma$) & 4.8 & 4.0 \\ 
FLASH-19 & B-grade & 11:56:00.4 -01:49:54.8 & 0.46 $\pm$  0.21 (2.2$\sigma$) & 0.4 & 0.8 \\ 
FLASH-20 & A-grade & 12:02:08.7 -02:34:14.4 & 0.31 $\pm$  0.16 (1.9$\sigma$) & 1.6 & 0.8 \\ 
\textit{FLASH-21} & A-grade &  & 3.87 $\pm$  0.29 (13.29$\sigma$) &  &  \\ 
FLASH-21C & ... & 12:14:25.5 -00:43:16.9 & 2.77 $\pm$  0.22 (12.8$\sigma$) & 0.7 & 1.4 \\ 
FLASH-21N & ... & 12:14:25.7 -00:43:16.3 & 1.1 $\pm$  0.19 (5.6$\sigma$) & 1.5 & 1.1 \\ 
FLASH-22SW & B-grade & 11:50:34.8 -00:54:24.0 & 3.41 $\pm$  0.36 (9.5$\sigma$) & 2.7 & 2.8 \\ 
FLASH-22S & ... & 11:50:34.7 -00:54:25.2 & 2.35 $\pm$  0.26 (9.1$\sigma$) & 4.9 & 4.2 \\ 
FLASH-23C & A-grade & 11:57:23.2 +01:13:12.7 & 1.92 $\pm$  0.38 (5.0$\sigma$) & 0.1 & 0.2 \\ 
FLASH-23S & ... & 11:57:22.6 +01:13:09.9 & 0.57 $\pm$  0.22 (2.6$\sigma$) & 8.5 & 8.4 \\ 
FLASH-24 & C-grade & 11:56:49.7 +00:35:41.4 & 2.78 $\pm$  0.62 (4.4$\sigma$) & 9.5 & 9.9 \\ 
FLASH-25 & C-grade & 11:50:55.6 -01:33:58.1 & 1.41 $\pm$  0.48 (3.0$\sigma$) & 7.0 & 6.0 \\ 
FLASH-26E & A-grade & 11:46:26.5 -01:45:29.5 & 5.64 $\pm$  0.86 (6.6$\sigma$) & 11.4 & 12.7 \\ 
FLASH-26C & ... & 11:46:26.7 -01:45:40.6 & 1.56 $\pm$  0.36 (4.4$\sigma$) & 0.2 & 1.4 \\ 
FLASH-26NW & ... & 11:46:27.0 -01:45:45.4 & 0.9 $\pm$  0.25 (3.6$\sigma$) & 6.6 & 5.1 \\ 
\textit{FLASH-27} & A-grade &  & 1.74 $\pm$  0.41 (4.27$\sigma$) &  &  \\ 
FLASH-27E & ... & 11:53:05.7 -00:37:44.0 & 0.68 $\pm$  0.2 (3.3$\sigma$) & 1.0 & 2.3 \\ 
FLASH-27N & ... & 11:53:05.7 -00:37:45.5 & 0.5 $\pm$  0.2 (2.5$\sigma$) & 0.7 & 2.0 \\ 
FLASH-27SW & ... & 11:53:05.8 -00:37:44.8 & 0.56 $\pm$  0.29 (1.9$\sigma$) & 0.8 & 1.3 \\ 
FLASH-28 & A-grade & 11:48:36.4 +01:01:10.7 & 1.74 $\pm$  0.21 (8.4$\sigma$) & 0.3 & 0.6 \\ 
FLASH-29E & C-grade & 12:04:29.2 -00:42:48.9 & 1.01 $\pm$  0.3 (3.4$\sigma$) & 6.3 & 6.5 \\ 
FLASH-29W & ... & 12:04:29.7 -00:42:36.9 & 1.07 $\pm$  0.34 (3.2$\sigma$) & 7.9 & 7.6 \\ 
FLASH-29C & ... & 12:04:29.5 -00:42:42.4 & 0.92 $\pm$  0.33 (2.8$\sigma$) & 2.1 & 1.6 \\ 
FLASH-30NW & A-grade & 12:14:41.3 +00:24:12.1 & 2.1 $\pm$  0.58 (3.7$\sigma$) & 5.8 & 5.0 \\ 
FLASH-30SE & ... & 12:14:41.8 +00:24:04.3 & 1.34 $\pm$  0.37 (3.6$\sigma$) & 4.5 & 5.4 \\ 
FLASH-31 & B-grade & 12:01:37.3 -00:04:19.8 & 0.82 $\pm$  0.35 (2.3$\sigma$) & 2.2 & 2.3 \\ 
FLASH-32 & C-grade & 11:59:37.8 -00:06:26.6 & 9.21 $\pm$  0.26 (36.0$\sigma$) & 3.3 & 0.8 \\ 
FLASH-33W & C-grade & 12:01:28.6 -01:10:12.3 & 2.0 $\pm$  0.28 (7.1$\sigma$) & 5.6 & 5.4 \\ 
FLASH-33SE & ... & 12:01:28.9 -01:10:24.4 & 3.77 $\pm$  0.77 (4.9$\sigma$) & 8.4 & 8.5 \\ 
FLASH-33C & ... & 12:01:28.7 -01:10:16.1 & 1.35 $\pm$  0.48 (2.8$\sigma$) & 2.7 & 2.6 \\ 
FLASH-34 & A-grade & 12:25:10.2 -00:18:05.8 & 0.89 $\pm$  0.27 (3.2$\sigma$) & 4.5 & 4.1 \\ 
FLASH-35S & B-grade & 12:06:08.5 -00:34:57.4 & 4.9 $\pm$  0.23 (20.9$\sigma$) & 3.3 & 3.9 \\ 
FLASH-35N & ... & 12:06:09.5 -00:34:52.9 & 5.99 $\pm$  1.01 (6.0$\sigma$) & 12.4 & 12.0 \\ 
FLASH-36SE & C-grade & 11:54:37.1 +00:59:39.4 & 2.78 $\pm$  0.67 (4.2$\sigma$) & 9.7 & 3.4 \\ 
FLASH-36W & ... & 11:54:37.7 +00:59:24.7 & 1.88 $\pm$  0.67 (2.8$\sigma$) & 9.2 & 13.6 \\ 
FLASH-36N & ... & 11:54:38.2 +00:59:35.3 & 0.67 $\pm$  0.26 (2.6$\sigma$) & 7.9 & 13.5 \\ 
FLASH-37 & A-grade & 12:17:23.4 -02:05:59.9 & 10.42 $\pm$  0.25 (41.4$\sigma$) & 1.0 & 0.8 \\ 
FLASH-38W & B-grade & 11:49:25.9 -02:07:29.0 & 4.33 $\pm$  0.34 (12.7$\sigma$) & 6.4 & 8.4 \\ 
FLASH-38SE & ... & 11:49:25.5 -02:07:16.5 & 3.15 $\pm$  0.39 (8.1$\sigma$) & 7.7 & 5.8 \\ 
FLASH-38S & ... & 11:49:25.3 -02:07:20.8 & 1.14 $\pm$  0.35 (3.3$\sigma$) & 6.9 & 6.3 \\ 
FLASH-39SW & C-grade & 12:19:50.3 +00:33:33.4 & 3.63 $\pm$  0.48 (7.6$\sigma$) & 8.8 & 4.4 \\ 
FLASH-39W & ... & 12:19:50.4 +00:33:34.2 & 0.98 $\pm$  0.33 (3.0$\sigma$) & 7.5 & 2.3 \\ 
\textit{FLASH-40} & A-grade &  & 5.37 $\pm$  0.55 (9.79$\sigma$) &  &  \\ 
FLASH-40CS & ... & 11:59:41.2 +00:02:41.2 & 1.77 $\pm$  0.19 (9.3$\sigma$) & 0.8 & 1.5 \\ 
FLASH-40NE & ... & 11:59:41.5 +00:02:45.6 & 1.47 $\pm$  0.31 (4.7$\sigma$) & 6.1 & 5.3 \\ 
FLASH-40CN & ... & 11:59:41.3 +00:02:41.0 & 0.9 $\pm$  0.21 (4.3$\sigma$) & 0.9 & 0.7 \\ 
FLASH-40SE & ... & 11:59:40.9 +00:02:49.2 & 1.23 $\pm$  0.35 (3.5$\sigma$) & 9.6 & 9.5 \\ 
FLASH-41 & B-grade & 12:13:48.8 -01:03:13.9 & 1.6 $\pm$  0.37 (4.3$\sigma$) & 0.1 & 2.7 \\ 
FLASH-42S & A-grade & 11:59:09.9 -01:20:57.7 & 2.4 $\pm$  0.27 (9.1$\sigma$) & 1.0 & 1.3 \\ 
FLASH-42N & ... & 11:59:10.4 -01:20:57.2 & 1.15 $\pm$  0.24 (4.7$\sigma$) & 5.6 & 5.3 \\ 
FLASH-43 & B-grade & 12:23:05.3 -01:13:07.9 & 5.54 $\pm$  0.3 (18.5$\sigma$) & 4.2 & 2.3 \\ 
FLASH-44S & B-grade & 12:00:47.4 -00:40:19.8 & 7.55 $\pm$  0.28 (27.2$\sigma$) & 2.7 & 3.1 \\ 
FLASH-44NW & ... & 12:00:48.1 -00:40:26.9 & 7.04 $\pm$  0.78 (9.1$\sigma$) & 11.0 & 10.0 \\ 
FLASH-45S & A-grade & 11:36:53.7 +01:16:34.9 & 0.78 $\pm$  0.28 (2.8$\sigma$) & 1.7 & 2.3 \\ 
FLASH-45NE & ... & 11:36:53.6 +01:16:33.6 & 1.0 $\pm$  0.37 (2.7$\sigma$) & 0.3 & 0.8 \\ 
FLASH-46 & A-grade & 12:14:12.1 -00:52:14.6 & 2.57 $\pm$  0.23 (11.0$\sigma$) & 1.4 & 1.0 \\ 
\textit{FLASH-47} & A-grade &  & 5.7 $\pm$  0.31 (18.66$\sigma$) &  &  \\ 
FLASH-47SW & ... & 12:20:19.4 -00:39:14.0 & 3.88 $\pm$  0.22 (17.2$\sigma$) & 2.8 & 3.3 \\ 
FLASH-47E & ... & 12:20:19.6 -00:39:11.5 & 1.82 $\pm$  0.21 (8.8$\sigma$) & 0.2 & 1.3 \\ 
\textit{FLASH-48} & A-grade &  & 7.69 $\pm$  0.44 (17.6$\sigma$) &  &  \\ 
FLASH-48NW & ... & 12:01:15.6 -01:27:24.9 & 6.11 $\pm$  0.34 (18.0$\sigma$) & 2.5 & 4.0 \\ 
FLASH-48SW & ... & 12:01:15.5 -01:27:22.7 & 1.58 $\pm$  0.27 (5.7$\sigma$) & 0.5 & 1.0 \\ 
FLASH-49 & C-grade & 11:58:06.0 -01:55:53.1 & 6.39 $\pm$  0.72 (8.8$\sigma$) & 10.7 & 9.7 \\ 
FLASH-50 & B-grade & 12:05:31.5 -00:37:04.3 & 6.47 $\pm$  0.26 (25.0$\sigma$) & 4.8 & 4.6 \\ 
FLASH-51 & C-grade & 12:01:42.1 -02:20:12.0 & 6.27 $\pm$  0.38 (16.4$\sigma$) & 7.8 & 6.8 \\ 
FLASH-52C & A-grade & 11:50:09.4 -00:36:52.1 & 4.12 $\pm$  0.3 (13.7$\sigma$) & 0.3 & 0.4 \\ 
FLASH-52E & ... & 11:50:09.6 -00:36:45.6 & 1.69 $\pm$  0.54 (3.1$\sigma$) & 6.6 & 6.9 \\ 
FLASH-53SE & B-grade & 11:49:21.5 -01:02:59.7 & 4.92 $\pm$  0.21 (23.3$\sigma$) & 2.9 & 3.1 \\ 
FLASH-53S & ... & 11:49:21.1 -01:03:03.1 & 5.86 $\pm$  0.4 (14.8$\sigma$) & 7.2 & 7.1 \\ 
FLASH-54N & A-grade & 12:08:06.2 +00:45:09.6 & 7.24 $\pm$  0.29 (25.0$\sigma$) & 1.8 & 0.9 \\ 
FLASH-54SE & ... & 12:08:05.7 +00:45:16.9 & 1.46 $\pm$  0.33 (4.5$\sigma$) & 8.6 & 9.3 \\ 
FLASH-55W & C-grade & 11:55:00.5 -00:07:29.2 & 9.34 $\pm$  0.39 (23.9$\sigma$) & 8.4 & 7.9 \\ 
FLASH-55NW & ... & 11:55:00.9 -00:07:24.9 & 2.44 $\pm$  0.28 (8.9$\sigma$) & 6.0 & 4.7 \\ 
FLASH-55S & ... & 11:55:00.4 -00:07:21.2 & 0.68 $\pm$  0.24 (2.9$\sigma$) & 4.3 & 5.3 \\ 
FLASH-56E & C-grade & 12:16:54.5 -01:27:24.6 & 7.29 $\pm$  0.3 (24.3$\sigma$) & 6.7 & 5.7 \\ 
% FLASH-56 & ... & 12:16:54.5 -01:27:25.0 & 1.84 $\pm$  0.41 (4.5$\sigma$) & 6.3 & 5.3 \\ 
FLASH-56S & ... & 12:16:53.8 -01:27:27.4 & 1.09 $\pm$  0.39 (2.8$\sigma$) & 9.5 & 8.5 \\ 
FLASH-56N & ... & 12:16:54.6 -01:27:29.7 & 0.95 $\pm$  0.38 (2.5$\sigma$) & 4.1 & 4.5 \\ 
FLASH-57 & A-grade & 11:46:10.0 -00:50:27.3 & 2.67 $\pm$  0.48 (5.6$\sigma$) & 0.1 & 1.7 \\ 
\textit{FLASH-58} & A-grade &  & 6.47 $\pm$  0.68 (9.48$\sigma$) &  &  \\ 
FLASH-58S & ... & 11:43:59.9 -00:16:00.7 & 4.07 $\pm$  0.46 (8.8$\sigma$) & 0.8 & 0.6 \\ 
FLASH-58N & ... & 11:44:00.0 -00:16:01.2 & 2.41 $\pm$  0.5 (4.8$\sigma$) & 0.2 & 1.6 \\ 
FLASH-59SW & C-grade & 12:22:11.2 -01:41:57.3 & 3.33 $\pm$  0.24 (14.1$\sigma$) & 4.3 & 5.2 \\ 
FLASH-59NE & ... & 12:22:11.7 -01:41:44.2 & 4.0 $\pm$  0.89 (4.5$\sigma$) & 11.4 & 10.7 \\ 
FLASH-60N & A-grade & 11:50:55.1 -00:44:06.0 & 7.51 $\pm$  0.25 (30.0$\sigma$) & 2.8 & 1.8 \\ 
FLASH-60C & ... & 11:50:55.0 -00:44:06.9 & 2.31 $\pm$  0.18 (12.7$\sigma$) & 0.9 & 0.9 \\ 
FLASH-61 & A-grade & 12:14:27.0 -02:24:47.1 & 6.89 $\pm$  0.23 (29.9$\sigma$) & 2.3 & 2.1 \\ 
FLASH-62C & B-grade & 12:14:02.6 -01:43:04.9 & 5.0 $\pm$  0.3 (16.9$\sigma$) & 0.1 & 2.8 \\ 
FLASH-62SW & ... & 12:14:02.4 -01:43:06.9 & 1.27 $\pm$  0.21 (6.0$\sigma$) & 3.3 & 4.0 \\ 
FLASH-63 & A-grade & 11:44:40.0 +00:54:30.9 & 1.64 $\pm$  0.19 (8.4$\sigma$) & 1.4 & 1.7 \\ 
FLASH-64NE & B-grade & 11:58:50.1 -00:57:02.0 & 1.57 $\pm$  0.47 (3.3$\sigma$) & 8.5 & 6.5 \\ 
FLASH-64E & ... & 11:58:49.8 -00:57:07.4 & 0.84 $\pm$  0.33 (2.5$\sigma$) & 1.1 & 3.7 \\ 
FLASH-64W & ... & 11:58:49.6 -00:57:13.4 & 0.56 $\pm$  0.23 (2.4$\sigma$) & 5.4 & 7.7 \\ 
FLASH-64NW & ... & 11:58:49.9 -00:57:11.1 & 0.33 $\pm$  0.16 (2.0$\sigma$) & 3.8 & 3.1 \\ 
\textit{FLASH-65} & A-grade &  & 9.05 $\pm$  0.47 (19.36$\sigma$) &  &  \\ 
FLASH-65E & ... & 12:10:58.0 -00:44:38.6 & 5.53 $\pm$  0.27 (20.8$\sigma$) & 1.4 & 0.5 \\ 
FLASH-65SE & ... & 12:10:57.9 -00:44:38.4 & 3.52 $\pm$  0.38 (9.1$\sigma$) & 1.6 & 1.1 \\ 
FLASH-66 & C-grade & 12:13:58.2 +01:10:43.9 & 7.54 $\pm$  0.3 (25.1$\sigma$) & 6.6 & 5.4 \\ 
FLASH-67NE & B-grade & 12:24:46.4 -01:52:36.6 & 6.11 $\pm$  0.34 (17.8$\sigma$) & 7.5 & 7.4 \\ 
FLASH-67NW & ... & 12:24:46.3 -01:52:50.4 & 7.14 $\pm$  1.02 (7.0$\sigma$) & 12.0 & 11.8 \\ 
FLASH-67N & ... & 12:24:46.2 -01:52:39.0 & 1.77 $\pm$  0.27 (6.5$\sigma$) & 3.9 & 3.9 \\ 
FLASH-68 & A-grade & 12:18:20.7 +01:13:48.6 & 1.91 $\pm$  0.2 (9.7$\sigma$) & 0.3 & 2.3 \\ 
FLASH-69S & A-grade & 11:44:40.3 +00:26:50.5 & 2.05 $\pm$  0.2 (10.2$\sigma$) & 1.7 & 2.1 \\ 
FLASH-69NW & ... & 11:44:40.7 +00:26:45.6 & 1.01 $\pm$  0.27 (3.7$\sigma$) & 6.5 & 5.8 \\ 
FLASH-70SE & A-grade & 12:13:57.4 +00:09:16.1 & 1.49 $\pm$  0.27 (5.4$\sigma$) & 6.2 & 6.1 \\ 
FLASH-70N & ... & 12:13:57.9 +00:09:09.6 & 1.35 $\pm$  0.28 (4.9$\sigma$) & 3.8 & 3.9 \\ 
FLASH-70C & ... & 12:13:57.7 +00:09:10.8 & 0.76 $\pm$  0.23 (3.3$\sigma$) & 0.3 & 0.4 \\ 
FLASH-71N & A-grade & 12:24:07.7 -00:33:31.4 & 7.09 $\pm$  0.23 (31.1$\sigma$) & 2.6 & 1.6 \\ 
FLASH-71NE & ... & 12:24:08.2 -00:33:26.2 & 3.02 $\pm$  0.8 (3.8$\sigma$) & 11.1 & 10.0 \\ 
FLASH-72NE & B-grade & 12:06:32.3 -00:33:11.1 & 3.93 $\pm$  0.24 (16.1$\sigma$) & 3.0 & 0.8 \\ 
FLASH-72E & ... & 12:06:32.1 -00:33:05.9 & 2.63 $\pm$  0.47 (5.6$\sigma$) & 6.5 & 5.5 \\ 
FLASH-72W & ... & 12:06:32.1 -00:33:11.0 & 0.95 $\pm$  0.25 (3.8$\sigma$) & 1.5 & 1.5 \\ 
\textit{FLASH-73} & A-grade &  & 14.26 $\pm$  0.63 (22.49$\sigma$) &  &  \\ 
FLASH-73N & ... & 12:05:53.4 -01:51:18.0 & 7.71 $\pm$  0.27 (28.4$\sigma$) & 1.1 & 1.5 \\ 
FLASH-73S & ... & 12:05:53.3 -01:51:18.1 & 3.85 $\pm$  0.47 (8.2$\sigma$) & 0.3 & 0.2 \\ 
FLASH-73E & ... & 12:05:53.4 -01:51:17.3 & 2.7 $\pm$  0.33 (8.2$\sigma$) & 0.9 & 1.2 \\ 
FLASH-74C & B-grade & 11:39:32.4 -01:54:58.9 & 2.88 $\pm$  0.19 (14.8$\sigma$) & 0.1 & 0.5 \\ 
FLASH-74SW & ... & 11:39:32.3 -01:55:02.1 & 0.77 $\pm$  0.24 (3.2$\sigma$) & 3.8 & 4.2 \\ 
FLASH-75NE & A-grade & 12:07:00.1 -01:13:01.2 & 3.98 $\pm$  0.22 (18.0$\sigma$) & 2.9 & 2.8 \\ 
FLASH-75S & ... & 12:07:00.0 -01:13:03.6 & 2.05 $\pm$  0.38 (5.4$\sigma$) & 3.0 & 2.9 \\ 
\textit{FLASH-76} & A-grade &  & 11.54 $\pm$  0.42 (27.52$\sigma$) &  &  \\ 
FLASH-76W & ... & 12:02:23.8 +01:03:14.3 & 8.68 $\pm$  0.37 (23.7$\sigma$) & 3.2 & 3.1 \\ 
FLASH-76E & ... & 12:02:23.8 +01:03:18.0 & 2.86 $\pm$  0.2 (14.0$\sigma$) & 0.6 & 0.8 \\ 
FLASH-77N & C-grade & 12:05:43.1 -00:38:33.6 & 8.34 $\pm$  0.24 (34.3$\sigma$) & 3.4 & 0.7 \\ 
FLASH-77NW & ... & 12:05:43.2 -00:38:36.4 & 3.92 $\pm$  0.43 (9.0$\sigma$) & 5.2 & 3.8 \\ 
FLASH-78E & A-grade & 11:39:01.1 -02:14:10.4 & 6.55 $\pm$  0.23 (28.5$\sigma$) & 1.4 & 1.9 \\ 
FLASH-78NW & ... & 11:39:01.3 -02:14:17.7 & 1.82 $\pm$  0.55 (3.3$\sigma$) & 7.1 & 5.9 \\ 
FLASH-78C & ... & 11:39:01.1 -02:14:11.6 & 1.05 $\pm$  0.37 (2.9$\sigma$) & 0.2 & 1.8 \\ 
FLASH-79NE & C-grade & 11:45:15.6 -00:08:37.9 & 9.59 $\pm$  0.41 (23.2$\sigma$) & 6.3 & 7.8 \\ 
FLASH-79W & ... & 11:45:15.3 -00:08:48.5 & 0.89 $\pm$  0.25 (3.5$\sigma$) & 6.3 & 4.0 \\ 
FLASH-79E & ... & 11:45:15.5 -00:08:37.3 & 1.29 $\pm$  0.49 (2.6$\sigma$) & 5.6 & 7.7 \\ 
\textit{FLASH-80} & A-grade &  & 13.24 $\pm$  0.45 (29.51$\sigma$) &  &  \\ 
FLASH-80N & ... & 11:38:16.3 -01:31:20.7 & 10.88 $\pm$  0.25 (44.3$\sigma$) & 2.2 & 4.9 \\ 
FLASH-80S & ... & 11:38:16.1 -01:31:20.5 & 2.36 $\pm$  0.38 (6.3$\sigma$) & 0.3 & 2.9 \\ 
FLASH-81SW & B-grade & 11:57:14.7 -00:28:43.3 & 10.32 $\pm$  0.23 (44.0$\sigma$) & 3.5 & 2.0 \\ 
FLASH-81N & ... & 11:57:15.2 -00:28:41.6 & 4.19 $\pm$  0.35 (12.0$\sigma$) & 5.6 & 6.5 \\ 
FLASH-82NW & A-grade & 11:52:51.4 -01:52:20.3 & 7.86 $\pm$  0.26 (29.9$\sigma$) & 1.3 & 2.0 \\ 
FLASH-82N & ... & 11:52:51.6 -01:52:21.6 & 3.07 $\pm$  0.44 (6.9$\sigma$) & 4.8 & 2.4 \\ 
FLASH-83NE & A-grade & 11:44:04.8 +00:38:46.8 & 3.94 $\pm$  0.36 (11.1$\sigma$) & 6.3 & 6.2 \\ 
FLASH-83C & ... & 11:44:04.4 +00:38:43.9 & 1.0 $\pm$  0.27 (3.7$\sigma$) & 0.3 & 6.1 \\ 
FLASH-83N & ... & 11:44:05.1 +00:38:48.3 & 3.03 $\pm$  0.9 (3.4$\sigma$) & 11.6 & 10.5 \\ 
FLASH-84S & C-grade & 11:36:16.5 +00:48:52.1 & 12.66 $\pm$  0.34 (36.9$\sigma$) & 6.3 & 2.7 \\ 
FLASH-84SE & ... & 11:36:16.6 +00:48:54.0 & 3.01 $\pm$  0.3 (9.9$\sigma$) & 6.2 & 0.6 \\ 
\textit{FLASH-85} & A-grade &  & 35.19 $\pm$  0.99 (35.41$\sigma$) &  &  \\ 
FLASH-85NW & ... & 11:37:39.2 -02:44:46.0 & 18.64 $\pm$  0.65 (28.8$\sigma$) & 1.2 & 2.7 \\ 
FLASH-85N & ... & 11:37:39.2 -02:44:46.0 & 7.39 $\pm$  0.51 (14.5$\sigma$) & 1.5 & 2.6 \\ 
FLASH-85NW & ... & 11:37:39.2 -02:44:45.4 & 2.97 $\pm$  0.24 (12.5$\sigma$) & 1.2 & 2.0 \\ 
FLASH-85SE & ... & 11:37:39.1 -02:44:44.7 & 4.44 $\pm$  0.39 (11.4$\sigma$) & 0.6 & 2.5 \\ 
FLASH-85W & ... & 11:37:39.2 -02:44:44.9 & 1.76 $\pm$  0.32 (5.5$\sigma$) & 1.2 & 1.5 \\ 
\textit{FLASH-86} & A-grade &  & 39.77 $\pm$  0.52 (77.12$\sigma$) &  &  \\ 
FLASH-86N & ... & 11:54:33.7 +00:50:42.0 & 30.82 $\pm$  0.39 (79.0$\sigma$) & 0.2 & 0.7 \\ 
FLASH-86S & ... & 11:54:33.6 +00:50:41.8 & 8.94 $\pm$  0.34 (26.5$\sigma$) & 0.9 & 0.6 \\ 
\end{longtable}
% \end{landscape}
\end{onecolumn}

%% file: mnras_template.bbl
\begin{thebibliography}{}
\makeatletter
\relax
\def\mn@urlcharsother{\let\do\@makeother \do\$\do\&\do\#\do\^\do\_\do\%\do\~}
\def\mn@doi{\begingroup\mn@urlcharsother \@ifnextchar [ {\mn@doi@}
  {\mn@doi@[]}}
\def\mn@doi@[#1]#2{\def\@tempa{#1}\ifx\@tempa\@empty \href
  {http://dx.doi.org/#2} {doi:#2}\else \href {http://dx.doi.org/#2} {#1}\fi
  \endgroup}
\def\mn@eprint#1#2{\mn@eprint@#1:#2::\@nil}
\def\mn@eprint@arXiv#1{\href {http://arxiv.org/abs/#1} {{\tt arXiv:#1}}}
\def\mn@eprint@dblp#1{\href {http://dblp.uni-trier.de/rec/bibtex/#1.xml}
  {dblp:#1}}
\def\mn@eprint@#1:#2:#3:#4\@nil{\def\@tempa {#1}\def\@tempb {#2}\def\@tempc
  {#3}\ifx \@tempc \@empty \let \@tempc \@tempb \let \@tempb \@tempa \fi \ifx
  \@tempb \@empty \def\@tempb {arXiv}\fi \@ifundefined
  {mn@eprint@\@tempb}{\@tempb:\@tempc}{\expandafter \expandafter \csname
  mn@eprint@\@tempb\endcsname \expandafter{\@tempc}}}

\bibitem[\protect\citeauthoryear{{ALMA Partnership} et~al.,}{{ALMA Partnership}
  et~al.}{2015}]{SV2015}
{ALMA Partnership} et~al., 2015, \mn@doi [\apjl] {10.1088/2041-8205/808/1/L4},
  \href {https://ui.adsabs.harvard.edu/abs/2015ApJ...808L...4A} {808, L4}

\bibitem[\protect\citeauthoryear{{Amvrosiadis} et~al.,}{{Amvrosiadis}
  et~al.}{2018}]{Amvrosiadis2018}
{Amvrosiadis} A.,  et~al., 2018, \mn@doi [\mnras] {10.1093/mnras/sty138}, \href
  {https://ui.adsabs.harvard.edu/abs/2018MNRAS.475.4939A} {475, 4939}

\bibitem[\protect\citeauthoryear{{Andrews} \& {Thompson}}{{Andrews} \&
  {Thompson}}{2011}]{Andrews2011}
{Andrews} B.~H.,  {Thompson} T.~A.,  2011, \mn@doi [\apj]
  {10.1088/0004-637X/727/2/97}, \href
  {https://ui.adsabs.harvard.edu/abs/2011ApJ...727...97A} {727, 97}

\bibitem[\protect\citeauthoryear{{Asboth} et~al.,}{{Asboth}
  et~al.}{2016}]{Asboth2016}
{Asboth} V.,  et~al., 2016, \mn@doi [\mnras] {10.1093/mnras/stw1769}, \href
  {https://ui.adsabs.harvard.edu/abs/2016MNRAS.462.1989A} {462, 1989}

\bibitem[\protect\citeauthoryear{{Bakx} \& {Dannerbauer}}{{Bakx} \&
  {Dannerbauer}}{2022}]{Bakx2022}
{Bakx} T. J.~L.~C.,  {Dannerbauer} H.,  2022, \mn@doi [\mnras]
  {10.1093/mnras/stac1306}, \href
  {https://ui.adsabs.harvard.edu/abs/2022MNRAS.515..678B} {515, 678}

\bibitem[\protect\citeauthoryear{{Bakx} et~al.,}{{Bakx} et~al.}{2018}]{bakx18}
{Bakx} T. J.~L.~C.,  et~al., 2018, \mn@doi [\mnras] {10.1093/mnras/stx2267},
  \href {https://ui.adsabs.harvard.edu/abs/2018MNRAS.473.1751B} {473, 1751}

\bibitem[\protect\citeauthoryear{{Bakx}, {Eales}  \& {Amvrosiadis}}{{Bakx}
  et~al.}{2020a}]{bakx2020}
{Bakx} T. J.~L.~C.,  {Eales} S.,   {Amvrosiadis} A.,  2020a, \mn@doi [\mnras]
  {10.1093/mnras/staa506}, \href
  {https://ui.adsabs.harvard.edu/abs/2020MNRAS.493.4276B} {493, 4276}

\bibitem[\protect\citeauthoryear{{Bakx} et~al.,}{{Bakx}
  et~al.}{2020b}]{Bakx2020Erratum}
{Bakx} T. J.~L.~C.,  et~al., 2020b, \mn@doi [\mnras] {10.1093/mnras/staa658},
  \href {https://ui.adsabs.harvard.edu/abs/2020MNRAS.494...10B} {494, 10}

\bibitem[\protect\citeauthoryear{{Bartelmann} \& {Schneider}}{{Bartelmann} \&
  {Schneider}}{2001}]{Bartelmann2001}
{Bartelmann} M.,  {Schneider} P.,  2001, \mn@doi [\physrep]
  {10.1016/S0370-1573(00)00082-X}, \href
  {https://ui.adsabs.harvard.edu/abs/2001PhR...340..291B} {340, 291}

\bibitem[\protect\citeauthoryear{{Baugh}, {Lacey}, {Frenk}, {Granato}, {Silva},
  {Bressan}, {Benson}  \& {Cole}}{{Baugh} et~al.}{2005}]{Baugh2005}
{Baugh} C.~M.,  {Lacey} C.~G.,  {Frenk} C.~S.,  {Granato} G.~L.,  {Silva} L.,
  {Bressan} A.,  {Benson} A.~J.,   {Cole} S.,  2005, \mn@doi [\mnras]
  {10.1111/j.1365-2966.2004.08553.x}, \href
  {https://ui.adsabs.harvard.edu/abs/2005MNRAS.356.1191B} {356, 1191}

\bibitem[\protect\citeauthoryear{{Bendo} et~al.,}{{Bendo}
  et~al.}{2023}]{Bendo2023}
{Bendo} G.~J.,  et~al., 2023, \mn@doi [\mnras] {10.1093/mnras/stac3771}, \href
  {https://ui.adsabs.harvard.edu/abs/2023MNRAS.522.2995B} {522, 2995}

\bibitem[\protect\citeauthoryear{{Bonavera} et~al.,}{{Bonavera}
  et~al.}{2019}]{Bonavera2019}
{Bonavera} L.,  et~al., 2019, \mn@doi [\jcap] {10.1088/1475-7516/2019/09/021},
  \href {https://ui.adsabs.harvard.edu/abs/2019JCAP...09..021B} {2019, 021}

\bibitem[\protect\citeauthoryear{{Bonavera}, {Cueli}, {Gonz{\'a}lez-Nuevo},
  {Ronconi}, {Migliaccio}, {Lapi}, {Casas}  \& {Crespo}}{{Bonavera}
  et~al.}{2021}]{Bonavera2021AnA}
{Bonavera} L.,  {Cueli} M.~M.,  {Gonz{\'a}lez-Nuevo} J.,  {Ronconi} T.,
  {Migliaccio} M.,  {Lapi} A.,  {Casas} J.~M.,   {Crespo} D.,  2021, \mn@doi
  [\aap] {10.1051/0004-6361/202141521}, \href
  {https://ui.adsabs.harvard.edu/abs/2021A&A...656A..99B} {656, A99}

\bibitem[\protect\citeauthoryear{{Bonavera}, {Cueli}  \&
  {Gonzalez-Nuevo}}{{Bonavera} et~al.}{2022}]{Bonavera2022}
{Bonavera} L.,  {Cueli} M.~M.,   {Gonzalez-Nuevo} J.,  2022, Proceedings of the
  MG16 Meeting on General Relativity, R. Ruffini \& G. Vereshchagin eds., World
  Scientific., \href {https://ui.adsabs.harvard.edu/abs/2021arXiv211202959B}
  {p. arXiv:2112.02959}

\bibitem[\protect\citeauthoryear{{Bourne} et~al.,}{{Bourne}
  et~al.}{2016}]{bourne2016}
{Bourne} N.,  et~al., 2016, \mn@doi [\mnras] {10.1093/mnras/stw1654}, \href
  {https://ui.adsabs.harvard.edu/abs/2016MNRAS.462.1714B} {462, 1714}

\bibitem[\protect\citeauthoryear{{Bussmann} et~al.,}{{Bussmann}
  et~al.}{2013}]{bussmann13}
{Bussmann} R.~S.,  et~al., 2013, \mn@doi [\apj] {10.1088/0004-637X/779/1/25},
  \href {https://ui.adsabs.harvard.edu/abs/2013ApJ...779...25B} {779, 25}

\bibitem[\protect\citeauthoryear{{Bussmann} et~al.,}{{Bussmann}
  et~al.}{2015}]{Bussmann2015}
{Bussmann} R.~S.,  et~al., 2015, \mn@doi [\apj] {10.1088/0004-637X/812/1/43},
  \href {https://ui.adsabs.harvard.edu/abs/2015ApJ...812...43B} {812, 43}

\bibitem[\protect\citeauthoryear{{Cai} et~al.,}{{Cai} et~al.}{2013}]{Cai2013}
{Cai} Z.-Y.,  et~al., 2013, \mn@doi [\apj] {10.1088/0004-637X/768/1/21}, \href
  {https://ui.adsabs.harvard.edu/abs/2013ApJ...768...21C} {768, 21}

\bibitem[\protect\citeauthoryear{{Casey} et~al.,}{{Casey}
  et~al.}{2013}]{Casey2013}
{Casey} C.~M.,  et~al., 2013, \mn@doi [\mnras] {10.1093/mnras/stt1673}, \href
  {https://ui.adsabs.harvard.edu/abs/2013MNRAS.436.1919C} {436, 1919}

\bibitem[\protect\citeauthoryear{{Chen}, {Cowie}, {Barger}, {Casey}, {Lee},
  {Sanders}, {Wang}  \& {Williams}}{{Chen} et~al.}{2013}]{Chen2013}
{Chen} C.-C.,  {Cowie} L.~L.,  {Barger} A.~J.,  {Casey} C.~M.,  {Lee} N.,
  {Sanders} D.~B.,  {Wang} W.-H.,   {Williams} J.~P.,  2013, \mn@doi [\apj]
  {10.1088/0004-637X/776/2/131}, \href
  {https://ui.adsabs.harvard.edu/abs/2013ApJ...776..131C} {776, 131}

\bibitem[\protect\citeauthoryear{Collaboration et~al.,}{Collaboration
  et~al.}{2018}]{planck2018}
Collaboration P.,  et~al., 2018, Planck 2018 results. VI. Cosmological
  parameters (\mn@eprint {arXiv} {1807.06209})

\bibitem[\protect\citeauthoryear{{Crespo}, {Gonz{\'a}lez-Nuevo}, {Bonavera},
  {Cueli}, {Casas}  \& {Goitia}}{{Crespo} et~al.}{2022}]{Crespo2022}
{Crespo} D.,  {Gonz{\'a}lez-Nuevo} J.,  {Bonavera} L.,  {Cueli} M.~M.,  {Casas}
  J.~M.,   {Goitia} E.,  2022, \mn@doi [\aap] {10.1051/0004-6361/202244016},
  \href {https://ui.adsabs.harvard.edu/abs/2022A&A...667A.146C} {667, A146}

\bibitem[\protect\citeauthoryear{{Cueli}, {Bonavera}, {Gonz{\'a}lez-Nuevo}  \&
  {Lapi}}{{Cueli} et~al.}{2021}]{Cueli2021}
{Cueli} M.~M.,  {Bonavera} L.,  {Gonz{\'a}lez-Nuevo} J.,   {Lapi} A.,  2021,
  \mn@doi [\aap] {10.1051/0004-6361/202039326}, \href
  {https://ui.adsabs.harvard.edu/abs/2021A&A...645A.126C} {645, A126}

\bibitem[\protect\citeauthoryear{{Cueli}, {Bonavera}, {Gonz{\'a}lez-Nuevo},
  {Crespo}, {Casas}  \& {Lapi}}{{Cueli} et~al.}{2022}]{Cueli2022}
{Cueli} M.~M.,  {Bonavera} L.,  {Gonz{\'a}lez-Nuevo} J.,  {Crespo} D.,  {Casas}
  J.~M.,   {Lapi} A.,  2022, \mn@doi [\aap] {10.1051/0004-6361/202142949},
  \href {https://ui.adsabs.harvard.edu/abs/2022A&A...662A..44C} {662, A44}

\bibitem[\protect\citeauthoryear{{Duivenvoorden} et~al.,}{{Duivenvoorden}
  et~al.}{2018}]{Duivenvoorden2018}
{Duivenvoorden} S.,  et~al., 2018, \mn@doi [\mnras] {10.1093/mnras/sty691},
  \href {https://ui.adsabs.harvard.edu/abs/2018MNRAS.477.1099D} {477, 1099}

\bibitem[\protect\citeauthoryear{{Dunne}, {Bonavera}, {Gonzalez-Nuevo},
  {Maddox}  \& {Vlahakis}}{{Dunne} et~al.}{2020}]{Dunne2020Lensing}
{Dunne} L.,  {Bonavera} L.,  {Gonzalez-Nuevo} J.,  {Maddox} S.~J.,   {Vlahakis}
  C.,  2020, \mn@doi [\mnras] {10.1093/mnras/staa2665}, \href
  {https://ui.adsabs.harvard.edu/abs/2020MNRAS.498.4635D} {498, 4635}

\bibitem[\protect\citeauthoryear{{Dye} et~al.,}{{Dye} et~al.}{2015}]{dye2015}
{Dye} S.,  et~al., 2015, \mn@doi [\mnras] {10.1093/mnras/stv1442}, \href
  {https://ui.adsabs.harvard.edu/abs/2015MNRAS.452.2258D} {452, 2258}

\bibitem[\protect\citeauthoryear{{Dye} et~al.,}{{Dye} et~al.}{2018}]{Dye2018}
{Dye} S.,  et~al., 2018, \mn@doi [\mnras] {10.1093/mnras/sty513}, \href
  {https://ui.adsabs.harvard.edu/abs/2018MNRAS.476.4383D} {476, 4383}

\bibitem[\protect\citeauthoryear{{Dyson}, {Eddington}  \& {Davidson}}{{Dyson}
  et~al.}{1920}]{Dyson1920}
{Dyson} F.~W.,  {Eddington} A.~S.,   {Davidson} C.,  1920, \mn@doi
  [Philosophical Transactions of the Royal Society of London Series A]
  {10.1098/rsta.1920.0009}, \href
  {https://ui.adsabs.harvard.edu/abs/1920RSPTA.220..291D} {220, 291}

\bibitem[\protect\citeauthoryear{{Eales}}{{Eales}}{2015}]{eales2015}
{Eales} S.~A.,  2015, \mn@doi [\mnras] {10.1093/mnras/stu2214}, \href
  {https://ui.adsabs.harvard.edu/abs/2015MNRAS.446.3224E} {446, 3224}

\bibitem[\protect\citeauthoryear{{Eales} et~al.,}{{Eales}
  et~al.}{2010}]{eales2010}
{Eales} S.,  et~al., 2010, \mn@doi [\pasp] {10.1086/653086}, \href
  {https://ui.adsabs.harvard.edu/abs/2010PASP..122..499E} {122, 499}

\bibitem[\protect\citeauthoryear{{Fernandez}, {Cueli}, {Gonz{\'a}lez-Nuevo},
  {Bonavera}, {Crespo}, {Casas}  \& {Lapi}}{{Fernandez}
  et~al.}{2022}]{Fernandez2022}
{Fernandez} L.,  {Cueli} M.~M.,  {Gonz{\'a}lez-Nuevo} J.,  {Bonavera} L.,
  {Crespo} D.,  {Casas} J.~M.,   {Lapi} A.,  2022, \mn@doi [\aap]
  {10.1051/0004-6361/202141905}, \href
  {https://ui.adsabs.harvard.edu/abs/2022A&A...658A..19F} {658, A19}

\bibitem[\protect\citeauthoryear{{Fleuren} et~al.,}{{Fleuren}
  et~al.}{2012}]{Fleuren2012}
{Fleuren} S.,  et~al., 2012, \mn@doi [\mnras]
  {10.1111/j.1365-2966.2012.21048.x}, \href
  {https://ui.adsabs.harvard.edu/abs/2012MNRAS.423.2407F} {423, 2407}

\bibitem[\protect\citeauthoryear{{Furlanetto} et~al.,}{{Furlanetto}
  et~al.}{2018}]{furlanetto2018}
{Furlanetto} C.,  et~al., 2018, \mn@doi [\mnras] {10.1093/mnras/sty151}, \href
  {https://ui.adsabs.harvard.edu/abs/2018MNRAS.476..961F} {476, 961}

\bibitem[\protect\citeauthoryear{{Garratt} et~al.,}{{Garratt}
  et~al.}{2023}]{Garratt2023}
{Garratt} T.~K.,  et~al., 2023, \mn@doi [\mnras] {10.1093/mnras/stad307}, \href
  {https://ui.adsabs.harvard.edu/abs/2023MNRAS.tmp..331G} {}

\bibitem[\protect\citeauthoryear{{Girelli}, {Pozzetti}, {Bolzonella},
  {Giocoli}, {Marulli}  \& {Baldi}}{{Girelli} et~al.}{2020}]{Girelli2020}
{Girelli} G.,  {Pozzetti} L.,  {Bolzonella} M.,  {Giocoli} C.,  {Marulli} F.,
  {Baldi} M.,  2020, \mn@doi [\aap] {10.1051/0004-6361/201936329}, \href
  {https://ui.adsabs.harvard.edu/abs/2020A&A...634A.135G} {634, A135}

\bibitem[\protect\citeauthoryear{{Gonz{\'a}lez-Nuevo}
  et~al.,}{{Gonz{\'a}lez-Nuevo} et~al.}{2012}]{GN2012}
{Gonz{\'a}lez-Nuevo} J.,  et~al., 2012, \mn@doi [\apj]
  {10.1088/0004-637X/749/1/65}, \href
  {https://ui.adsabs.harvard.edu/abs/2012ApJ...749...65G} {749, 65}

\bibitem[\protect\citeauthoryear{{Gonz{\'a}lez-Nuevo}
  et~al.,}{{Gonz{\'a}lez-Nuevo} et~al.}{2014}]{GN2014}
{Gonz{\'a}lez-Nuevo} J.,  et~al., 2014, \mn@doi [\mnras]
  {10.1093/mnras/stu1041}, \href
  {https://ui.adsabs.harvard.edu/abs/2014MNRAS.442.2680G} {442, 2680}

\bibitem[\protect\citeauthoryear{{Gonz{\'a}lez-Nuevo}
  et~al.,}{{Gonz{\'a}lez-Nuevo} et~al.}{2017}]{GN2017}
{Gonz{\'a}lez-Nuevo} J.,  et~al., 2017, \mn@doi [\jcap]
  {10.1088/1475-7516/2017/10/024}, \href
  {https://ui.adsabs.harvard.edu/abs/2017JCAP...10..024G} {2017, 024}

\bibitem[\protect\citeauthoryear{{Gonz{\'a}lez-Nuevo}
  et~al.,}{{Gonz{\'a}lez-Nuevo} et~al.}{2019}]{GN2019}
{Gonz{\'a}lez-Nuevo} J.,  et~al., 2019, \mn@doi [\aap]
  {10.1051/0004-6361/201935475}, \href
  {https://ui.adsabs.harvard.edu/abs/2019A&A...627A..31G} {627, A31}

\bibitem[\protect\citeauthoryear{{Gonz{\'a}lez-Nuevo}, {Cueli}, {Bonavera},
  {Lapi}, {Migliaccio}, {Arg{\"u}eso}  \& {Toffolatti}}{{Gonz{\'a}lez-Nuevo}
  et~al.}{2021}]{GN2021}
{Gonz{\'a}lez-Nuevo} J.,  {Cueli} M.~M.,  {Bonavera} L.,  {Lapi} A.,
  {Migliaccio} M.,  {Arg{\"u}eso} F.,   {Toffolatti} L.,  2021, \mn@doi [\aap]
  {10.1051/0004-6361/202039043}, \href
  {https://ui.adsabs.harvard.edu/abs/2021A&A...646A.152G} {646, A152}

\bibitem[\protect\citeauthoryear{{Grillo}, {Lombardi}  \& {Bertin}}{{Grillo}
  et~al.}{2008}]{Grillo2008}
{Grillo} C.,  {Lombardi} M.,   {Bertin} G.,  2008, \mn@doi [\aap]
  {10.1051/0004-6361:20077534}, \href
  {https://ui.adsabs.harvard.edu/abs/2008A&A...477..397G} {477, 397}

\bibitem[\protect\citeauthoryear{{Gruppioni} et~al.,}{{Gruppioni}
  et~al.}{2013}]{Gruppioni2013}
{Gruppioni} C.,  et~al., 2013, \mn@doi [\mnras] {10.1093/mnras/stt308}, \href
  {https://ui.adsabs.harvard.edu/abs/2013MNRAS.432...23G} {432, 23}

\bibitem[\protect\citeauthoryear{{Harrington} et~al.,}{{Harrington}
  et~al.}{2021}]{Harrington2021}
{Harrington} K.~C.,  et~al., 2021, \mn@doi [\apj] {10.3847/1538-4357/abcc01},
  \href {https://ui.adsabs.harvard.edu/abs/2021ApJ...908...95H} {908, 95}

\bibitem[\protect\citeauthoryear{{Hughes} et~al.,}{{Hughes}
  et~al.}{1998}]{Hughes1998}
{Hughes} D.~H.,  et~al., 1998, \mn@doi [\nat] {10.1038/28328}, \href
  {https://ui.adsabs.harvard.edu/abs/1998Natur.394..241H} {394, 241}

\bibitem[\protect\citeauthoryear{{Ivison}, {Smail}, {Le Borgne}, {Blain},
  {Kneib}, {Bezecourt}, {Kerr}  \& {Davies}}{{Ivison}
  et~al.}{1998}]{Ivison1998}
{Ivison} R.~J.,  {Smail} I.,  {Le Borgne} J.~F.,  {Blain} A.~W.,  {Kneib}
  J.~P.,  {Bezecourt} J.,  {Kerr} T.~H.,   {Davies} J.~K.,  1998, \mn@doi
  [\mnras] {10.1046/j.1365-8711.1998.01677.x}, \href
  {https://ui.adsabs.harvard.edu/abs/1998MNRAS.298..583I} {298, 583}

\bibitem[\protect\citeauthoryear{{Ivison} et~al.,}{{Ivison}
  et~al.}{2016}]{ivison16}
{Ivison} R.~J.,  et~al., 2016, \mn@doi [\apj] {10.3847/0004-637X/832/1/78},
  \href {https://ui.adsabs.harvard.edu/abs/2016ApJ...832...78I} {832, 78}

\bibitem[\protect\citeauthoryear{{Kamieneski} et~al.,}{{Kamieneski}
  et~al.}{2023}]{Kamieneski2023}
{Kamieneski} P.~S.,  et~al., 2023, \mn@doi [arXiv e-prints]
  {10.48550/arXiv.2301.09746}, \href
  {https://ui.adsabs.harvard.edu/abs/2023arXiv230109746K} {p. arXiv:2301.09746}

\bibitem[\protect\citeauthoryear{{Kelly} et~al.,}{{Kelly}
  et~al.}{2018}]{Kelly2018}
{Kelly} P.~L.,  et~al., 2018, \mn@doi [Nature Astronomy]
  {10.1038/s41550-018-0430-3}, \href
  {https://ui.adsabs.harvard.edu/abs/2018NatAs...2..334K} {2, 334}

\bibitem[\protect\citeauthoryear{{Kneib} \& {Natarajan}}{{Kneib} \&
  {Natarajan}}{2011}]{Kneib2011}
{Kneib} J.-P.,  {Natarajan} P.,  2011, \mn@doi [\aapr]
  {10.1007/s00159-011-0047-3}, \href
  {https://ui.adsabs.harvard.edu/abs/2011A&ARv..19...47K} {19, 47}

\bibitem[\protect\citeauthoryear{{Kochanek}}{{Kochanek}}{1992}]{Kochanek1992}
{Kochanek} C.~S.,  1992, \mn@doi [\apj] {10.1086/170845}, \href
  {https://ui.adsabs.harvard.edu/abs/1992ApJ...384....1K} {384, 1}

\bibitem[\protect\citeauthoryear{{Kochanek}}{{Kochanek}}{1996}]{Kochanek1996}
{Kochanek} C.~S.,  1996, \mn@doi [\apj] {10.1086/178175}, \href
  {https://ui.adsabs.harvard.edu/abs/1996ApJ...473..595K} {473, 595}

\bibitem[\protect\citeauthoryear{{Lagos}, {da Cunha}, {Robotham}, {Obreschkow},
  {Valentino}, {Fujimoto}, {Magdis}  \& {Tobar}}{{Lagos}
  et~al.}{2020}]{Lagos2020}
{Lagos} C. d.~P.,  {da Cunha} E.,  {Robotham} A. S.~G.,  {Obreschkow} D.,
  {Valentino} F.,  {Fujimoto} S.,  {Magdis} G.~E.,   {Tobar} R.,  2020, \mn@doi
  [\mnras] {10.1093/mnras/staa2861}, \href
  {https://ui.adsabs.harvard.edu/abs/2020MNRAS.499.1948L} {499, 1948}

\bibitem[\protect\citeauthoryear{{Lapi} et~al.,}{{Lapi}
  et~al.}{2011}]{Lapi2011}
{Lapi} A.,  et~al., 2011, \mn@doi [\apj] {10.1088/0004-637X/742/1/24}, \href
  {https://ui.adsabs.harvard.edu/abs/2011ApJ...742...24L} {742, 24}

\bibitem[\protect\citeauthoryear{{Lapi}, {Negrello}, {Gonz{\'a}lez-Nuevo},
  {Cai}, {De Zotti}  \& {Danese}}{{Lapi} et~al.}{2012}]{Lapi2012}
{Lapi} A.,  {Negrello} M.,  {Gonz{\'a}lez-Nuevo} J.,  {Cai} Z.~Y.,  {De Zotti}
  G.,   {Danese} L.,  2012, \mn@doi [\apj] {10.1088/0004-637X/755/1/46}, \href
  {https://ui.adsabs.harvard.edu/abs/2012ApJ...755...46L} {755, 46}

\bibitem[\protect\citeauthoryear{{Maddox} et~al.,}{{Maddox}
  et~al.}{2018}]{maddox2018}
{Maddox} S.~J.,  et~al., 2018, \mn@doi [\apjs] {10.3847/1538-4365/aab8fc},
  \href {https://ui.adsabs.harvard.edu/abs/2018ApJS..236...30M} {236, 30}

\bibitem[\protect\citeauthoryear{{Narayanan} et~al.,}{{Narayanan}
  et~al.}{2015}]{narayanan2015}
{Narayanan} D.,  et~al., 2015, \mn@doi [\nat] {10.1038/nature15383}, \href
  {https://ui.adsabs.harvard.edu/abs/2015Natur.525..496N} {525, 496}

\bibitem[\protect\citeauthoryear{{Negrello} et~al.,}{{Negrello}
  et~al.}{2010}]{negrello2010}
{Negrello} M.,  et~al., 2010, \mn@doi [Science] {10.1126/science.1193420},
  \href {https://ui.adsabs.harvard.edu/abs/2010Sci...330..800N} {330, 800}

\bibitem[\protect\citeauthoryear{{Negrello} et~al.,}{{Negrello}
  et~al.}{2014}]{Negrello2014}
{Negrello} M.,  et~al., 2014, \mn@doi [\mnras] {10.1093/mnras/stu413}, \href
  {https://ui.adsabs.harvard.edu/abs/2014MNRAS.440.1999N} {440, 1999}

\bibitem[\protect\citeauthoryear{{Negrello} et~al.,}{{Negrello}
  et~al.}{2017}]{negrello2017}
{Negrello} M.,  et~al., 2017, \mn@doi [\mnras] {10.1093/mnras/stw2911}, \href
  {https://ui.adsabs.harvard.edu/abs/2017MNRAS.465.3558N} {465, 3558}

\bibitem[\protect\citeauthoryear{{Oguri} et~al.,}{{Oguri}
  et~al.}{2012}]{Oguri2012}
{Oguri} M.,  et~al., 2012, \mn@doi [\aj] {10.1088/0004-6256/143/5/120}, \href
  {https://ui.adsabs.harvard.edu/abs/2012AJ....143..120O} {143, 120}

\bibitem[\protect\citeauthoryear{{Oliver} et~al.,}{{Oliver}
  et~al.}{2012}]{oliver2012}
{Oliver} S.~J.,  et~al., 2012, \mn@doi [\mnras]
  {10.1111/j.1365-2966.2012.20912.x}, \href
  {https://ui.adsabs.harvard.edu/abs/2012MNRAS.424.1614O} {424, 1614}

\bibitem[\protect\citeauthoryear{{Oteo} et~al.,}{{Oteo}
  et~al.}{2017}]{Oteo2017}
{Oteo} I.,  et~al., 2017, \mn@doi [\apj] {10.3847/1538-4357/aa8ee3}, \href
  {https://ui.adsabs.harvard.edu/abs/2017ApJ...850..170O} {850, 170}

\bibitem[\protect\citeauthoryear{{Overzier}}{{Overzier}}{2016}]{Overzier2016}
{Overzier} R.~A.,  2016, \mn@doi [\aapr] {10.1007/s00159-016-0100-3}, \href
  {https://ui.adsabs.harvard.edu/abs/2016A&ARv..24...14O} {24, 14}

\bibitem[\protect\citeauthoryear{{Pearson} et~al.,}{{Pearson}
  et~al.}{2013}]{pearson13}
{Pearson} E.~A.,  et~al., 2013, \mn@doi [\mnras] {10.1093/mnras/stt1369}, \href
  {https://ui.adsabs.harvard.edu/abs/2013MNRAS.435.2753P} {435, 2753}

\bibitem[\protect\citeauthoryear{{Pearson} et~al.,}{{Pearson}
  et~al.}{2023}]{Pearson2023}
{Pearson} J.,  et~al., 2023, \mn@doi [arXiv e-prints]
  {10.48550/arXiv.2309.00888}, \href
  {https://ui.adsabs.harvard.edu/abs/2023arXiv230900888P} {p. arXiv:2309.00888}

\bibitem[\protect\citeauthoryear{{Rizzo}, {Vegetti}, {Powell}, {Fraternali},
  {McKean}, {Stacey}  \& {White}}{{Rizzo} et~al.}{2020}]{Rizzo2020}
{Rizzo} F.,  {Vegetti} S.,  {Powell} D.,  {Fraternali} F.,  {McKean} J.~P.,
  {Stacey} H.~R.,   {White} S.~D.~M.,  2020, \mn@doi [\nat]
  {10.1038/s41586-020-2572-6}, \href
  {https://ui.adsabs.harvard.edu/abs/2020Natur.584..201R} {584, 201}

\bibitem[\protect\citeauthoryear{{Roseboom} et~al.,}{{Roseboom}
  et~al.}{2010}]{Roseboom2010}
{Roseboom} I.~G.,  et~al., 2010, \mn@doi [\mnras]
  {10.1111/j.1365-2966.2010.17634.x}, \href
  {https://ui.adsabs.harvard.edu/abs/2010MNRAS.409...48R} {409, 48}

\bibitem[\protect\citeauthoryear{{Rowan-Robinson} et~al.,}{{Rowan-Robinson}
  et~al.}{2016}]{RowanRobinson2016}
{Rowan-Robinson} M.,  et~al., 2016, \mn@doi [\mnras] {10.1093/mnras/stw1169},
  \href {https://ui.adsabs.harvard.edu/abs/2016MNRAS.461.1100R} {461, 1100}

\bibitem[\protect\citeauthoryear{{Rybak}, {McKean}, {Vegetti}, {Andreani}  \&
  {White}}{{Rybak} et~al.}{2015a}]{Rybak2015}
{Rybak} M.,  {McKean} J.~P.,  {Vegetti} S.,  {Andreani} P.,   {White} S.~D.~M.,
   2015a, \mn@doi [\mnras] {10.1093/mnrasl/slv058}, \href
  {https://ui.adsabs.harvard.edu/abs/2015MNRAS.451L..40R} {451, L40}

\bibitem[\protect\citeauthoryear{{Rybak}, {McKean}, {Vegetti}, {Andreani}  \&
  {White}}{{Rybak} et~al.}{2015b}]{Rybak2015MNRAS.451L..40R}
{Rybak} M.,  {McKean} J.~P.,  {Vegetti} S.,  {Andreani} P.,   {White} S.~D.~M.,
   2015b, \mn@doi [\mnras] {10.1093/mnrasl/slv058}, \href
  {https://ui.adsabs.harvard.edu/abs/2015MNRAS.451L..40R} {451, L40}

\bibitem[\protect\citeauthoryear{{Scudder}, {Oliver}, {Hurley}, {Griffin},
  {Sargent}, {Scott}, {Wang}  \& {Wardlow}}{{Scudder}
  et~al.}{2016}]{Scudder2016MNRAS.460.1119S}
{Scudder} J.~M.,  {Oliver} S.,  {Hurley} P.~D.,  {Griffin} M.,  {Sargent}
  M.~T.,  {Scott} D.,  {Wang} L.,   {Wardlow} J.~L.,  2016, \mn@doi [\mnras]
  {10.1093/mnras/stw1044}, \href
  {https://ui.adsabs.harvard.edu/abs/2016MNRAS.460.1119S} {460, 1119}

\bibitem[\protect\citeauthoryear{{Serjeant}}{{Serjeant}}{2012}]{serjeant2012}
{Serjeant} S.,  2012, \mn@doi [\mnras] {10.1111/j.1365-2966.2012.20761.x},
  \href {https://ui.adsabs.harvard.edu/abs/2012MNRAS.424.2429S} {424, 2429}

\bibitem[\protect\citeauthoryear{{Shirley} et~al.,}{{Shirley}
  et~al.}{2021}]{Shirley2021}
{Shirley} R.,  et~al., 2021, \mn@doi [\mnras] {10.1093/mnras/stab1526}, \href
  {https://ui.adsabs.harvard.edu/abs/2021MNRAS.507..129S} {507, 129}

\bibitem[\protect\citeauthoryear{{Smail}, {Ivison}  \& {Blain}}{{Smail}
  et~al.}{1997}]{Smail1997}
{Smail} I.,  {Ivison} R.~J.,   {Blain} A.~W.,  1997, \mn@doi [\apjl]
  {10.1086/311017}, \href
  {https://ui.adsabs.harvard.edu/abs/1997ApJ...490L...5S} {490, L5}

\bibitem[\protect\citeauthoryear{{Spilker}, {Bezanson}, {Marrone}, {Weiner},
  {Whitaker}  \& {Williams}}{{Spilker} et~al.}{2016}]{Spilker2016}
{Spilker} J.~S.,  {Bezanson} R.,  {Marrone} D.~P.,  {Weiner} B.~J.,  {Whitaker}
  K.~E.,   {Williams} C.~C.,  2016, \mn@doi [\apj]
  {10.3847/0004-637X/832/1/19}, \href
  {https://ui.adsabs.harvard.edu/abs/2016ApJ...832...19S} {832, 19}

\bibitem[\protect\citeauthoryear{{Sutherland} \& {Saunders}}{{Sutherland} \&
  {Saunders}}{1992}]{Sutherland1992}
{Sutherland} W.,  {Saunders} W.,  1992, \mn@doi [\mnras]
  {10.1093/mnras/259.3.413}, \href
  {https://ui.adsabs.harvard.edu/abs/1992MNRAS.259..413S} {259, 413}

\bibitem[\protect\citeauthoryear{{Tamura}, {Oguri}, {Iono}, {Hatsukade},
  {Matsuda}  \& {Hayashi}}{{Tamura} et~al.}{2015}]{Tamura2015}
{Tamura} Y.,  {Oguri} M.,  {Iono} D.,  {Hatsukade} B.,  {Matsuda} Y.,
  {Hayashi} M.,  2015, \mn@doi [\pasj] {10.1093/pasj/psv040}, \href
  {https://ui.adsabs.harvard.edu/abs/2015PASJ...67...72T} {67, 72}

\bibitem[\protect\citeauthoryear{{Treu}}{{Treu}}{2010}]{Treu2010}
{Treu} T.,  2010, \mn@doi [\araa] {10.1146/annurev-astro-081309-130924}, \href
  {https://ui.adsabs.harvard.edu/abs/2010ARA&A..48...87T} {48, 87}

\bibitem[\protect\citeauthoryear{{Trombetti}, {Burigana}, {Bonato}, {Herranz},
  {De Zotti}, {Negrello}, {Galluzzi}  \& {Massardi}}{{Trombetti}
  et~al.}{2021}]{Trombetti2021}
{Trombetti} T.,  {Burigana} C.,  {Bonato} M.,  {Herranz} D.,  {De Zotti} G.,
  {Negrello} M.,  {Galluzzi} V.,   {Massardi} M.,  2021, \mn@doi [\aap]
  {10.1051/0004-6361/202140830}, \href
  {https://ui.adsabs.harvard.edu/abs/2021A&A...653A.151T} {653, A151}

\bibitem[\protect\citeauthoryear{{Valiante} et~al.,}{{Valiante}
  et~al.}{2016}]{valiante2016}
{Valiante} E.,  et~al., 2016, \mn@doi [\mnras] {10.1093/mnras/stw1806}, \href
  {https://ui.adsabs.harvard.edu/abs/2016MNRAS.462.3146V} {462, 3146}

\bibitem[\protect\citeauthoryear{{Vieira} et~al.,}{{Vieira}
  et~al.}{2013}]{Vieira2013}
{Vieira} J.~D.,  et~al., 2013, \mn@doi [\nat] {10.1038/nature12001}, \href
  {https://ui.adsabs.harvard.edu/abs/2013Natur.495..344V} {495, 344}

\bibitem[\protect\citeauthoryear{{Ward} et~al.,}{{Ward}
  et~al.}{2022}]{Ward2022}
{Ward} B.~A.,  et~al., 2022, \mn@doi [\mnras] {10.1093/mnras/stab3300}, \href
  {https://ui.adsabs.harvard.edu/abs/2022MNRAS.510.2261W} {510, 2261}

\bibitem[\protect\citeauthoryear{{Welch} et~al.,}{{Welch}
  et~al.}{2022}]{Welch2022}
{Welch} B.,  et~al., 2022, \mn@doi [\apjl] {10.3847/2041-8213/ac9d39}, \href
  {https://ui.adsabs.harvard.edu/abs/2022ApJ...940L...1W} {940, L1}

\bibitem[\protect\citeauthoryear{{Wright} et~al.,}{{Wright}
  et~al.}{2019}]{Wright2019}
{Wright} A.~H.,  et~al., 2019, \mn@doi [\aap] {10.1051/0004-6361/201834879},
  \href {https://ui.adsabs.harvard.edu/abs/2019A&A...632A..34W} {632, A34}

\bibitem[\protect\citeauthoryear{{Zavala} et~al.,}{{Zavala}
  et~al.}{2017}]{Zavala2017}
{Zavala} J.~A.,  et~al., 2017, \mn@doi [\mnras] {10.1093/mnras/stw2630}, \href
  {https://ui.adsabs.harvard.edu/abs/2017MNRAS.464.3369Z} {464, 3369}

\makeatother
\end{thebibliography}
